\documentclass[useAMS,usenatbib]{mn2e}

\usepackage{epsfig,color}
\usepackage{natbib}
\bibpunct{(}{)}{;}{a}{}{,}
\def\spose#1{\hbox to 0pt{#1\hss}}
\def\ltsimm{\mathrel{\spose{\lower 3pt\hbox{$\sim$}}
        \raise 2.0pt\hbox{$<$}}}
\def\gtsimm{\mathrel{\spose{\lower 3pt\hbox{$\sim$}}
        \raise 2.0pt\hbox{$>$}}}
\def\Mdot{\hbox{${\dot M}$}}

\def\km{{\rm\thinspace km}}
\def\cm{{\rm\thinspace cm}}

\def\s{{\rm\thinspace s}}
\def\yr{{\rm\thinspace yr}}
\def\g{{\rm\thinspace g}}
\def\kmps{\hbox{${\rm\km\s^{-1}\,}$}}

\def\erg{{\rm\thinspace erg}}

\def\Hz{{\rm\thinspace Hz}}

\def\ster{{\rm\thinspace ster}}
\def\ergps{\hbox{${\rm\erg\s^{-1}\,}$}}
\def\Rsol{\hbox{${\rm\thinspace R_{\odot}}$}}
\def\Msol{\hbox{${\rm\thinspace M_{\odot}}$}}

\def\Msolpyr{\hbox{${\rm\Msol\yr^{-1}\,}$}}
\def\pcm{\hbox{${\rm\cm^{-1}\,}$}}
\def\pcm2{\hbox{${\rm\cm^{-2}\,}$}}
\def\pcm3{\hbox{${\rm\cm^{-3}\,}$}}
\def\ergpscm3Hz{\hbox{${\rm\ergps\cm^{-3}\Hz^{-1}\,}$}}
\def\ergpscm3Hzster{\hbox{${\rm\ergps\cm^{-3}\Hz^{-1}\ster^{-1}\,}$}}
\def\gpcm3{\hbox{${\rm\g\cm^{-3}\,}$}}
\def\ergpcm2{\hbox{${\rm\erg\cm^{-2}\,}$}}
\def\ergpcm3{\hbox{${\rm\erg\cm^{-3}\,}$}}
\def\phpscm2{\hbox{${\rm photons\s^{-1}\cm^{-2}\,}$}}

\title[X-ray emission models of colliding wind binaries]{3D models of radiatively driven colliding winds in massive O+O star binaries - III. Thermal X-ray emission} 
\author[J.~M.~Pittard and E.~R.~Parkin] {J.~M.~Pittard\thanks{E-mail:
jmp@ast.leeds.ac.uk} and E.~R.~Parkin\\School of Physics and Astronomy, The
University of Leeds, Leeds LS2 9JT, UK\\ }

\begin{document}

\date{Accepted 2009 September 23.  Received 2009 September 23; in original form
2009 August 4}

\pagerange{\pageref{firstpage}--\pageref{lastpage}} \pubyear{2009}

\maketitle

\label{firstpage}

\begin{abstract}
The X-ray emission from the wind-wind collision in short-period
massive O+O-star binaries is investigated. The emission is calculated
from three-dimensional hydrodynamical models which incorporate
gravity, the driving of the winds, orbital motion of the stars, and
radiative cooling of the shocked plasma.  Changes in the amount of
stellar occultation and circumstellar attenuation introduce
phase-dependent X-ray variability in systems with circular orbits,
while strong variations in the intrinsic emission also occur in
systems with eccentric orbits. The X-ray emission in eccentric systems
can display strong hysteresis, with the emission softer after
periastron than at corresponding orbital phases prior to periastron,
reflecting the physical state of the shocked plasma at these times.

Our simulated X-ray lightcurves bear many similarities to observed
lightcurves. In systems with circular orbits the lightcurves show two
minima per orbit, which are identical (although not symmetric) if the
winds are identical. The maxima in the lightcurves are produced near
quadrature, with a phase delay introduced due to the aberration and
curvature of the wind collision region. Circular systems with unequal
winds produce minima of different depths and duration. In systems with
eccentric orbits the maxima in the lightcurves may show a very sharp
peak (depending on the orientation of the observer), followed by a
precipitous drop due to absorption and/or cooling. We show that the
rise to maximum does not necessarily follow a $1/d_{\rm sep}$ law.
Our models further demonstrate that the effective circumstellar column
can be highly energy dependent. Therefore, spectral fits which assume
energy independent column(s) are overly simplified and may compromise
the interpretation of observed data.

To better understand observational analyzes of such systems we apply
{\em Chandra} and {\em Suzaku} response files, plus poisson noise, to
the spectra calculated from our simulations and fit these using
standard {\sc XSPEC} models.  We find that the recovered temperatures
from two or three-temperature mekal fits are comparable to those from
fits to the emission from real systems with similar stellar and
orbital parameters/nature. We also find that when the global abundance
is thawed in the spectral fits, sub-solar values are exclusively
returned, despite the calculations using solar values as input. This
highlights the problem of fitting oversimplified models to data, and
of course is of wider significance than just the work presented here.

Further insight into the nature of the stellar winds and the WCR in
particular systems will require dedicated hydrodynamical modelling,
the results of which will follow in due course.
\end{abstract}


\section{Introduction}
\label{sec:intro}
The shock-heated plasma in the wind-wind collision region (WCR) of
massive stellar binaries can produce copious X-ray emission.  The
emission often displays orbital variability, which can result from
changes to the occultation of the emitting region by the stars, to the
attenuation through the stellar winds, and to the separation of the
stars \citep[e.g.][]{Rauw:2002,Schild:2004,Sana:2004,DeBecker:2006,Linder:2006,
Pollock:2006,Naze:2007,Hamaguchi:2007,Sana:2008}. In wider systems, the post-shock
plasma may exhibit signs of non-equilibrium ionization
\citep{Pollock:2005}, and non-equilibrium electron and ion
temperatures \citep{Zhekov:2000}.

The X-ray emission is a useful probe of the underlying wind
parameters. The hardness of the emission is related to the post-shock
temperatures within the WCR, which in turn depends on the pre-shock
wind speed. The X-ray brightness depends on the pre-shock density of
the winds, while the absorption of soft X-rays through the
circumstellar environment depends on the integrated density along
sight lines to the WCR
\citep*[e.g.][]{Stevens:1992,Stevens:1996,Pittard:1997,Pittard:1998b,Pittard:2002,Parkin:2008,Parkin:2009}. Hence both the brightness
and the degree of absorption provide information about the stellar
mass-loss rates. Although the X-ray emissivity is proportional to the
square of the density, inhomogeneties can be rapidly smoothed out
within adiabatic WCRs: thus the resulting X-ray emission may be
relatively insensitive to the presence of clumps
\citep{Pittard:2007}. Since mass-loss rate estimates are often
uncertain due to unknown wind clumping factors, an insensitivity to
clumping potentially allows the X-ray emission from CWBs to provide
accurate determinations of stellar mass-loss rates.
Recent observations in the UV have highlighted the uncertainty
which still exists, with mass-loss rate estimates differing by
factors of up to 100 with respect to other methods
\citep*[e.g.][]{Bouret:2005,Martins:2005b,Fullerton:2006}. A recent
review of the current situation can be found in \citet*{Puls:2008}.

The WCR can also be a site of particle acceleration. The energetic
particles produce non-thermal radio emission via the synchrotron
process \citep[e.g.][]{Dougherty:2003,Pittard:2006}, and non-thermal
X-ray and $\gamma$-ray emission from inverse Compton cooling, neutral
pion decay, and relativistic bremsstrahlung
\citep{Pittard:2006b,Leyder:2008}.  The non-thermal radio emission can
sometimes be spatially resolved
\citep[e.g.][]{Williams:1997,Dougherty:2000b,Dougherty:2005}, and
can also undergo dramatic variations in flux
\citep[e.g.][]{Williams:1992,White:1995,Rauw:2002b,DeBecker:2004c,Blomme:2005,Blomme:2007,vanloo:2008}. If
the particle acceleration efficiency is sufficiently high, the thermal
structure of the WCR may be affected, resulting in softer X-ray
emission as the plasma becomes cooler and denser. In this way the
characteristics of the thermal X-ray emission may also constrain the
efficiency of particle acceleration.

Models of the X-ray emission from colliding wind systems based on
hydrodynamical simulations have, to date, been almost entirely
performed in two-dimensions, with an underlying assumption of
axissymmetry. While this approach is perfectly reasonable for wide
systems with long orbital periods, axissymmetry is a poor assumption
in shorter period systems where orbital effects become
important. Though three-dimensional hydrodynamical simulations have
been presented by \citet{Walder:1998} and \citet*{Lemaster:2007}, these
works also assumed that the winds were instantaneously accelerated to
their terminal velocities. In reality, the winds in short-period
systems collide prior to reaching their terminal velocities, so
realistic simulations must also account for the acceleration of each
wind. In a new advance, three-dimensional models with radiatively
driven winds were recently presented by \citet[][hereafter
Paper~I]{Pittard:2009a}.  In addition to the acceleration of the winds,
these models also account for orbital motion of the stars, gravity,
and cooling in the post-shock plasma.

In this work we examine the {\em thermal} X-ray properties of the WCR
from these models. We produce synthetic X-ray spectra and lightcurves,
and examine how these change with the viewing
angle. Section~\ref{sec:setup} describes details of the models and
summarizes the method of calculating the X-ray emission and
absorption. This section also notes the procedure adopted for folding
the theoretical spectra through the response files of current X-ray
observatories to simulate ``fake'' observations, which are
subsequently fit using standard analysis techniques.  We present our
results in Section~\ref{sec:results}. Comparisons to previous
numerical models and observations are made in
Sections~\ref{sec:discuss_models} and~\ref{sec:discuss}, respectively.
Section~\ref{sec:summary} summarizes and concludes this work.

\section{Details of the Calculations}
\label{sec:setup}
\subsection{The Numerical Models}
The X-ray calculations in this paper are based on the
three-dimensional hydrodynamical models described in Paper~I. The
models incorporate the radiative driving of the stellar winds (based
on the \citet{Castor:1975} formalism, with the finite disk correction
factor of \citet*{Pauldrach:1986}), gravity, orbital effects, and cooling. 
The models were not designed to be of particular systems.
Rather, the aim was to obtain a better understanding of how the nature of
the collision region depends on various key parameters. The models are
summarized in Tables~\ref{tab:models}
and~\ref{tab:stellar_params}.  The assumption of main sequence stars
minimizes the effects of tidal distortions, which are not
modelled. The winds are also assumed to be spherically
symmetric. Further details about the models can be found in Paper~I.

\begin{table*}
\begin{center}
\caption[]{Assumed binary parameters for the models calculated in
Paper~I. The semi-major axis is $34.26\;\Rsol$ in model cwb1,
$76.3\;\Rsol$ in models cwb2 and cwb3, and $55\;\Rsol$ in model cwb4.
$e$ is the orbital eccentricity and $\eta$ is the (terminal velocity)
momentum ratio of the winds. $v_{\rm orb}$ and $v_{\rm w}$ are the
orbital speeds of the stars and the preshock wind speeds along the
line of centres. $\chi$ is the ratio of the cooling time to the
characteristic flow time of the hot shocked plasma. $\chi \ltsimm 1$
indicates that the shocked gas rapidly cools, while $\chi \gtsimm 1$
indicates that the plasma in the WCR remains hot as it flows out of
the system. Larger values of the ratio $v_{\rm orb}/v_{\rm w}$ produce
a greater aberration angle, $\theta_{\rm ab}$, and tighter downstream
curvature, of the WCR.  The degree of downstream curvature of the WCR
in the orbital plane is given by $\alpha_{\rm coriolis}$, where the
curvature is assumed to trace an Archimedean spiral which in polar
coordinates is described by $r = \alpha_{\rm coriolis}\theta$. The
value of $\alpha_{\rm coriolis}$ corresponds to the approximate downstream
distance (in units of $d_{\rm sep}$) along the WCR for each radian of
arc it sweeps out in the orbital plane. Smaller values indicate
tighter curvature.  The leading and trailing arms of the WCR in model
cwb3 display differing degrees of curvature, so the value quoted for
this model is an average.  The pre-shock orbital and wind speeds in
model cwb3 are also different for each star/wind - the first (second)
value is for the primary (secondary) star/wind. The values of $\chi$,
$v_{\rm orb}/v_{\rm w}$, $\theta_{\rm ab}$ and $\alpha_{\rm coriolis}$
are phase dependent in model cwb4, because of its eccentric orbit -
values at periastron and apastron are quoted. The values for
$\alpha_{\rm coriolis}$ are calculated after comparing the orbital
speeds at periastron and apastron against those in models cwb1 and
cwb2, and represent the ``instantaneous'' curvature at these phases.}
\label{tab:models}
\begin{tabular}{lllllllllll}
\hline
\hline
Model & Stars & Period & $e$ & $\eta$ & $v_{\rm orb}$ & $v_{\rm w}$ & $\chi$ & $v_{\rm orb}/v_{\rm w}$ & $\theta_{\rm ab}$ & $\alpha_{\rm coriolis}$ \\
 & & (d) &  & & ($\kmps$) & ($\kmps$) & & &($^{\circ}$) & ($d_{\rm sep}\,{\rm rad}^{-1}$) \\
\hline
cwb1 & O6V+O6V & 3 & 0.0 & 1 & 290 & 730 & 0.34 & 0.40 & 17 & 3.5\\
cwb2 & O6V+O6V & 10 & 0.0 & 1 & 225 & 1630 & 19 & 0.14 & $3-4$ & 6.5\\
cwb3 & O6V+O8V & 10.74 & 0.0 & 0.4 & 152,208 & 1800,1270 & 28,14 & $0.16-0.084$ & $\sim 2$ & 4.5\\
cwb4 & O6V+O6V & 6.1 & 0.36 & 1 & $334-156$ & $710-1665$ & $0.34-19$ & $0.47-0.09$ & 21-4 & 3-10 \\
\hline
\end{tabular}
\end{center}
\end{table*}

In model cwb1 two identical O6V stars move around each other in a
circular orbit with a period of 3 days. The stellar separation is
$34.26\;\Rsol$, and each star has an orbital velocity $v_{\rm orb} =
290\;\kmps$. The thermal behaviour of the WCR can be described by the
ratio of the cooling time to the characteristic flow time of the hot
shocked plasma, $\chi\approx\frac{v_{8}^{4}d_{12}}{\Mdot_{-7}}$, where
$v_{8}$ is the pre-shock wind speed in units of $1000\kmps$, $d_{12}$
is the separation of the stars, and $\Mdot_{-7}$ is the stellar
mass-loss rate in units of $10^{-7}\Msolpyr$
\citep[c.f.][]{Stevens:1992}. In model cwb1, the WCR is highly
radiative ($\chi << 1$), and significantly distorted by orbital
effects, showing strong aberration and downstream curvature.  Model
cwb1 is similar to many real systems, including HD\,215835
\citep[DH\,Cep; see][and references therein]{Linder:2007}, HD\,165052
\citep{Arias:2002,Linder:2007}, and HD\,159176
\citep{DeBecker:2004b,Linder:2007}. All of these systems have near
identical main-sequence stars of spectral type O6$-$O7, and circular
or near-circular orbits with periods near 3 days.

In model cwb2 the orbital period is increased to 10 days, with the
stellar separation becoming $76.3\;\Rsol$.  The winds collide at
significantly higher speeds than in model cwb1, and the postshock gas
is largely adiabatic. The aberration and downstream curvature of the
WCR are both lessened compared to model cwb1. Model cwb2 is similar to
HD\,93161A, an O8V + O9V system with a circular orbit and an orbital
period of 8.566 days \citep{Naze:2005}, albeit with slightly more
massive stars and powerful winds. Another system which is not too
dissimilar in its properties is Plaskett's star
\citep{Linder:2006,Linder:2008}, though this object contains stars
which have evolved off the main sequence.

Model cwb3 examines the interaction of unequal winds in a hypothetical
O6V+O8V binary. The stars in this model have the same separation as
those in model cwb2. The primary wind collides at higher speed than
the secondary wind, and its postshock plasma is slightly more
adiabatic.

Model cwb4 investigates the effect of an eccentric orbit which takes
the stars through a separation of $34.26-76.3\;\Rsol$ (i.e.  the
separations of the stars in the circular orbits of models cwb1 and
cwb2). The WCR is radiative at periastron and adiabatic at apastron,
and the aberration and downstream curvature are phase dependent. A
surprising finding from Paper~I is that dense cold clumps formed in
the WCR at periastron persist near the apex of the WCR until almost
apastron. This is because the clumps have relatively high inertia, 
and flow out of the system much more slowly than the hotter gas which
streams past them. Some well-known O+O binaries with eccentric orbits
include (in order of increasing orbital period) HD\,152248 \citep[$e =
0.127$;][]{Sana:2004}, HD\,93205 \citep[$e=0.46$;][]{Morrell:2001},
HD\,93403 \citep[$e=0.234$;][]{Rauw:2002}, Cyg\,OB2\#8A
\citep[$e=0.24$;][]{DeBecker:2004,DeBecker:2006} and $\iota$\,Orionis
\citep[$e=0.764$;][]{Bagnuolo:2001}. 

\subsection{Modelling the X-ray emission and absorption}
To calculate the X-ray emission we read our hydrodynamical models into
a radiative transfer ray-tracing code, and calculate appropriate
emission and absorption coefficients for each cell using the
temperature and density values. A synthetic image on the plane of the
sky is then generated by solving the radiative transfer equation along
suitable lines of sight through the grid.  Since non-equilibrium
effects are small in short period O+O systems (see Paper~I), the X-ray
emissivity is calculated using the mekal emission code \citep[][and
references therein]{Mewe:1995} for optically thin thermal plasma in
collisional ionization equilibrium. Solar abundances
\citep{Anders:1989} are assummed throughout this work. The emissivity
is stored in look-up tables containing 200 logarithmic energy bins
between $0.1-10\;$keV, and 91 logarithmic temperature bins between
$10^{4}-10^{9}\;$K. Line emission dominates the cooling at
temperatures below $10^{7}\;$K, with thermal bremsstrahlung dominating
at higher temperatures. The hydrodynamical grid is large enough to
capture the majority of the X-ray emission from each of the models.

The main contributors to the absorption of keV X-rays are the K shells
of the CNO elements. The photoelectric absorption is calculated using
{\em Cloudy} \citep{Ferland:2000}. The opacity is stored in look-up
tables containing 26 temperatures between $10^{4}-10^{9}\;$K. As in
previous works
\citep*[e.g.][]{Luo:1990,Stevens:1992,Myasnikov:1993,Pittard:1997},
electron scattering is neglected. Electron scattering becomes
important once the optical depth to this process nears unity,
i.e. when $\tau_{\rm e} = N_{\rm e}\sigma_{\rm e} \gtsimm 1$, where
$N_{\rm e}$ is the column density of free electrons along a line of
sight and $\sigma_{\rm e}$ is the Thomson cross-section. In the
ionized winds, the proton and electron column densities are
approximately equal. In our models, $\tau_{\rm e} < 1$ is indeed
satisified. For example, in the dense circumstellar environment of
model cwb1, an observer viewing the system pole on ($i=0^{\circ}$)
sees an average ``effective'' hydrogen column density to high energy
($2-10\,$keV) X-rays of $\approx 2\times10^{22}\,{\rm cm^{-2}}$ (see
Fig.~\ref{fig:cwb123_xray_circum}(a) and
Section~\ref{sec:cwb1_spectra} for further details). Since occulation
is minimal, the ``effective'' column density in this case reflects the
true column density of the circumstellar environment. The electron
scattering optical depth is then $\tau_{\rm e} \approx 0.13$.  Note
that the higher ``effective'' column densities shown in
Fig.~\ref{fig:cwb123_xray_circum}(a) for an observer in the orbital
plane ($i=90^{\circ}$) at phase 0.0 are instead a reflection of the
occultation that takes place at this time, and do not indicate that
electron scattering becomes optically thick (see
Section~\ref{sec:cwb1_spectra} for further details). 
Electron scattering will, however, be important in systems with denser winds,
where very high column densities can be reached. Such systems,
include, for example, the supermassive system $\eta$\,Car \citep[see,
e.g.,][]{Parkin:2009}. The likely effect is that abrupt changes in the
emission (e.g. in lightcurves and spectra) will be somewhat
smoothed/blurred, though we leave a study of this effect to
future work.

The present calculations also have an interstellar absorption column
of $10^{21}\;{\rm cm^{-2}}$ added to them, and each model is assumed
to be at a distance of 1\,kpc from an observer.  The X-ray
spectra/lightcurves were calculated from a single ``frame'' (i.e.
changing the orientation) for the circular orbit models (cwb1, cwb2,
and cwb3), and from a sequence of snapshots for the eccentric model
(cwb4). The lightcurve for $i=0^{\circ}$ is invariant for models cwb1,
cwb2, and cwb3.

\subsection{Generating and analyzing ``fake'' spectra}
In the following section we ``observe'' the theoretical spectra
generated from our ray-tracing code with the {\em Chandra} and {\em
Suzaku} X-ray observatories. This involves convolving the theoretical
spectra with the energy response and effective area of these
telescopes, to generate spectra in counts/energy bin. Counting
statistics are included in this process. The resulting ``fake''
spectra are then analyzed using {\sc XSPEC}, and fitted with standard
spectral models, in an analogous manner to the analysis of real data
(the only difference is that a background component does not need to
be subtracted). The aim is to study how the fit parameters compare
with those from the analysis of real data, and how they compare to
what is known about the theoretical input spectra. This type of
analysis remains very novel, having been applied to colliding wind 
binaries only by \citet{Stevens:1996}, \citet{Pittard:1997},
\citet{Zhekov:2000}, and \citet{Pittard:2002}.

The majority of our analysis is concentrated on simulated {\em Suzaku}
XIS spectra. To generate these we used the XIS0 ancillary response
file (ARF) and redistribution matrix file (RMF) for an on-axis point
source downloaded from the HEASARC
website\footnote{http://heasarc.gsfc.nasa.gov/docs/heasarc/caldb/data/\\suzaku/xis/index.html}. Although
these are old (2006) calibrations, they are fine for our purpose,
which is to investigate the values and variation of the best-fit
parameters, and the corresponding fluxes of the best-fit models.  A
small number of simulated {\em Chandra} ACIS-I spectra were also
computed. These used the Cycle 11 ACIS-I aimpoint ARF and RMF,
downloaded from the {\em Chandra}
website\footnote{http://cxc.harvard.edu/caldb/prop\_plan/imaging/index.html}.

The spectra were binned with the {\sc FTOOLS} task {\em grppha} so
that each energy bin contained a minimum of 20 counts. The ``fake''
spectra were fitted using {\sc XSPEC} version 12.5.0ac, distributed
with {\sc HEASoft6.6.3}.  Since an ISM column of $10^{21}\,{\rm
cm^{-2}}$ was added to our theoretical spectra, we force the absorbing
column to each model component to be at least as large. However, we
note that if this restriction is relaxed, the best fit often returned
lower columns. The theoretical spectra were generated using
emissivities from the mekal thermal emission code, so for
consistency we also fit the data in {\sc XSPEC} using the mekal thermal
model.

\begin{table}
\begin{center}
\caption[]{Assumed stellar parameters for the models.}
\label{tab:stellar_params}
\begin{tabular}{lllll}
\hline
\hline
Parameter/Star & O6V & O8V \\
\hline
Mass ($\Msol$) & 30 & 22 \\
Radius ($\Rsol$) & 10 & 8.5 \\
Effective temperature (K) & 38000 & 34000 \\ 
Mass-loss rate ($\Msolpyr$) & $2 \times 10^{-7}$ & $10^{-7}$ \\
Terminal wind speed ($\kmps$) & 2500 & 2000 \\
\hline
\end{tabular}
\end{center}
\end{table}

\section{Results}
\label{sec:results}

\begin{figure*}
\psfig{figure=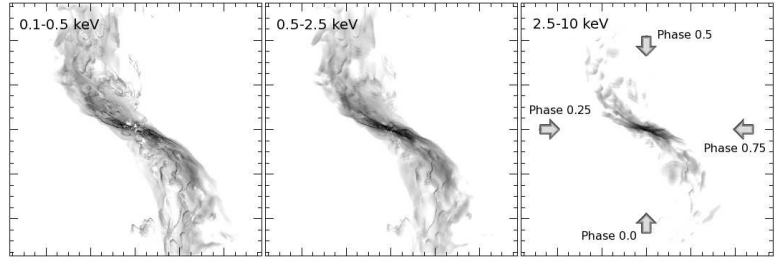,width=10.2cm}
\caption[]{Broad-band intensity images from model cwb1 at $i=0^{\circ}$. From
left to right the images are in the $0.1-0.5$\,keV, $0.5-2.5$\,keV, and $2.5-10$\,keV bands.
The grayscale covers 4 orders of magnitude in brightness, with black corresponding to a
maximum intensity of $10^{7}\,{\rm erg\,cm^{-2}\,s^{-1}\,keV^{-1}\,ster^{-1}}$ in the
left and right panels, and $10^{8}\,{\rm erg\,cm^{-2}\,s^{-1}\,keV^{-1}\,ster^{-1}}$ in
the central panel. The major ticks on each axis mark out 0.2\,mas. The stars are at
$\pm0.08$\,mas north and south of the image centre. The arrows mark the direction in
which an observer looks into the system at the indicated orbital phases. We also
define an azimuthal viewing angle, $\phi$, which increases anti-clockwise from the
bottom of the image, so that at phase 0.0 $\phi=0^{\circ}$. $\phi=90^{\circ}$ 
corresponds to phase 0.75, and $\phi=180^{\circ}$ and $\phi=270^{\circ}$ correspond
to phases 0.5 and 0.25, respectively.}
\label{fig:cwb1_xray_images1}
\end{figure*}

\begin{figure*}
\psfig{figure=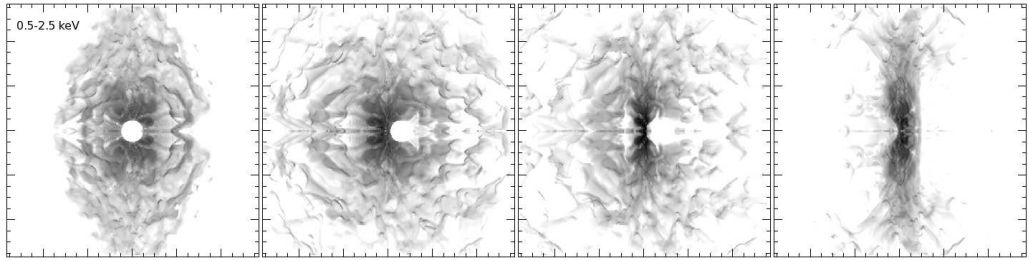,width=13.6cm}
\caption[]{Broad-band 0.5-2.5\,keV intensity images from model cwb1 at
$i=90^{\circ}$.  From left to right the phase of the observation increases
from 0.0 (conjunction, $\phi = 0^{\circ}$), to 0.125 ($\phi=315^{\circ}$),
to 0.25 (quadrature, $\phi=270^{\circ}$), to 0.375 ($\phi=225^{\circ}$). The
grayscale covers 4 orders of magnitude in brightness, with black
corresponding to a maximum intensity of $10^{8}\,{\rm
erg\,cm^{-2}\,s^{-1}\,keV^{-1}\,ster^{-1}}$. The major ticks on each
axis mark out 0.2\,mas.}
\label{fig:cwb1_xray_images2}
\end{figure*}

\subsection{Model cwb1}
\label{sec:xray_cwb1}
\subsubsection{Images}
Fig.~\ref{fig:cwb1_xray_images1} shows broad-band images from model
cwb1 for an observer located directly above the orbital plane ($i =
0^{\circ}$). The stars are oriented north-south in these images
\citep[cf. the images of the thermal radio emission in][hereafter
Paper~II]{Pittard:2009b}, and the orbital induced aberration and
downstream curvature of the WCR is clearly visible. The emission
morphology reflects the underlying structure and clumpiness of the
WCR, resulting from the powerful dynamical instabilities present in
this system. A detailed discussion of the hydrodynamics can be found
in Paper~I. The projected emission from different inhomogeneities
merges together near the apex of the WCR, but individual clumps and
bowshocks can be identified further downstream. It is clear that a
small amount of emission, particularly at the lowest energies, is not
captured due to the finite extent of the hydrodynamical grid used in
the model, but this loss should not be significant. The hard
($2-10$\,keV) emission predominantly arises from the apex of the
WCR. Although there are regions of hot gas further downstream (see
Paper~I), the density there is too low for these regions to contribute
significantly to the emission. The spatial scale of the emission
is far too small to be resolved with current X-ray telescopes: WR\,147
is likely to remain the only system with a spatially resolved WCR
\citep{Pittard:2002c} for some time to come.

Fig.~\ref{fig:cwb1_xray_images2} shows broad-band images from model
cwb1 for an observer located in the orbital plane ($i =
90^{\circ}$). The clumpy nature of the WCR and the bowshocks around
some of the denser regions are visible. At particular phases/viewing
angles the emission from bright parts of the WCR is occulted by the
foreground star.  Additional foreground emission is sometimes seen in
front of the stellar disc at these moments.

\begin{figure*}
\psfig{figure=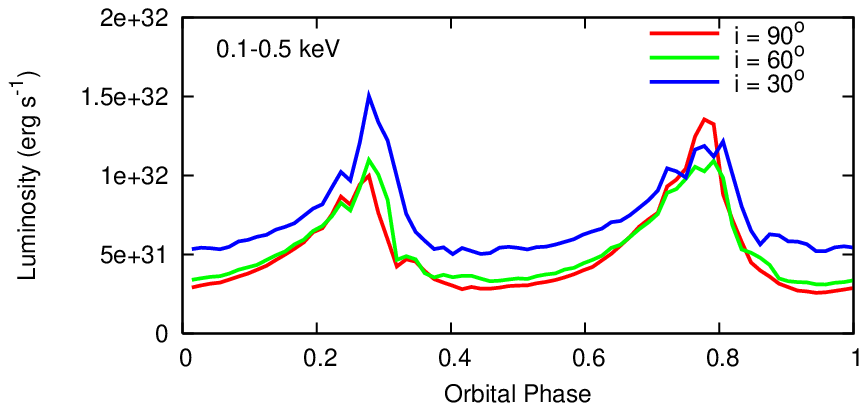,width=5.67cm}
\psfig{figure=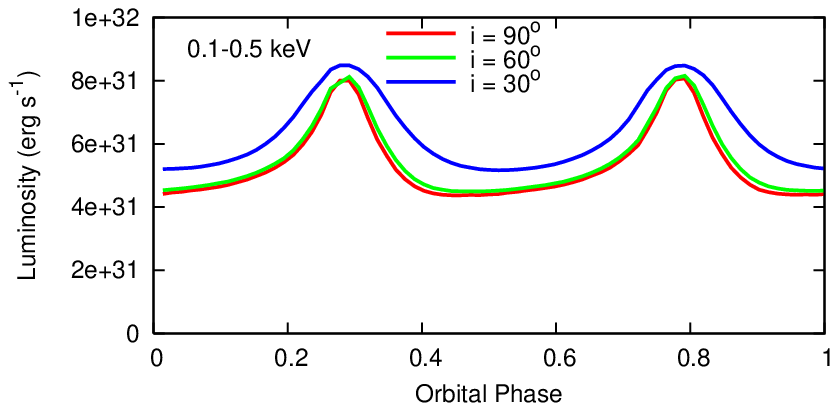,width=5.67cm}
\psfig{figure=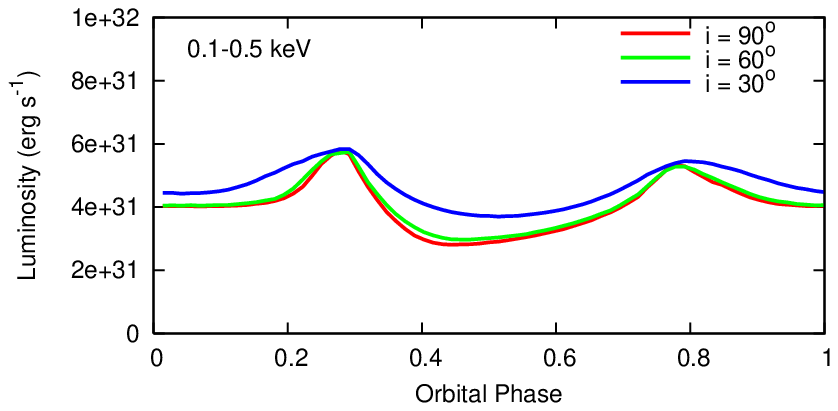,width=5.67cm}
\psfig{figure=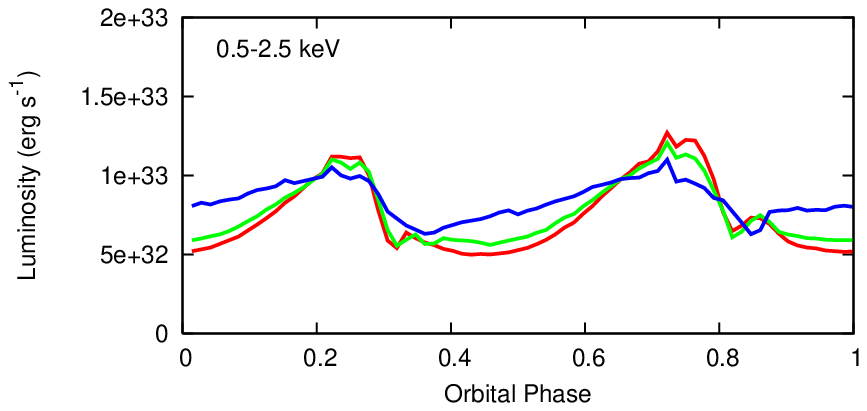,width=5.67cm}
\psfig{figure=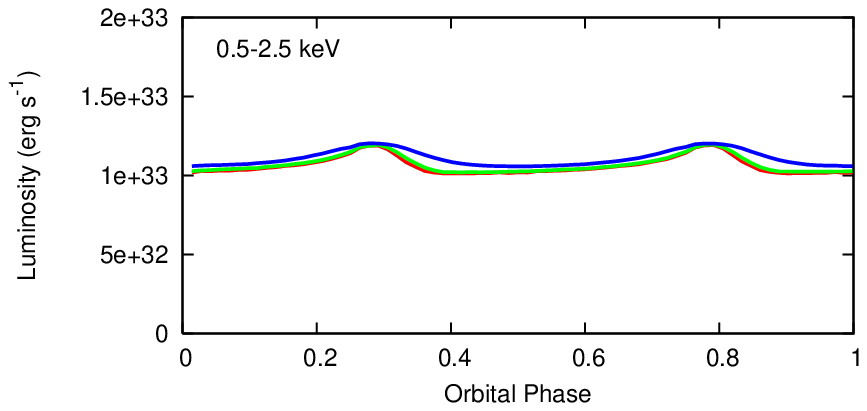,width=5.67cm}
\psfig{figure=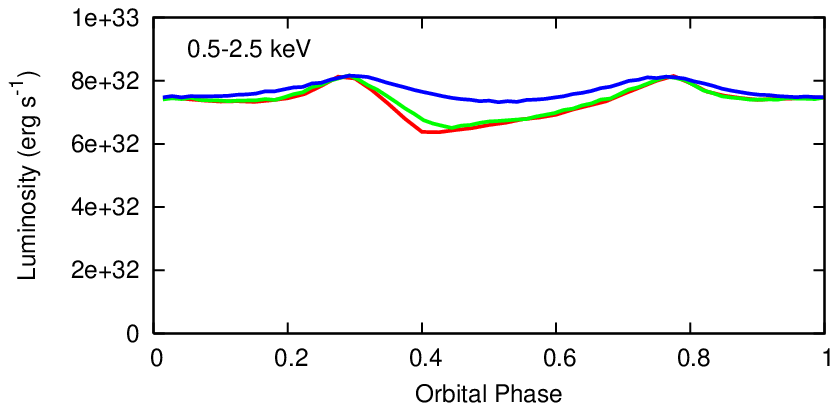,width=5.67cm}
\psfig{figure=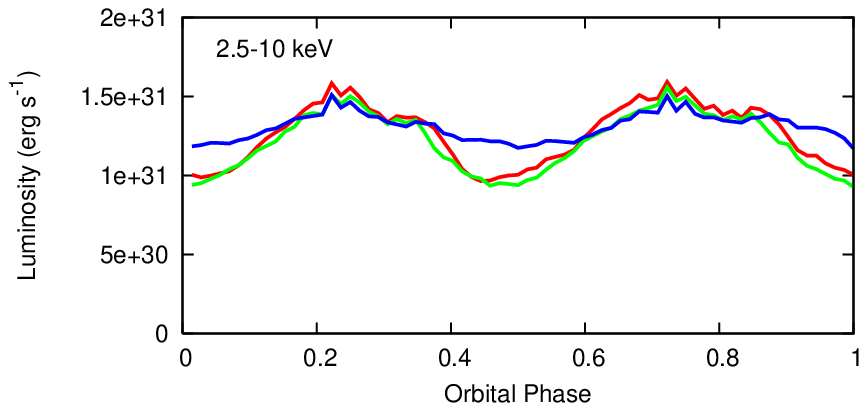,width=5.67cm}
\psfig{figure=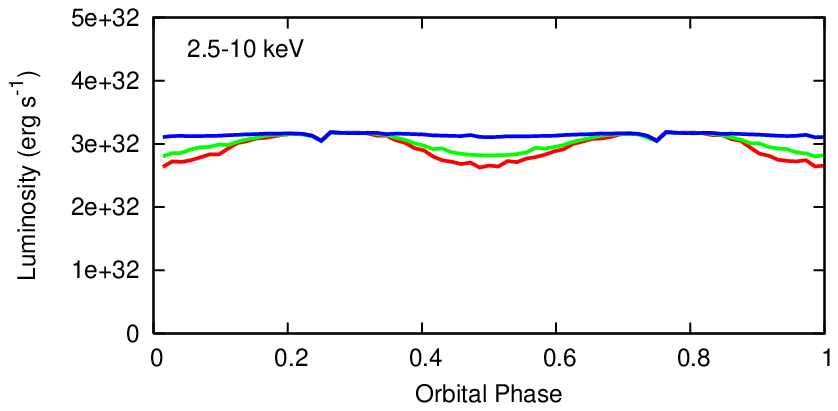,width=5.67cm}
\psfig{figure=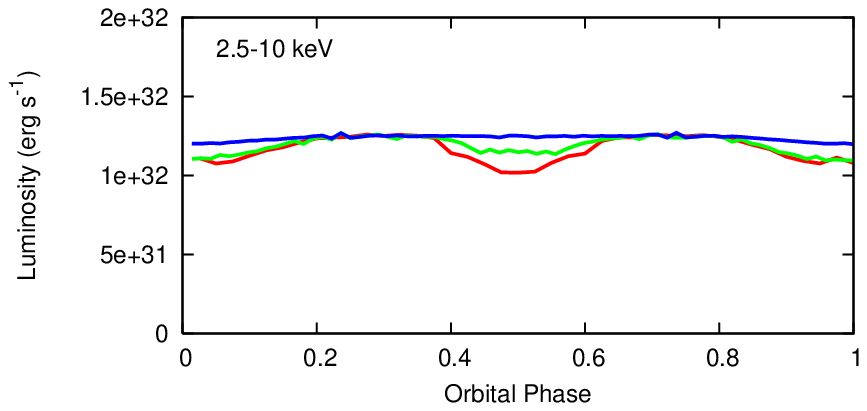,width=5.67cm}
\caption[]{X-ray lightcurves for models cwb1 (left), cwb2 (middle), and cwb3
(right), calculated over the energy bands
$0.1-0.5\;$keV (top), $0.5-2.5\;$keV (middle), and $2.5-10\;$keV (bottom)
for inclination angles $i = 30^{\circ}, 60^{\circ}$ and $90^{\circ}$.
In all cases, the observer is located along a direction vector
specified by $\phi=0^{\circ}$ (i.e. the longitude of periastron, 
$\omega=270^{\circ}$). The stars are at conjunction at phases 0.0 and 0.5,
and quadrature at phases 0.25 and 0.75.}
\label{fig:cwb123_xray_lc}
\end{figure*}

\subsubsection{Lightcurves}
Lightcurves from model cwb1, computed over the energy bands
$0.1-0.5\;$keV, $0.5-2.5\;$keV, and $2.5-10\;$keV, are shown in the
left column of Fig.~\ref{fig:cwb123_xray_lc}. The stars pass in front
of each other at phases 0.0 and 0.5, and are at quadrature at phases
0.25 and 0.75.  As the orbit is circular, the intrinsic emission is
constant, so the variations displayed in the lightcurves in
Fig.~\ref{fig:cwb123_xray_lc} are entirely due to changes in the
occultation and circumstellar absorption as a function of
phase. If there were no orbital induced effects on the WCR (i.e. no
aberration or curvature of the WCR), the lightcurves would display
dual symmetry about phases corresponding to both quadrature (0.25,
0.75) and conjunction (0, 0.5) of the stars to the line of sight
\citep*[c.f.][]{Pittard:1997,Antokhin:2004}, because of the equal
winds and constant stellar separation. When orbital effects are
included, the symmetry about quadrature is broken, and the lightcurves
instead are expected to show a double periodicity in the case of
identical winds.  However, close examination of
Fig.~\ref{fig:cwb123_xray_lc} reveals that in fact this dual
periodicty is also broken (note that the peaks in the $0.1-0.5\;$keV
lightcurve have different heights).  Clearly, the dynamical
instabilities which form in the WCR develop independently in each arm,
and break this symmetry too.

The variation of flux with orbital phase is largest in all lightcurves
when the inclination angle $i=90^{\circ}$, and decreases with
decreasing $i$. The amplitude of variation is also largest in the
softest band. Both of these findings are expected: the soft emission
is more easily absorbed by the stellar winds, while lines of sight to
the WCR pass, on average, through a greater attenuation column, and
there is also greater occultation of the emission region by the stars,
when the observer is in the orbital plane.

Minima in the lightcurves occur near phases 0.0 and 0.5, when the
emission from the WCR suffers the greatest reduction by stellar
occultation and wind absorption. However, close examination reveals
that the minima actually occur slightly before each conjunction. This
reflects the aberration of the WCR. In model cwb1 the
abberation angle is $\approx 17^{\circ}$, which corresponds to 0.05 in
orbital phase, and is therefore similar to the observed lead.

Short, sharp dips are also seen in the lightcurves near orbital phases
0.32 and 0.82 (being most visible in the $0.5-2.5$\,keV lightcurve for
$i=90^{\circ}$). This is due to absorption from the thin dense
layer of cooled post-shock gas (see Paper~I). The dips are similar to
those seen in Fig.~15 of \citet{Antokhin:2004}, but are broader and
less obvious due to the orbital-induced curvature of the WCR. They
also display a phase lead consistent with the phase lead of the main minima.

%The observed X-ray luminosity is $5.79\times10^{32}\,\ergps$,
%$1.21\times10^{33}$, and $5.73\times10^{32}$ at viewing angles of
%$i,\phi = 90^{\circ},0^{\circ}$, $90^{\circ},90^{\circ}$, and
%$0,0$ (pole-on), giving $L_{\rm x}/L_{\rm bol} =  4\times10^{-7}$,
%$8.5\times10^{-7}$, and $4\times10^{-7}$, respectively.
The ISM corrected $0.5-10$\,keV luminosity is $6.75\times10^{32}\,\ergps$,
$1.41\times10^{33}\,\ergps$, and $6.54\times10^{32}\,\ergps$ at viewing angles of
$(i,\phi) = (90^{\circ},0^{\circ})$, $(90^{\circ},90^{\circ})$, and
$(0,0)$ (pole-on), giving $L_{\rm x}/L_{\rm bol} =  4.7\times10^{-7}$,
$9.9\times10^{-7}$, and $4.6\times10^{-7}$, respectively.
%The intrinsic 0.5-10\,keV luminosity is $3.11\times10^{33}\,\ergps$, which 
%gives $L_{\rm x}/L_{\rm bol} =  2.2\times10^{-6}$. 
These values are all significantly above the scaling law 
(log\,$L_{\rm x}/L_{\rm bol}$)$=-6.912\pm0.153$) 
%Lx = 1.22e-7 Lbol
determined by \citet{Sana:2004}, and are thus 
indicative of a binary system with strong colliding winds emission. 

\begin{figure*}
\psfig{figure=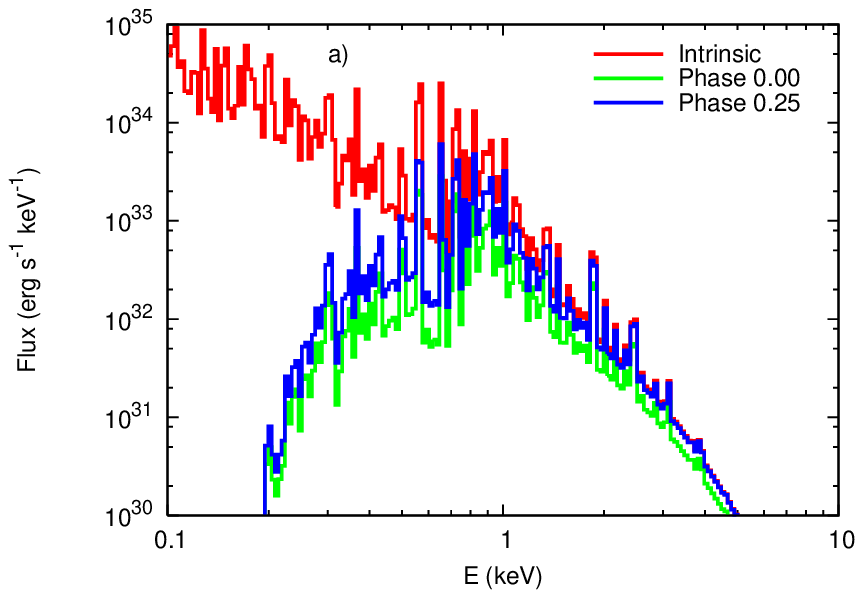,width=5.67cm}
\psfig{figure=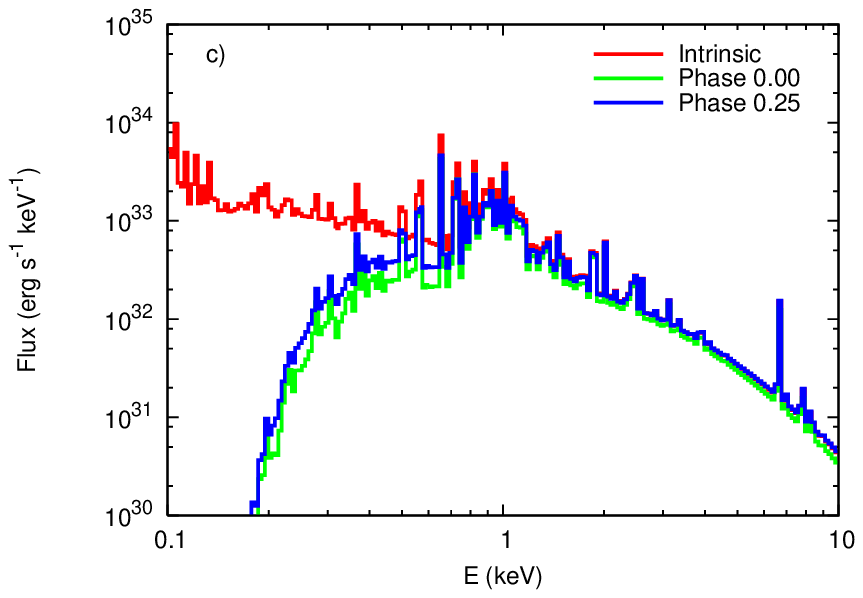,width=5.67cm}
\psfig{figure=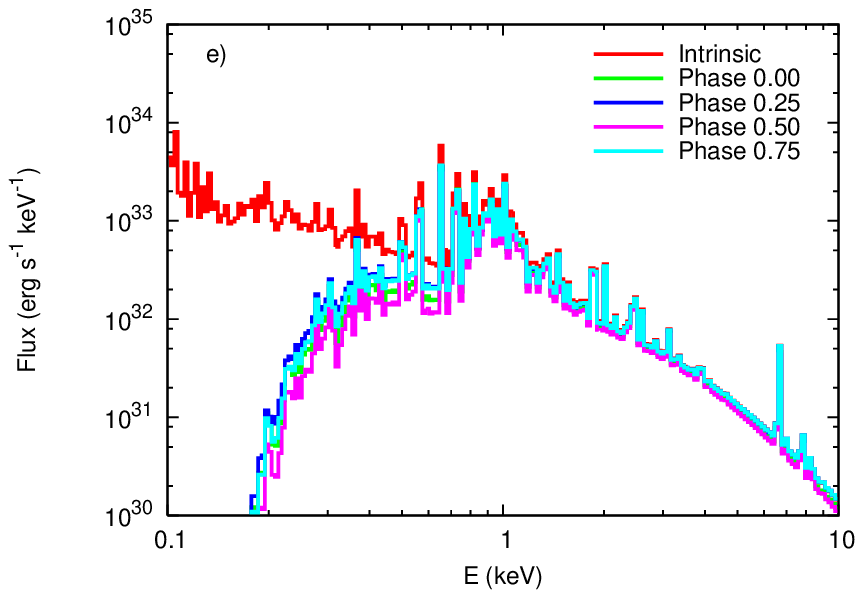,width=5.67cm}
\psfig{figure=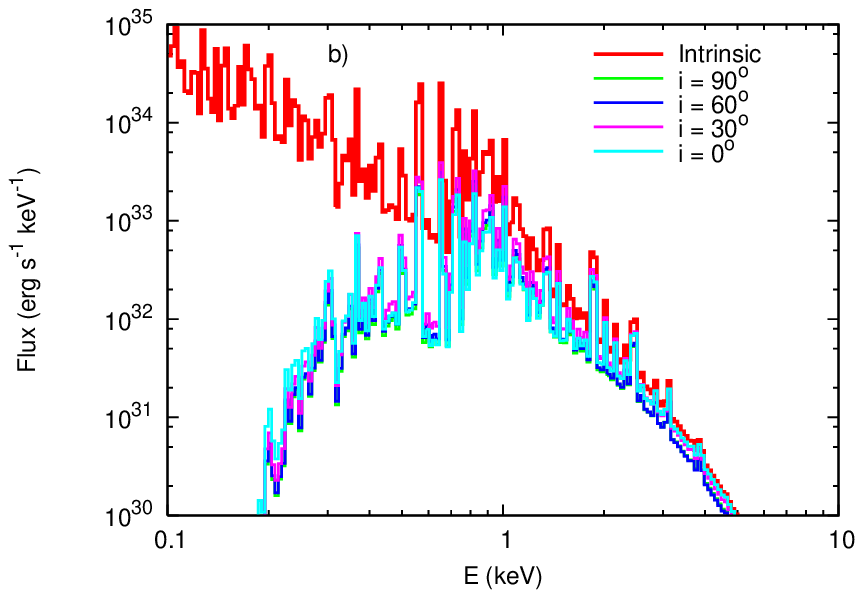,width=5.67cm}
\psfig{figure=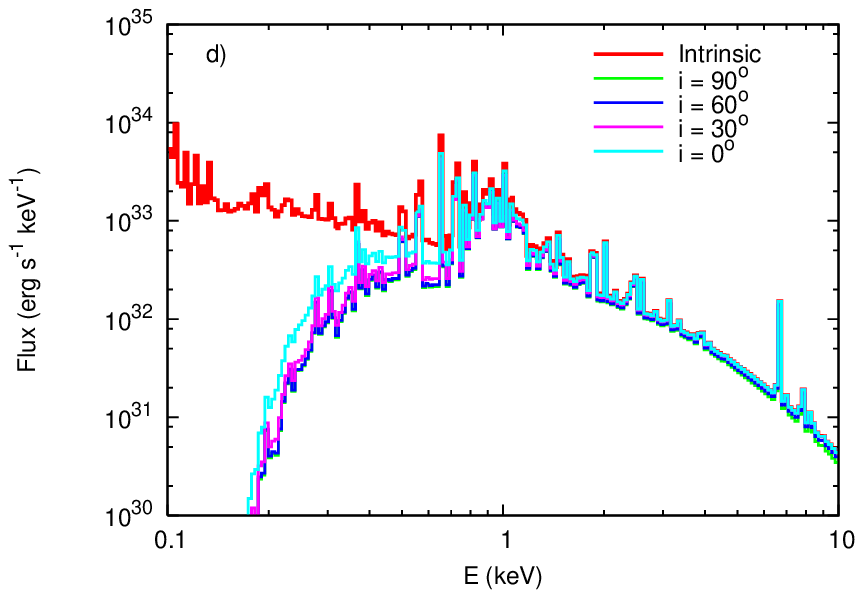,width=5.67cm}
\psfig{figure=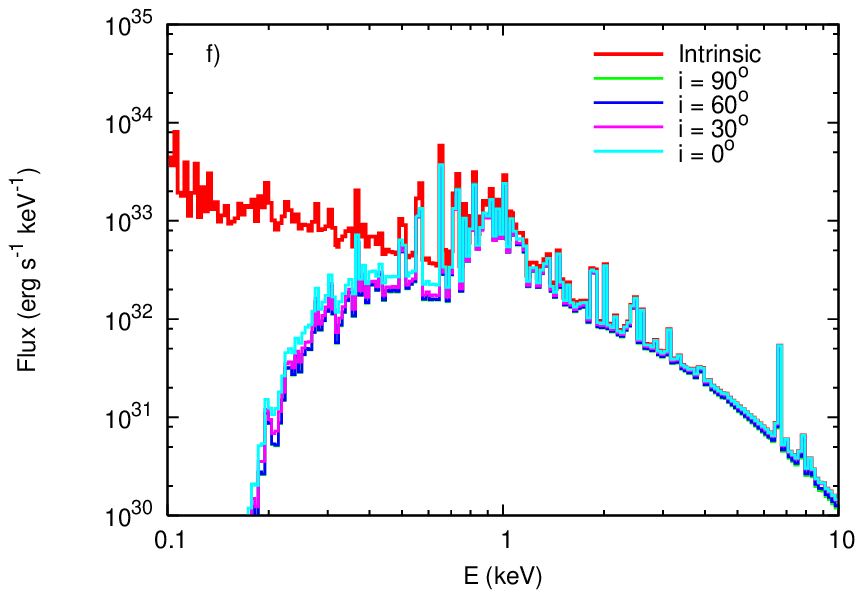,width=5.67cm}
\caption[]{(Top) X-ray spectra from models cwb1 (left), cwb2 (middle),
and cwb3 (right) for an observer in the orbital plane
($i=90^{\circ}$) as a function of orbital phase.  (Bottom) As top
but as a function of $i$ for an observer with $\phi=0^{\circ}$.}
\label{fig:cwb123_xray_spec}
\end{figure*}

\begin{figure*}
\psfig{figure=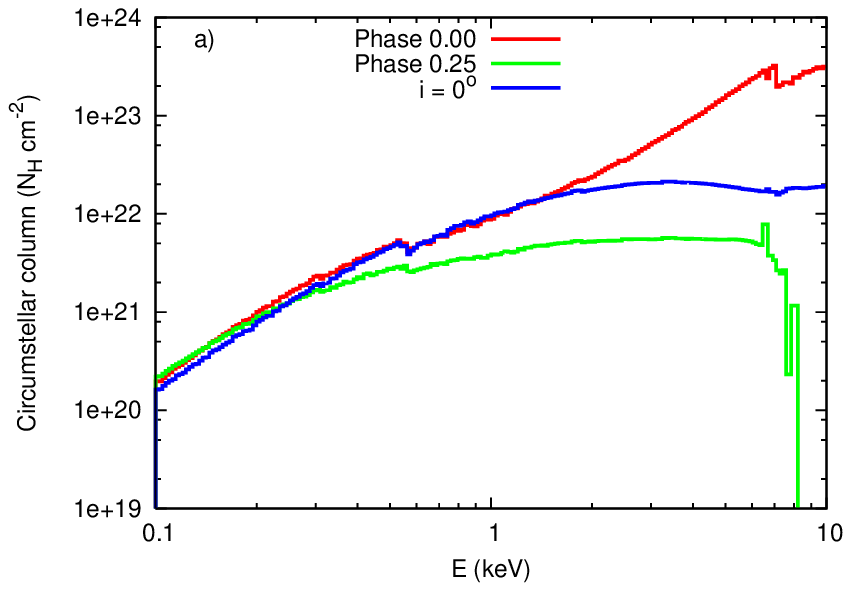,width=5.67cm}
\psfig{figure=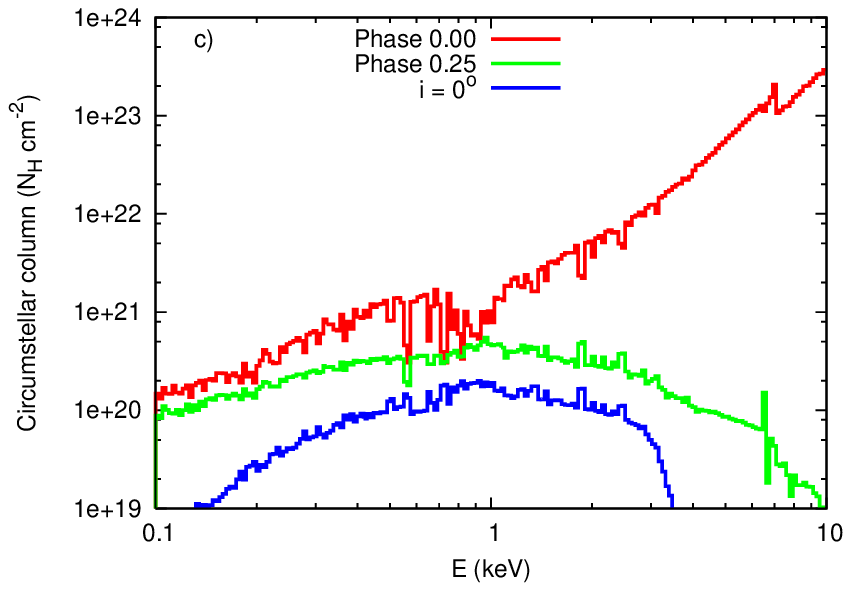,width=5.67cm}
\psfig{figure=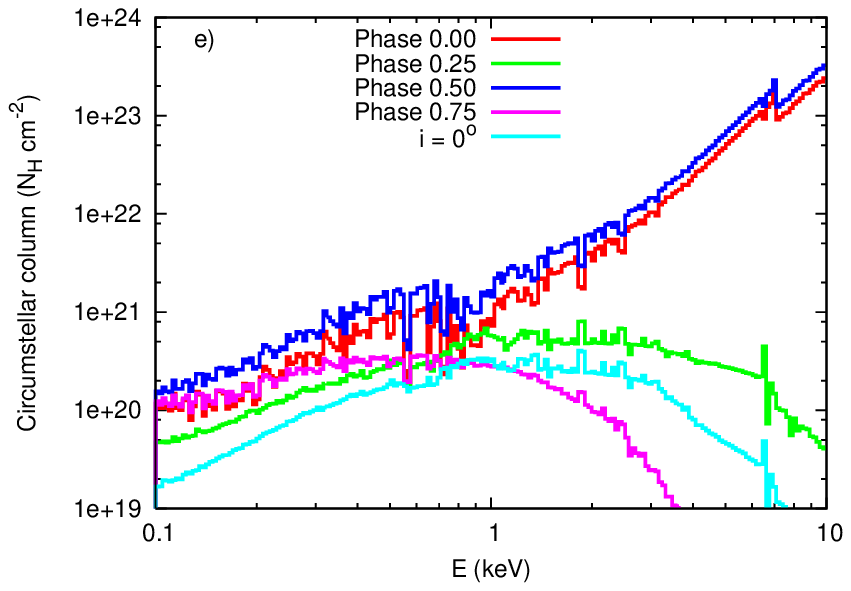,width=5.67cm}
\psfig{figure=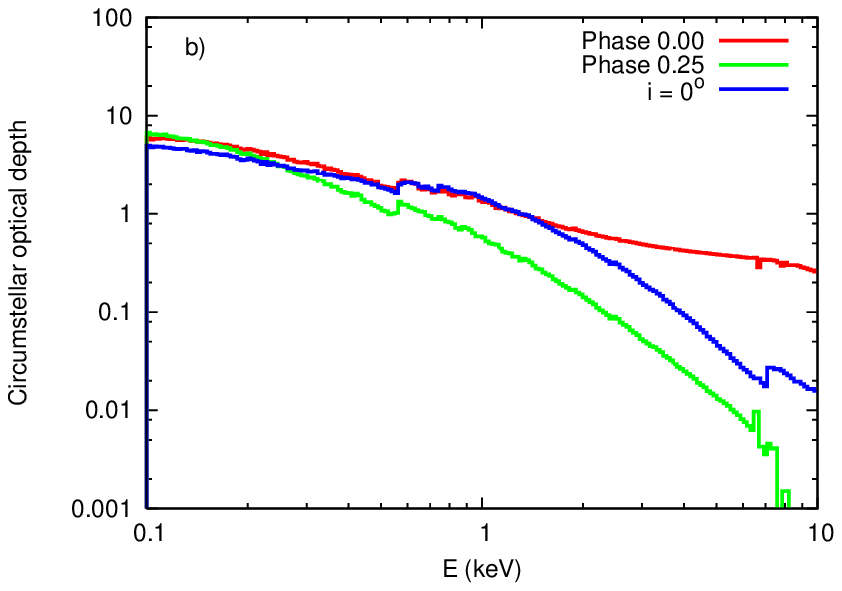,width=5.67cm}
\psfig{figure=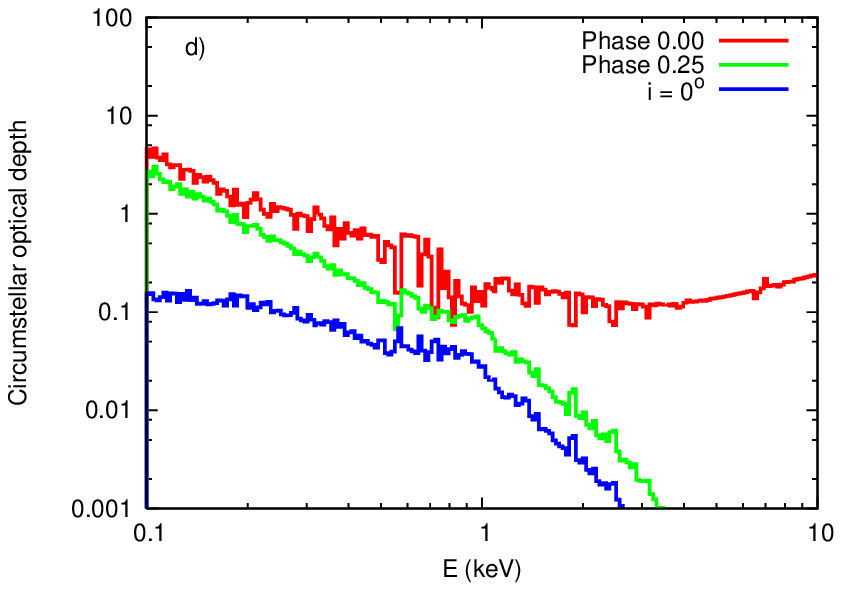,width=5.67cm}
\psfig{figure=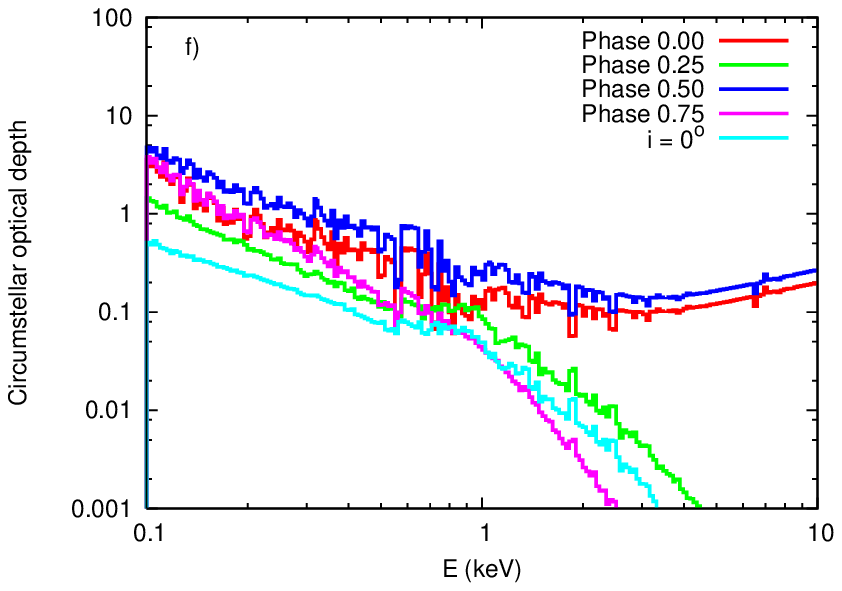,width=5.67cm}
\caption[]{``Effective'' circumstellar column (top) and optical depth
(bottom) from models cwb1 (left), cwb2 (middle), and cwb3 (right) for
an observer in the orbital plane ($i=90^{\circ}$) as a function
of energy and orbital phase. The circumstellar column and optical
depth for an observer at $i = 0^{\circ}$ are also shown. Note that
in some cases the ``effective'' column can be strongly weighted by
occultation effects - at these times it is not a good indicator of the 
actual circumstellar column that the observed X-rays 
travel through.}
\label{fig:cwb123_xray_circum}
\end{figure*}

\subsubsection{Spectra}
\label{sec:cwb1_spectra}
The emission from model cwb1 is overall very soft, reflecting the
rapid cooling of the hot plasma in the
WCR. Fig.~\ref{fig:cwb123_xray_spec}(a) displays synthetic X-ray
spectra as a function of orbital phase for an observer in the orbital
plane. The spectra at phase 0.5 and 0.75 are almost identical to those
at phase 0.0 and 0.25, differing only due to the time dependence of
the dynamic instabilities present in each arm of the WCR, and so are
not shown.  The hard emission is lower at conjunction (phases 0.0 and
0.5) when one of the stars passes in front of the apex of the WCR,
than at quadrature (phases 0.25 and 0.75). This is a combination of
occultation and wind absorption, with the former dominating at hard
energies (see below). The softer emission is also reduced at these
phases as it is attenuated by a greater amount of unshocked stellar
wind between the WCR and the observer. In contrast, at quadrature the
emission suffers less attenuation and occultation, and it more easily
escapes the system.  Fig.~\ref{fig:cwb123_xray_spec}(b) shows that the
variation of the soft emission with the inclination angle, $i$, is
small.  The soft emission arises from relatively low temperature
plasma, comprising a relatively large volume of the WCR. Consequently
the loss of flux due to circumstellar absorption and stellar
occultation is relatively independent of the orientation. As we
shall see, interstellar absorption is a large contribution to the
overall absorption of soft X-rays.

The variation of the hard emission with $i$ (see
Fig.~\ref{fig:cwb123_xray_spec}b) is of a similar magnitude to the
change between conjunction and quadrature when the observer is in the
orbital plane (see Fig.~\ref{fig:cwb123_xray_spec}a). This is not
surprising since in both cases the variation is mainly due to changes
in the occultation, and these changes are comparable: more of the apex
of the WCR is revealed as the observer's sight line moves away from
the eclipse at $i=90^{\circ}$, $\phi = 0^{\circ}$.

Figs.~\ref{fig:cwb123_xray_circum}(a) and (b) show the ``effective''
{\em circumstellar} column and optical depth as a function of energy
for an observer in the orbital plane. These were calculated by
comparing the ray-traced intrinsic and attenuated spectra (the latter
prior to the addition of the interstellar column). Their ratio gives
an energy dependent optical depth
(Fig.~\ref{fig:cwb123_xray_circum}b), which can be converted into an
``effective'' column (Fig.~\ref{fig:cwb123_xray_circum}a) by
considering the energy dependent opacity of the cold plasma in the
winds. Note that this ``effective'' column does not always
provide a good measure of the circumstellar column for an observed
X-ray, since it can be substantially weighted by occultation effects
(see below). Instead it is intended as an indicator of how 
realistic the usual method of using
energy-independent columns with, for example mekal fits, is.

In Fig.~\ref{fig:cwb123_xray_circum}(a) we see that the circumstellar
column from the phase 0.0 calculation monotonically increases with
energy (except at a few noteable absorption edges), reflecting the
fact that the hardest emission is on average generated closest to the
stars (see Fig.~\ref{fig:cwb1_xray_images1}). The effective
attenuation resulting from circumstellar absorption and stellar
occultation exceeds that due to interstellar absorption (assumed to be
$N_{\rm H} = 10^{21}\,{\rm cm^{-2}}$) at energies above 0.2\,keV.  The
effective circumstellar column exceeds $10^{23}\,{\rm cm^{-2}}$ for $E
\gtsimm 4$\,keV. Columns this high cannot be produced by the winds
themselves, and instead indicate substantial occultation of the
intrinsic emission by the stars. The strong energy dependence of the
column at phase 0.0 as shown in Fig.~\ref{fig:cwb123_xray_circum}(a)
also hints at weaknesses that are inherent in simple analyses of X-ray
data of colliding wind binaries in which energy-independent columns
are applied in model fits. While individual columns to multiple emission
components clearly provide some level of energy-dependent $N_{\rm H}$, this
is of course achieved in a rather crude and clumsy way.

The equivalent optical depth from circumstellar absorption and
occultation is shown in Fig.~\ref{fig:cwb123_xray_circum}(b). It is
highest at the lowest energies. At phase 0.0, the optical depth
$\tau=6$ at 0.1\,keV, dropping to $\tau=0.26$ at 10\,keV. In
comparison, the ISM column provides an optical depth of 30 at
0.1\,keV, declining to $10^{-3}$ at 10\,keV.

Occultation of the harder X-ray emission is much reduced at phase 0.25
(see the spectrum in Fig.~\ref{fig:cwb123_xray_spec}a), and this is
manifest as a marked drop in the ``effective'' circumstellar column
and optical depth values in Figs.~\ref{fig:cwb123_xray_circum}(a)
and~(b). In fact, the column is now roughly constant over the energy
range $0.5-7$\,keV. In this case a simple analysis using
energy-independent columns would probably not be too bad an
approximation at this phase. However, it is clear that such analyses
have major short-comings in short period binaries at phases when
occultation is likely to be significant (for instance, the intrinsic
luminosity can be significantly underestimated - see the following
section).  At the very highest energies ($E
\gtsimm 7$\,keV) the effective circumstellar column again declines.
This is due to the fact that the highest temperature plasma exists in
localized bowshocks around clumps far downstream in the WCR, as
discussed in Paper~I.

Figs.~\ref{fig:cwb123_xray_spec}(a) and~(b) also show the effective
circumstellar column and optical depth for an observer directly
above/below the centre of mass in the orbital plane (i.e. at an
inclination angle $i = 0^{\circ}$). The degree of absorption and
occultation are intermediate between the conjunction and quadrature
phases of an observer in the orbital plane.

The deduced values of the energy-dependent circumstellar column shown
in Fig.~\ref{fig:cwb123_xray_circum}(a) can be compared to analytical
estimates. Using Eq.~11 in \citet{Stevens:1992}, which gives the
column density at quadrature from the stagnation point (assumed to be
on the line of centres in the axisymmetric case) through the
undisturbed terminal speed wind, we obtain $N_{\rm H}\sim
3.4\times10^{21}\,{\rm cm^{-2}}$. While this is in good agreement with
the column to the higher energy emission (which should be the best
proxy to the apex of the WCR) at phase 0.25 (see
Fig.~\ref{fig:cwb123_xray_circum}a), it would seem to be a somewhat
fortuitous coincidence. For instance, \citet{Stevens:1992} note that
under the assumption of axisymmetry the column at quadrature is
independent of the system inclination. However,
Fig.~\ref{fig:cwb123_xray_circum}(a) shows that the circumstellar
column for a pole-on observer ($i=0^{\circ}$) is significantly greater
at most energies than the phase 0.25 column.  This is because with
equal strength winds all the material along the line-of-sight from the
apex of the WCR to an observer at $i=0^{\circ}$ has been processed
through the WCR, and is denser than the surrounding unshocked
winds. In contrast, the line-of-sight to an observer in the orbital
plane at phase 0.25 passes mostly through unshocked wind material. So
there are actually large differences in the density structure along
these two sightlines, yet this is not accounted for in Eq.~11 of
\citet{Stevens:1992}. 

\begin{figure*}
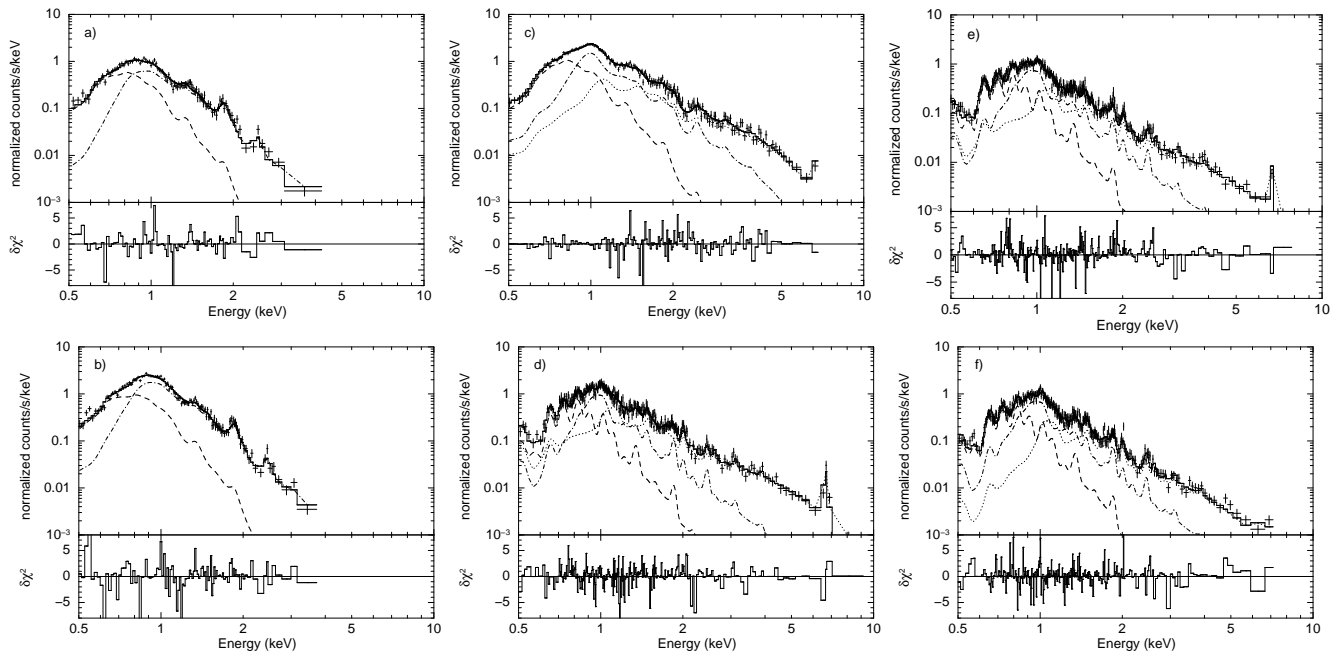

\psfig{figure=cwb1_fit3_good.ps,width=4.2cm,angle=-90}
\psfig{figure=cwb2_fit13_good.ps,width=4.2cm,angle=-90}
\psfig{figure=cwb3_suzaku_fit3_good.ps,width=4.39cm,angle=-90}
\psfig{figure=cwb1_fit14_good.ps,width=4.2cm,angle=-90}
\psfig{figure=cwb2_suzaku4_good.ps,width=4.2cm,angle=-90} %Suzaku
\psfig{figure=cwb3_suzaku_fit5_good.ps,width=4.2cm,angle=-90}
\caption[]{Fits to ``fake'' spectra from the models for different
viewing angles. a) Two-temperature fit to a fake {\em Chandra}
spectrum of model cwb1 at $i=90^{\circ}$ and phase 0.0
(conjunction). b) As a) except the system is viewed at quadrature
(phase 0.25). c) Three-temperature fit to a fake {\em Chandra}
spectrum of model cwb2 for an observer in the orbital plane viewing
the system at conjunction (phase 0.0). d) As c) except to a fake {\em
Suzaku} spectrum.  Note the much better spectral resolution in the
{\em Suzaku} spectrum. The spectra in panels a)-d) have effective
``exposures'' of 10\,ksec.  e) Three-temperature fit to a fake {\em
Suzaku} spectrum of model cwb3 for an observer in the orbital plane
viewing the system at phase 0.0 (weaker O8V wind in front). f) As e)
but at phase 0.5 (stronger O6V wind in front).  Note the difference in
the strength of the Fe\,K line. The spectra in panels e) and f) have
effective ``exposures'' of 20\,ksec.}
\label{fig:cwb123_fits}
\end{figure*}

\subsubsection{Spectral fits}
\label{sec:spectralfits}
In this section we ``observe'' the theoretical spectra generated from
model cwb1 with the {\em Chandra} and {\em Suzaku} X-ray
observatories. Table~\ref{tab:fitresults}
notes the results of the subsequent spectral fits, where an exposure 
time of 10\,ksec has been assumed. At least three mekal
components are needed to obtain satisfactory fits to the simulated
{\em Suzaku} spectra, while the simulated {\em Chandra} spectra 
require at least two mekal components.

Fig.~\ref{fig:cwb123_fits}(a) shows the fake {\em Chandra} spectrum from
model cwb1 at $i=90^{\circ}$ and phase 0.0, with Poisson statistics
added. A single temperature wabs(mekal) model is a very poor fit to
the simulated data ($\chi^{2}_{\nu}=3.91$). However, an acceptable fit
is achieved with the addition of another mekal component
($\chi^{2}_{\nu}=1.27$). Fig.~\ref{fig:cwb123_fits}(b) shows the
corresponding spectrum and fit at phase 0.25. The hotter temperature
component has a slightly reduced temperature, a higher absorbing
column, and a lower normalization at phase 0.0 (conjunction) compared
to phase 0.25 (quadrature), in line with expectations 
(see Figs.~\ref{fig:cwb123_xray_spec}a and~\ref{fig:cwb123_xray_circum}a). 

Two mekal components return reasonable fits to simulated {\em Suzaku}
spectra with an exposure time of 10\,ksec ($\chi^{2}_{\nu}=1.37$, at
phase 0.0 when $i=90^{\circ}$), but fail to provide a good fit when
the exposure time is increased to 40\,ksec ($\chi^{2}_{\nu}=1.64$),
notably failing to fit the line at 0.55\,keV. Adding a third mekal
component does not significantly improve the fit ($\chi^{2}_{\nu}=1.50$).  
Irrespective of the orbital phase, the
temperatures of the three mekal components are all below 0.75\,keV,
reflecting the relatively cool shocked plasma created by the
relatively low preshock wind speeds in this model. Furthermore,
the best-fit absorbing columns are often significantly higher than
the ISM value, indicating that the circumstellar absorption of X-rays
is significant in this model.

The circumstellar columns returned from the spectral fits are
consistent with the trend shown in
Fig.~\ref{fig:cwb123_xray_circum}(a) of higher effective column with
energy. Also of note is that the derived $N_{\rm H}$ to the hottest
mekal component of the spectral models is greater at phase 0.0
(conjunction) than at phase 0.25 (quadrature). In addition, the
hottest mekal component is both hotter and brighter at phase
0.25. These are consistent with the changes to the circumstellar
absorption and stellar occultation of the WCR with the orientation of
the observer, as noted earlier.

A good fit to the simulated {\em Chandra} spectrum can also be
obtained with a single mekal component if the global metal abundance
is allowed to vary. In this case, a metal abundance of
$0.13^{+0.01}_{-0.02}\,Z_\odot$ provides a good fit ($\chi^{2}_{\nu}=1.11$) to
the data. This is very interesting, since we know that the actual
plasma has solar abundances. Clearly, there is great opportunity for
the analysis of low spectral resolution data to return unphysical fit
parameters when fitting emission from inherently multi-temperature
plasma with simpler (e.g. single temperature) spectral models. In an
identical (10\,ksec) analysis to a fake {\em Suzaku} spectrum we again find that
a single temperature fit is very poor ($\chi^{2}_{\nu}=2.69$), but
this time it remains poor ($\chi^{2}_{\nu}=2.13$) when the global
metal abundance is allowed to vary (fitting at $z=0.079\pm0.011\,Z_\odot$). 
This nicely illustrates the advantage of having higher spectral resolution.
The story is more complicated for fits with two mekal components.
Tying the global abundances of the mekal components together, one
finds that the returned value from the analysis of the phase 0.0
fake {\em Chandra} spectrum is very poorly constrained, with its
value depending on how the model approaches its minimum (e.g. the
initial values entered into the model). The 90 per cent confidence
interval typically extends from metallicities of $0.4-2.0\times$ solar,
with a ``best-fit'' value of $z=0.79\,Z_\odot$. A three-temperature
fit to the fake {\em Suzaku} phase 0.25 spectrum returns a global abundance of
$z=0.71\,Z_\odot$ (since $\chi^{2}_{\nu} > 2$, the uncertainty on this
value cannot be estimated using the ``error'' command in {\sc XSPEC}).
It would therefore appear that the fits return more accurate abundances the
more complex the spectral model is. These findings agree with the 
earlier work of \citet{Strickland:1998} who were examining {\em ROSAT}
data in a related context.

Table~\ref{tab:lumerror} shows the intrinsic luminosity of model cwb1,
plus the intrinsic luminosity returned from the spectral fits
(i.e. the observed luminosity, corrected for the interstellar and
circumstellar absorption determined by the fit). We find that the
intrinsic luminosities returned from the fits in all cases
underestimate the true intrinsic luminosity from the system, by
factors of up to 2. This discrepancy arises because direct occultation
of the emission is a significant factor in close binaries like cwb1,
yet no account is made for occultation in the fits. It again
highlights problems which can ensue when fitting too-simple models to
data. The discrepancy is larger at conjunction than at quadrature, as
expected. We further note that the size of the discrepancy is not
dependent on the resolution of the two spectra (i.e. {\em Chandra}
versus {\em Suzaku}). The discrepancy reduces in wider systems (see
model cwb2, next), but could be even larger in yet closer systems.

\begin{figure*}
\psfig{figure=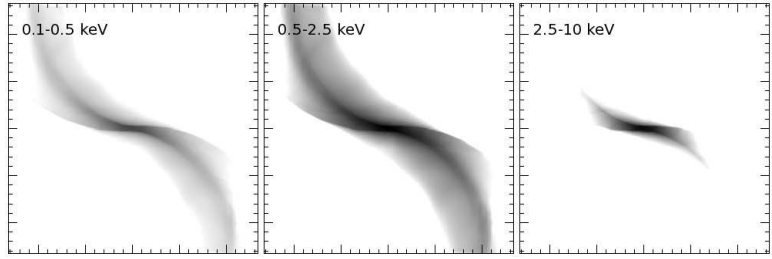,width=10.2cm}
\caption[]{Broad-band intensity images from model cwb2 at
$i=0^{\circ}$. From left to right the images are in the
$0.1-0.5$\,keV, $0.5-2.5$\,keV, and $2.5-10$\,keV bands.  The maximum
intensity in the images is $10^{7}\,{\rm
erg\,cm^{-2}\,s^{-1}\,keV^{-1}\,ster^{-1}}$, and the grayscale covers
4 orders of magnitude in brightness. The major ticks on each axis mark
out 0.5\,mas. The stars are located north and south of the image centre
by 0.18\,mas.}
\label{fig:cwb2_xray_images1}
\end{figure*}

\begin{figure*}
\psfig{figure=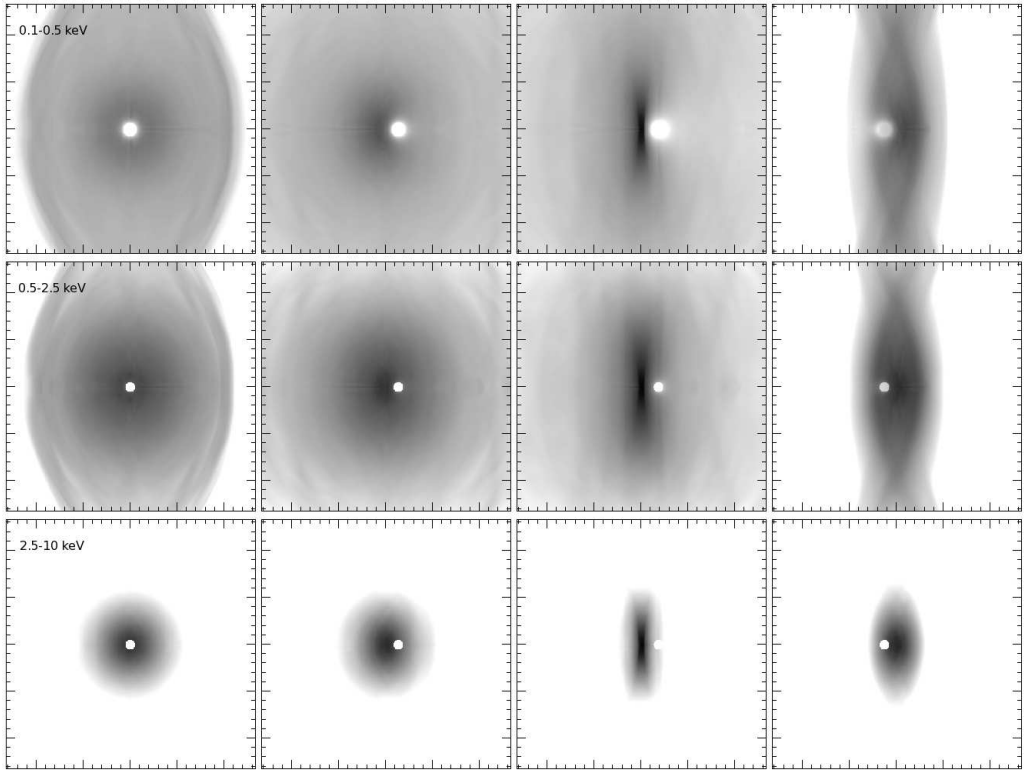,width=13.6cm}
\caption[]{Broad-band intensity images from model cwb2 at $i=90^{\circ}$. From
top to bottom the images are in the $0.1-0.5$\,keV, $0.5-2.5$\,keV, and $2.5-10$\,keV bands.
From left to right the phase of the observation increase from 0.0 (conjunction, 
$\phi = 0^{\circ}$), to 0.125 ($\phi = 315^{\circ}$), to 0.25 
(quadrature, $\phi = 270^{\circ}$), to 0.375 ($\phi = 225^{\circ}$).
The maximum intensity in the $0.1-0.5$\,keV images is $10^{6}\,{\rm
erg\,cm^{-2}\,s^{-1}\,keV^{-1}\,ster^{-1}}$, while in the
$0.5-2.5$\,keV, and $2.5-10$\,keV images it is $10^{7}\,{\rm
erg\,cm^{-2}\,s^{-1}\,keV^{-1}\,ster^{-1}}$. The grayscale covers
4 orders of magnitude in brightness, and the major ticks on each axis mark
out 0.5\,mas.}
\label{fig:cwb2_xray_images2}
\end{figure*}

\begin{figure*}
\psfig{figure=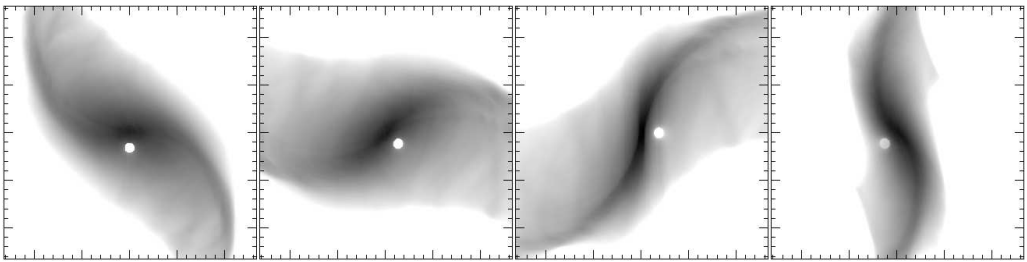,width=13.6cm}
\caption[]{Broad-band $0.5-2.5$\,keV intensity images from model cwb2
at $i=30^{\circ}$.  From left to right the orbital phases are 0.0,
0.125, 0.25, and 0.375.  The stars and the WCR again rotate
anti-clockwise, while the maximum intensity of the images is
$10^{7}\,{\rm erg\,cm^{-2}\,s^{-1}\,keV^{-1}\,ster^{-1}}$, and the
major ticks on each axis mark out 0.5\,mas.}
\label{fig:cwb2_xray_images3}
\end{figure*}

\subsection{Model cwb2}
\label{sec:xray_cwb2}
\subsubsection{Images}
Fig.~\ref{fig:cwb2_xray_images1} shows the broad-band intensity images
from model cwb2 for an observer located directly above/below the
orbital plane. The hardest emission is again confined to a region
close to the apex of the WCR, and the curved shape of the WCR is
clearly seen. The leading edge of each arm of the WCR is sharper and
more distinct. The trailing edge is blurred because the hot plasma
inside the WCR displays an increasing phase lag as one goes further
from the orbital plane. This reflects the fact that the motion of this
plasma is strongly influenced by the prior (rather than the current)
positions of the stars. The extent of the low surface brightness
emission from the WCR is affected by the size of the numerical
grid. Calculations with a bigger grid would reveal that, for instance,
the left-hand edge of the emission in the bottom right corner of the
$0.5-2.5$\,keV image in Fig.~\ref{fig:cwb2_xray_images1} would extend
further to the left. However, we remain confident that the majority of
the emission is captured in this (and the other) models, since the
surface brightness is 4 orders of magnitude lower than the peak
surface brightness obtained at the apex of the WCR.

Broad-band intensity images from model cwb2 for an observer in the
orbital plane are shown in Fig.~\ref{fig:cwb2_xray_images2}. The
images bear substantial similarities to the radio images shown in
Paper~II.  The morphology of these images is determined by the
relative orientation of the observer to the WCR and the stars. The
foreground star eclipses the emission from those parts of the WCR
which lie behind it.  The double-helix-like structure seen in the soft
and medium-band images when the system is half-way between conjunction
and quadrature (at phase 0.375) is due to limb-brightened
emission. The vertical curvature again illustrates the increasing
phase-lag of the shocked gas with distance above/below the orbital
plane.  The increasing confinement of emission to the apex of the WCR
at higher energies means that the hard-band images instead show a
disc-like structure. In all 3 energy bands the surface brightness of
the emission is highest in the images generated at quadrature, when
the X-rays from the apex of the WCR initially escape through the hot,
low opacity, gas within the WCR. This is also true of radio images
when $\nu \ltsimm 100$\,GHz (see Paper~II).

Images for an observer viewing the system with an inclination angle
$i=30^{\circ}$ are shown in Fig.~\ref{fig:cwb2_xray_images3}.
The brightest parts of the WCR are again those which are limb-brightened.
The overall morphology of the WCR is ``S''-shaped, and the position of the
foreground star is again clear through its occultation of background emission.
Comparison to radio images again reveals significant similarities (see
Fig.~8 in Paper~II).

\subsubsection{Lightcurves}
Lightcurves from model cwb2 are shown in the central column
of Fig.~\ref{fig:cwb123_xray_lc}.
The longer orbital period and the higher speed of the wind-wind collision
changes the lightcurves compared to those from model cwb1 
in several important ways:

\begin{enumerate}
\item The luminosity in the $2.5-10$\,keV lightcurve is over an
order of magnitude higher. This reflects the harder emission resulting
from the higher postshock temperatures created by the faster wind
collision speeds ($1630 \kmps$ versus $730 \kmps$ at the apex of the
WCR).
\item The amplitude of variation with orbital phase is much reduced in the
$0.5-2.5$\,keV and $2.5-10$\,keV lightcurves. This reflects weaker
circumstellar attenuation due to the lower wind densities surrounding 
the WCR, and reduced occultation effects due to the larger size of
the WCR relative to the stars.
\item The lightcurves display clear symmetry, with 2 identical cycles
per orbit. Dynamical instabilities are weak in this model, because the
WCR is largely adiabatic and the winds have equal speeds, and do not
appreciably disrupt the inherent symmetry between each arm of the WCR.
\item The sharp absorption features seen in the lightcurves from
model cwb1 have disappeared, since there is no longer a dense thin layer
of post-shock gas to absorb the X-rays in this way.
\end{enumerate}

The ISM corrected $0.5-10$\,keV luminosity is $1.55\times10^{33}\,\ergps$,
$1.73\times10^{33}\,\ergps$, and $1.77\times10^{33}\,\ergps$ at viewing angles of
$(i,\phi) = (90^{\circ},0^{\circ})$, $(90^{\circ},90^{\circ})$, and
$(0,0)$ (pole-on), giving $L_{\rm x}/L_{\rm bol} =  1.09\times10^{-6}$,
$1.21\times10^{-6}$, and $1.24\times10^{-6}$, respectively. The luminosities
and $L_{\rm x}/L_{\rm bol}$ values are slightly greater than from
model cwb1, and are again consistent with strong colliding winds emission.

\subsubsection{Spectra}
X-ray spectra from model cwb2 are shown in
Figs.~\ref{fig:cwb123_xray_spec}(c) and~(d). It has already been noted
that the continuum emission from model cwb2 is much harder than from
model cwb1, due to the higher pre-shock speeds which the winds attain
before they collide in this system. However, the line emission also
reflects the higher temperatures in model cwb2: a strong Fe\,K line is
visible at approximately $6.7\;$keV, but this line is much weaker
relative to the continuum in model cwb1. In contrast, most of the
other spectral lines in model cwb2, particularly those with $E <
1$\,keV, display weaker emission relative to the continuum than in
model cwb1, reflecting the lack of strong cooling in model cwb2
compared to model cwb1. There is significantly more soft emission at
$i = 0^{\circ}$ than at higher inclination, which reflects the fact
that the sight line to the stagnation point of the WCR is entirely
through the hot lower opacity WCR. This enhancement does not occur in
model cwb1, due to the different nature of the WCR (specifically the
relative lack of hot gas within it) in this model. The soft and hard
emission is again lower at conjunction due respectively to enhanced
circumstellar absorption and occultation, as was also the case for
model cwb1.

Figs.~\ref{fig:cwb123_xray_circum}(c) and~(d) show the ``effective''
circumstellar column and optical depth as a function of energy from
the ray-traced calculation for a variety of viewing angles. The column
at conjunction (phase 0.0) shows again a general rise with increasing
energy. The columns (and hence the optical depths) at 0.1 and 10\,keV
are similar to those from model cwb1, though they are substantially
reduced in value at intermediate energies (e.g. at 1\,keV the
circumstellar column and optical depth is now an order of magnitude
lower).  Lower circumstellar columns and optical depths are expected,
of course, because of the wider stellar separation. In model cwb2 the
circumstellar column at phase 0.25 peaks at around 1\,keV, whereas in
model cwb1 the peak column at phase 0.25 is to emission near 4\,keV
(ignoring the high column to line emission at 6.7\,keV).
Fig.~\ref{fig:cwb123_xray_circum}(c) also shows that the circumstellar
column to an observer with $i=0^{\circ}$ is lower than the columns
obtained for an observer in the orbital plane, consistent with the
higher flux of soft X-rays at this orientation (see
Fig.~\ref{fig:cwb123_xray_spec}d). This contrasts with model cwb1
where the column for a pole-on system lies (for $E > 0.25\,$keV)
inbetween the columns to observers with $i=90^{\circ}$ at phase
0.0 and 0.25.

Another noticeable difference to model cwb1 is that the effective
columns show significant variability between adjacent energy
bins.  This is caused by the different temperatures (and thus
locations) at which line emission and the adjacent continuum are
formed. Finally, we can again make a comparison between the
circumstellar column obtained from Eq.~11 in \citet{Stevens:1992}
and Fig.~\ref{fig:cwb123_xray_circum}(c). The former gives
$N_{\rm H} \sim 1.5\times10^{21}\,{\rm cm^{-2}}$, which is substantially
greater than the values shown in Fig.~\ref{fig:cwb123_xray_circum}(c)
to an observer in the orbital plane at quadrature (phase 0.25) and to 
an observer with $i=0^{\circ}$. This further highlights that
the formula in \citet{Stevens:1992} is unsuitable for use in
short-period systems. We therefore make no further comparisons to it
in this work.

\subsubsection{Spectral fits}
Fig.~\ref{fig:cwb123_fits}(c) shows a simulated 10\,ksec {\em Chandra}
spectrum from model cwb2 at $i=90^{\circ}$ and phase 0.0. Folding the
same theoretical spectrum through the {\em Suzaku} response and exposing for
10\,ksec yields the
``fake'' spectrum shown in Fig.~\ref{fig:cwb123_fits}(d). The higher
spectral resolution of the latter observatory is clearly
evident. Two-temperature fits to these spectra are uniformally poor
(typically $\chi^{2}_{\nu} \gtsimm 2$), with the flux at low and high
energies underestimated. Another mekal component is clearly required.
As expected three-temperature fits are more acceptable. Importantly
the fits to both the {\em Chandra} and {\em Suzaku} spectra return
consistent parameter values. At phase 0.0, both sets of fits find that
the normalization of the mekal components increases monotonically with
the temperature of the component (see Table~\ref{tab:fitresults}).
Compared to the results from model cwb1, the returned temperatures are
significantly higher, and the absorbing columns significantly lower,
both of which are consistent with expectations.
%However, there are some differences. Although the individual uncertainties
%are large, the {\em Chandra} fit returns columns above 1e21, whereas the Suzaku
%fit does not. Does this also happen at phase 0.25 and 0.5? No, not at phase 0.25.
%Why does the Chandra 3T fit to cwb2 produce columns above 1e21, whereas the Suzaku
%fit does not? 

Another finding is that the mekal components consistently return
cooler temperatures at phase 0.25 compared to phase 0.0. This reflects
the greater ease at which low energy X-rays can escape absorption by
the circumstellar environment at this phase (see
Fig.~\ref{fig:cwb123_xray_spec}c), and is further manifest by the
consistently lower columns to the mekal components at phase 0.25
compared to phase 0.0.  In all four of the fits (to the ``fake'' {\em
Chandra} and {\em Suzaku} spectra at phases 0.0 and 0.25) the
absorbing column to the lowest temperature mekal component (at $kT
\approx 0.32$\,keV) is indicative of {\em only} ISM
absorption. Fig.~\ref{fig:cwb123_xray_circum}(c) reveals
that the additional circumstellar column is indeed small in comparison
($\approx 2 \times 10^{20}\,{\rm cm^{-2}}$, versus $10^{21}\,{\rm
cm^{-2}}$ for the assumed ISM column). The additional (above ISM)
columns to the medium and hard mekal components returned from the fit
to the phase 0.0 {\em Chandra} spectrum (respectively
$1.5^{+0.9}_{-1.5}\times10^{21}\,{\rm
cm^{-2}}$ and $2.1^{+2.5}_{-2.1} \times10^{21}\,{\rm
cm^{-2}}$), though not particularly well constrained, are comparable
to the columns shown in Fig.~\ref{fig:cwb123_xray_circum}(c) obtained
at the energy of the individual mekal components. This is a pleasing
result.  Surprisingly, although the implied circumstellar columns to
the medium mekal components obtained from the phase 0.0 and 0.25 {\em
Suzaku} fits are also comparable to those shown in
Fig.~\ref{fig:cwb123_xray_circum}(c), the fits find that there is no
need for additional circumstellar absorption to the hard mekal
components. In this respect, the fits to the ``fake'' {\em Chandra}
spectra are better at recovering the actual circumstellar columns than
the higher spectral resolution {\em Suzaku} spectra. Having said this,
we note that the columns to the hard component from the {\em Suzaku}
fits have upper limits which are still consistent with
Fig.~\ref{fig:cwb123_xray_circum}(c).

We again note that when fitting a three-temperature mekal model to
medium-resolution spectra, relaxing the global abundance in the model
can lead to erroneous results. For the $i=90^{\circ}$ phase 0.0 case,
the fit to the {\em Chandra} spectrum returns
$z=0.60^{+0.23}_{-0.18}\,Z_\odot$. From the {\em Suzaku} spectrum we obtain
$z=0.73^{+0.18}_{-0.12}\,Z_\odot$. Both results are significantly below the solar
abundances used in our models. Finally, we again compare the real
intrinsic luminosities calculated directly from the models and the
inferred intrinsic luminosities from the spectral
fits. Table~\ref{tab:lumerror} shows that while luminosity differences
still exist, occultation effects are now largely insignificant.
Indeed, the returned intrinsic luminosity is now often greater than
the intrinsic value. Such differences, including overestimates, result
from the imperfect nature of the fit and poisson noise in the count
rate.

\begin{figure}
\begin{center}
\psfig{figure=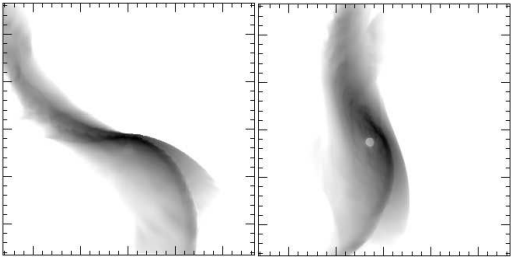,width=6.8cm}
\end{center}
\caption[]{Broad-band $0.5-2.5$\,keV intensity images from model cwb3.
The left panel shows the emission at $i=0^{\circ}$. The right panel
displays the emission at $i=30^{\circ}$ and $\phi = 45^{\circ}$ (phase 0.875). 
The maximum intensity is $10^{7}\,{\rm
erg\,cm^{-2}\,s^{-1}\,keV^{-1}\,ster^{-1}}$, and the major ticks on
each axis mark out 0.5\,mas. $\phi=0^{\circ}$ corresponds to the
O8V star in front.}
\label{fig:cwb3_xray_images1}
\end{figure}

\begin{figure*}
\psfig{figure=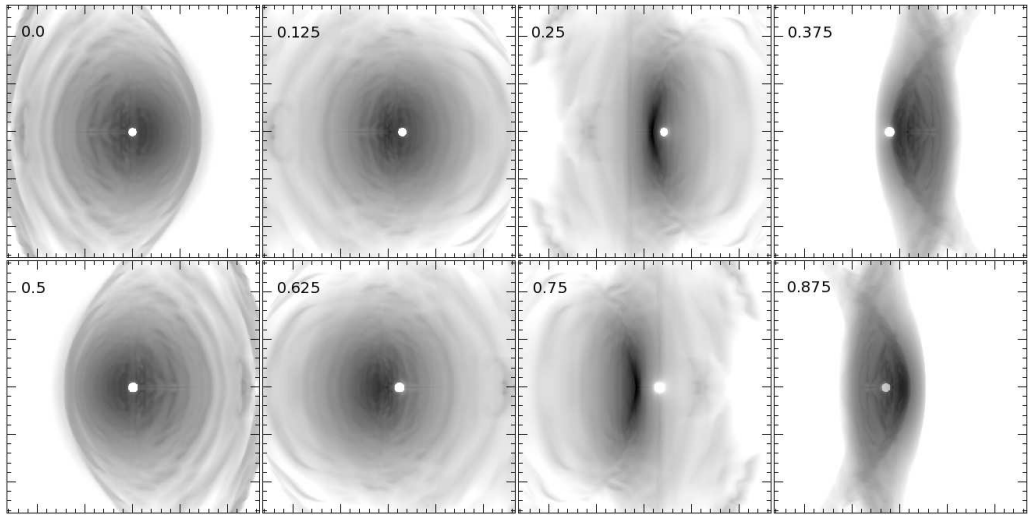,width=13.6cm}
\caption[]{Broad-band $0.5-2.5$\,keV intensity images from model cwb3
at $i=90^{\circ}$. From top to bottom and left to right the orbital
phase is 0.0 ($\phi = 0^{\circ}$, conjunction, O8V star in front), 0.125, 0.25
($\phi=270^{\circ}$, quadrature), 0.375, 0.5 ($\phi=180^{\circ}$, conjunction, 
O6V star in front), 0.625, 0.75 ($\phi=90^{\circ}$, quadrature), and 0.875. 
The maximum intensity is $10^{7}\,{\rm
erg\,cm^{-2}\,s^{-1}\,keV^{-1}\,ster^{-1}}$, and the major ticks on
each axis mark out 0.5\,mas.}
\label{fig:cwb3_xray_images2}
\end{figure*}

\subsection{Model cwb3}
\label{sec:xray_cwb3}
\subsubsection{Images}
The stellar winds in model cwb3 are of unequal strength, being blown
from hypothetical O6V and O8V stars. The stronger wind from the O6V
star pushes the WCR closer towards the O8V star, and, compared to
model cwb2 where the winds are of equal strength, bends the arms of
the WCR inwards towards the weaker wind.  The left panel of
Fig.~\ref{fig:cwb3_xray_images1} shows broad-band $0.5-2.5$\,keV
intensity images from model cwb3 for an observer directly above/below
the orbital plane. A comparison against the corresponding image in
Fig.~\ref{fig:cwb2_xray_images1} reveals several differences. Firstly,
the brightest part of the image (at the apex of the WCR) is located
closer to the O8V star (which is to the south in these
images). Secondly, the downstream positions of the arms of the WCR are
also in different locations, due to the different ram-pressure balance
of the winds. Thirdly, there are differences in the brightness
contrast across the contact discontinuity. At the apex of the WCR, ram
pressure balance requirements mean that since the O8V wind has a lower
pre-shock velocity than the O6V wind, the O8V material must have a
higher pre-shock density. This directly translates into a higher
post-shock density (and lower post-shock temperature), and thus into
greater X-ray emissivity on the O8V side of the contact
discontinuity. A similar effect is also seen from the thermal radio
emission (see Paper~II). This emission contrast is then amplified in
the leading arm of the WCR, but reduced in the trailing arm, as a
result of the dynamics of the WCR (see Paper~I). In comparison, there
is initially no contrast in the emission across the contact
discontinuity at the apex of the WCR in model cwb2, though such an
effect subsequently develops in the downstream arms.  Finally, we note
that due to velocity shear Kelvin-Helmholtz instabilities occur along
the contact discontinuity in model cwb3, and these are also visible
in the images. The clearest sign occurs at phases 0.375 and 0.875 in Fig.~\ref{fig:cwb3_xray_images2},
where an oscillation of the contact discontinuity separating the bright and
fainter parts of the WCR can be seen.  The right panel in
Fig.~\ref{fig:cwb2_xray_images1} shows the intensity image for an
observer with $i=30^{\circ}$ and $\phi=45^{\circ}$.  The O8V star is
silhouetted against the WCR. The emission from the shocked O8V wind is
clearly brighter, for the reasons already described above.

Fig.~\ref{fig:cwb3_xray_images2} shows broad-band $0.5-2.5$\,keV
intensity images from model cwb3 for an observer in the orbital plane
with $\phi=0^{\circ}$ as a function of phase. The O8V star is
in front at phase 0.0, while the larger O6V star is in front at
phase 0.5. Significant differences to the images from model
cwb2 are again apparent (cf. Fig.~\ref{fig:cwb2_xray_images2}). For
instance, at conjunction when the weaker O8V wind is in front
(phase 0.0), the limb brightened edge of the leading arm of the
WCR (to the right side of the image) shows greater curvature and is
projected closer to the centre of the image, while the limb brightened
part of the trailing arm is located at the far left side of the
image. In addition, the O8V star occults a smaller region of the WCR
than the larger O6V star does in model cwb2.

Other differences are also apparent. At phase 0.375, the
double-helix-like structure seen from model cwb2 is replaced by a
more imbalanced morphology, where the brightest regions trace the
limb-brightened edge of the dense, shocked O8V gas on the trailing
edge of the leading arm. Interestingly, compared to simulated radio
images (see Fig.~12 in Paper~II), at phase 0.875 the shocked O6V gas is much more
visible on the right side of the image. Obviously the size of the
occulted region differs depending on which star is in front, but we
note that the position of this relative to the limb-brightened part of
the WCR is also different at phase 0.375 and 0.875, reflecting
the different amounts of downstream curvature imparted to the leading
and trailing arms of the WCR (see Paper~I for further details).
Finally, we note that the vertical curvature of the WCR is also
apparent in the images at quadrature (phase 0.25 and 0.75).

\subsubsection{Lightcurves}
The right column of Fig.~\ref{fig:cwb123_xray_lc} shows the orbital
phase variation of the X-ray emission from model cwb3. The O8V star is
in front at phase 0.0, with the O6V star in front at phase 0.5. The
unequal wind strengths in model cwb3 is manifest in the unequal depths
of the two minima around the orbit.  The deeper minimum occurs when
the O6V star (which has the denser wind) is in front of the WCR apex,
as expected. A phase lead to the bottom of the minimum is present in
some lightcurves (e.g. the $0.5-2.5$\,keV curve), though not in others
(e.g. the $2.5-10$\,keV curve). The emisison at quadrature (phases
0.25 and 0.75) is brighter than at conjunction, as seen in the other
simulations.  The luminosity in the softest lightcurve is slightly
higher near phase 0.25 than near 0.75. This is because lines of sight
to the apex at phases near 0.25 initially pass through the hot, low
opacity, gas in the WCR because of orbital aberration.

The X-ray luminosity is somewhat lower in this model compared to model
cwb2, reflecting the reduced wind power of the O8V star. The reduction
in luminosity is greatest in the harder energy bands (the
$2.5-10$\,keV luminosity in model cwb3 is only 40 per cent of the
luminosity in model cwb2), reflecting the slower speed of the O8V wind
and the increased obliquity of the shock in the primary wind with
off-axis distance relative to model cwb2.  The ISM corrected
$0.5-10$\,keV luminosity is $1.07\times10^{33}\,\ergps$,
$1.14\times10^{33}\,\ergps$, and $1.16\times10^{33}\,\ergps$ at
viewing angles of $(i,\phi) = (90^{\circ},0^{\circ})$,
$(90^{\circ},90^{\circ})$, and $(0,0)$ (pole-on). The slightly reduced
X-ray luminosities compared to model cwb2 are somewhat offset by a
corresponding reduction in $L_{\rm bol}$, giving $L_{\rm x}/L_{\rm
bol} \approx 10^{-6}$ for all 3 of the viewing angles indicated, again
consistent with strong colliding winds emission.

\begin{figure}
\begin{center}
\psfig{figure=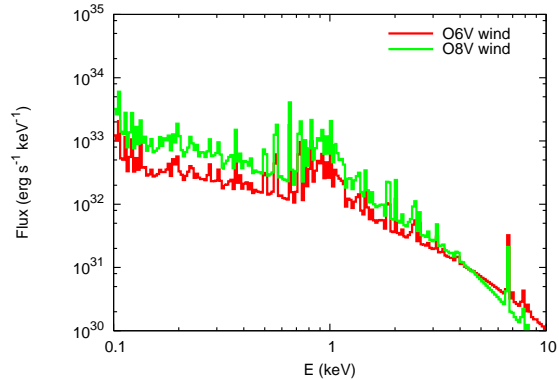,width=7.5cm}%5.67cm
\caption[]{Intrinsic X-ray spectra from each shocked wind in model cwb3.}
\label{fig:cwb3_xray_spec}
\end{center}
\end{figure}

\subsubsection{Spectra}
The right column of Fig.~\ref{fig:cwb123_xray_spec} shows intrinsic
and attenuated spectra from model cwb3. Because of the unequal winds,
the emission at phases 0.5 (0.75) is no longer identical to that at
phases 0.0 (0.25), so all conjunction and quadrature phases are
shown.  There is now a clear difference in the strength of the low
energy absorption at orbital phases 0.0 and 0.5, reflecting changes to
the wind density along sight lines when either the O8V or O6V stars
are in front.

In models cwb1 and cwb2, the winds were of equal strength and the
intrinsic emission from their shocked plasma was identical,
and contributed equally to the total. However, 
Fig.~\ref{fig:cwb3_xray_spec} shows that the intrinsic emission
from the postshock wind of the O6V star is harder than the intrinsic
emission from the postshock wind of the O8V star.  This reflects the
higher velocity at which the O6V wind encounters the WCR, due to the
greater distance over which it can accelerate and its higher terminal
velocity (although this increase in the preshock velocity is somewhat
offset by the greater shock obliquity further downstream). But while
the shocked O6V wind dominates the hard emission, the shocked O8V wind 
dominates the overall intrinsic emission, as is clearly apparent in
Fig.~\ref{fig:cwb3_xray_spec},
contributing approximately two-thirds to the total luminosity. This is
despite the total wind power of the O8V star being just one third of
that of the O6V star. The fact that it dominates the X-ray luminosity is due
to a greater fraction of it being processed through the
WCR, and it subsequently radiating energy more efficiently
\citep[c.f.][]{Pittard:2002b}.

Figs.~\ref{fig:cwb123_xray_circum}(e) and~(f) show the ``effective''
circumstellar column and optical depth as a function of energy for
various viewing angles into model cwb3. We find similar columns at
phases 0.0 and 0.5 to those reported previously from models cwb1 and
cwb2. The column at phase 0.5 (when the larger O6V star and its denser
wind is in front) is slightly higher than at phase 0.0 (when the
smaller O8V star is in front) as expected. The slope of the column
with energy and its bin-to-bin variation are similar to those from
model cwb2 shown in Fig.~\ref{fig:cwb123_xray_circum}(c). As
previously noted, the different wind strengths in model cwb3 break the
symmetry that models cwb1 and cwb2 display at quadrature.
Fig.~\ref{fig:cwb123_xray_circum}(e) shows that the columns at phase
0.75 are very low at high energies, whereas they are significantly
higher at phase 0.25. In contrast, the columns at $E < 0.25$\,keV at
phase 0.0 and 0.75 are identical, whereas they are lower at phase
0.25.

The effective circumstellar column to an observer at $i = 0^{\circ}$ is
$1.7 \times 10^{19}\;{\rm cm^{-2}}$ at 0.1\,keV, rises an order of
magnitude to $\approx 3 \times 10^{20}\;{\rm cm^{-2}}$ by 1\,keV, and
then drops below $ 10^{19}\;{\rm cm^{-2}}$ by 7.5\,keV. The $i=0^{\circ}$
circumstellar column and optical depth are greater than those from model
cwb2, since the unequal wind strengths bend the WCR around the O8V star,
so that only a relatively short segment along the line-of-sight from the
apex of the WCR passes through hot plasma, with the rest of it through 
wind material from the O6V star.

\subsubsection{Spectral fits}
For model cwb3 we simulate {\em Suzaku} spectra with an exposure time
of 20\,ksec. Two-temperature mekal fits are poor ($\chi^{2}_{\nu}
\approx 2.0$), and underestimate the hard X-ray
flux. Three-temperature fits are much more acceptable (see
Fig.~\ref{fig:cwb123_fits}(e) and (f), and Table~\ref{tab:fitresults}).
Significantly higher absorption is found to the hot component at phase
0.5 compared to phase 0.0, consistent with expectations given that the
denser O6V wind is in front at phase 0.5. The temperature returned to
the hottest mekal component is also significantly lower at phase 0.5
compared to phase 0.0.  This reflects the much weaker Fe\,K line
emission at phase 0.5, due to greater occultation (by a larger star)
of the hottest part of the WCR at phase 0.5 (the surface brightness of
the high energy emission falls off very rapidly from the apex of the
WCR, and thus the observed flux of high energy X-rays is highly
susceptible to the amount of occultation).  No extra absorbing column
(above the ISM value) is required to the separate mekal components in
many of the fits. This is roughly consistent with
Fig.~\ref{fig:cwb123_xray_circum}(c) where it can be seen that the effective
circumstellar column is typically low, due to the relatively small
mass-loss rates and wide stellar separations in the model.

Having said this, significant additional absorption above the ISM
value is returned from the fit made to the phase 0.5 spectrum. The
degree of extra absorption increases with the temperature of the mekal
component (no extra absorption is required to the lowest temperature
mekal component in the fit). The additional absorption to the
0.85\,keV component ($4^{+10}_{-4}\times10^{20}\,{\rm cm^{-2}}$) is a
bit lower than the effective circumstellar column at this energy in
Fig.~\ref{fig:cwb123_xray_circum}(e) (though within the 90~per~cent
confidence range the fitted value is reasonable). In contrast, the
additional absorption to the 1.66\,keV component
($3.6^{+1.2}_{-1.5}\times10^{21}\,{\rm cm^{-2}}$) is both in better
agreement with Fig.~\ref{fig:cwb123_xray_circum}(e) and is more tightly
constrained.

Interestingly, if the exposure time is reduced to 10\,ksec, the fit to
the $i=90^{\circ}$ phase 0.0 spectrum returns $kT_{3} =
1.68^{+0.57}_{-0.16}$\,keV. Although formally this is still consistent with
the value returned from the fit to the 20\,ksec exposure spectrum
($kT_{3} = 2.09^{+0.12}_{-0.40}$\,keV), it highlights that a lack of counts
at high energies due to relatively short exposures can bias the resulting fits
towards lower temperatures.

\begin{figure*}
\psfig{figure=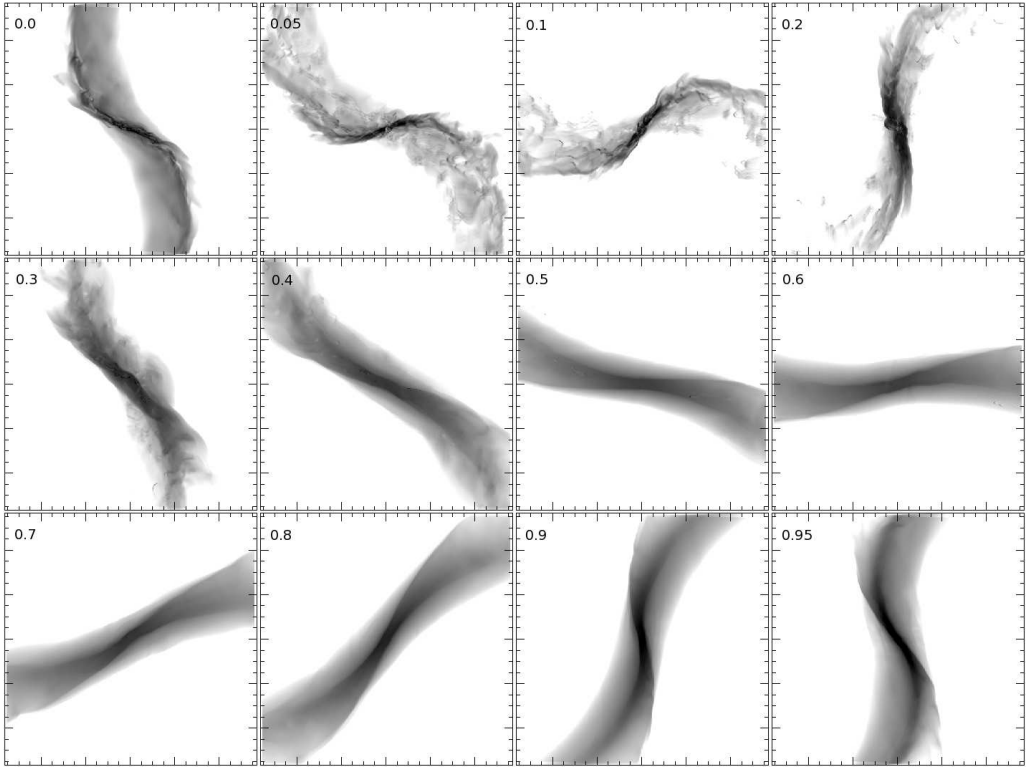,width=13.6cm}
\caption[]{Broad-band ($0.5-2.5$\,keV) intensity images from model cwb4 for an observer
with $i=0^{\circ}$. The orbital phase is noted on each panel.
The maximum intensity (black in the images) is
$10^{8}\,{\rm erg\,cm^{-2}\,s^{-1}\,keV^{-1}\,ster^{-1}}$. 
The major ticks on each axis mark out 0.2\,mas.}
\label{fig:cwb4_xray_image_bb2}
\end{figure*}

\begin{figure*}
\psfig{figure=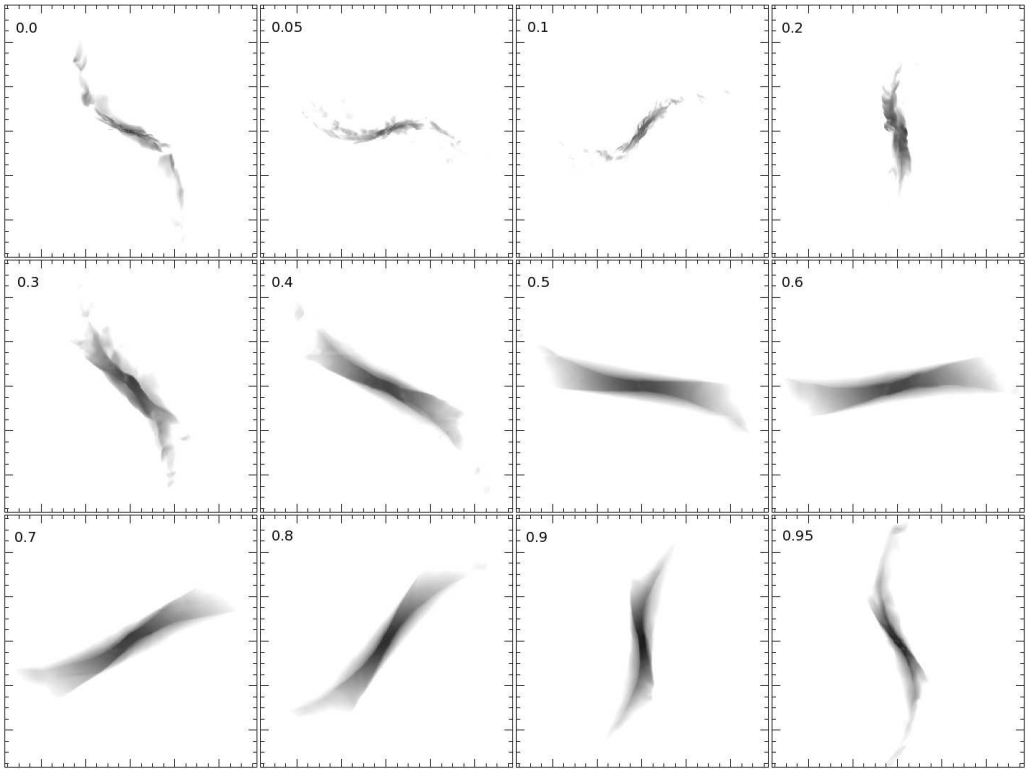,width=13.6cm}
\caption[]{As Fig.~\ref{fig:cwb4_xray_image_bb2} but showing broad-band ($2.5-10$\,keV) intensity images.
The grey scale and tick marks are the same as in Fig.~\ref{fig:cwb4_xray_image_bb2}.}
\label{fig:cwb4_xray_image_bb3}
\end{figure*}

\subsection{Model cwb4}
\label{sec:xray_cwb4}
\subsubsection{Images}
In contrast to the previous models which all had circular orbits,
model cwb4 simulates a CWB with an eccentric orbit
($e=0.36$).  This introduces a time-dependence to the
intrinsic emission, which now varies with phase, whereas it was
constant in the circular orbit models cwb1$-$cwb3.
Fig.~\ref{fig:cwb4_xray_image_bb2} shows intensity images of the
$0.5-2.5$\,keV broad-band emission from model cwb4 for an observer
directly above the orbital plane. The images show striking variations
in their brightness and morphology as a function of orbital phase,
reflecting the dramatic changes in the WCR during the orbit (see
Paper~I for full details of the hydrodynamics). Dramatic changes were
also seen in the thermal radio-to-sub-mm emission (see Paper~II for
further details).

There is a smooth morphology to the emission from the WCR at phase 0.5
(apastron), when the WCR is adiabatic and instabilities are rare.  As
the stars progress in phase the WCR rotates in the images. The WCR
becomes brighter and shows increasing curvature as the snapshots move
towards the time of periastron passage (the maximum surface brightness
of $1.9\times10^{8}\,{\rm erg\,cm^{-2}\,s^{-1}\,keV^{-1}\,ster^{-1}}$
occurs at phase 0.95. At periastron there is a distinct change in the
morphology of the images, with instabilities now clearly visible.  This
reflects the sudden cooling and formation of dense clumps within the
WCR (see Paper~I for further details). This morphology persists until
phase 0.2, at which point the ratio of the cooling time to the flow
time of the shocked gas near the apex of the WCR becomes significant
again. Dense cold clumps which formed during the periastron passage
are still present in the WCR, but there is now also a substantial
volume of hot gas. These cold clumps are gradually destroyed or
cleared out of the system, so that by phase 0.7 there is no longer any
cold, post-shock, gas on the hydrodynamical grid (note, however, that
even at phase 0.5 this process is far from complete). Careful
examination reveals that individual clumps and their ablated tails are
visible in these images, although the level of detail is not great
enough for this to be apparent in Fig.~\ref{fig:cwb4_xray_image_bb2}.
Since the WCR rotates to follow the motion of the stars, while the
dense clumps flow out on almost ballistic trajectories, the clumps are
often seen exiting the WCR through its trailing
shock. They are then exposed to the full ``fury'' of whichever
high-speed wind they find themselves in, and are enveloped by a
bowshock and high temperature plasma.  In Paper~I it was speculated
that this additional contribution to the overall amount of hot plasma
in the system may be significant in terms of the resulting X-ray
luminosity. However, it is now clear from
Fig.~\ref{fig:cwb4_xray_image_bb2} that such regions have a negligible
effect in this regard (although careful inspection reveals that there
are in fact noticeable features in the intensity images).
 
Fig.~\ref{fig:cwb4_xray_image_bb3} shows intensity images of the
$2.5-10$\,keV emission. The emission is clearly less extended and is
more concentrated towards the central part of the WCR. Otherwise the
morphology and brightness of the emission behaves in a rather similar
way to that in the $0.5-2.5$\,keV images. The highest surface
brightness of $3.5\times10^{7}\,{\rm
erg\,cm^{-2}\,s^{-1}\,keV^{-1}\,ster^{-1}}$ occurs at phase 0.9.

\subsubsection{Lightcurves and spectra}
\begin{figure*}
\psfig{figure=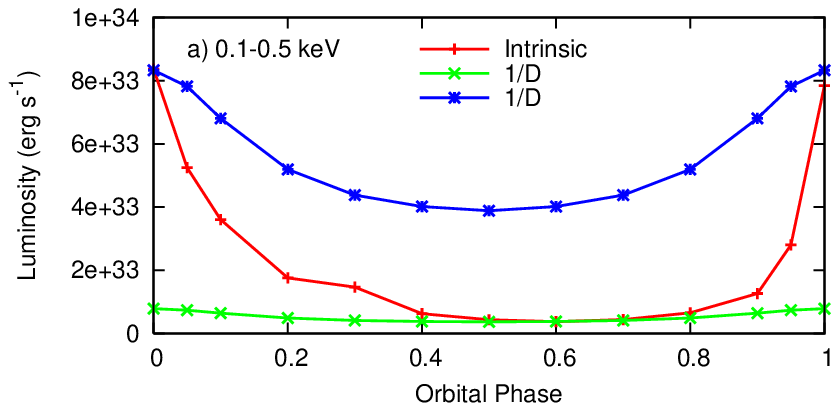,width=5.67cm}
\psfig{figure=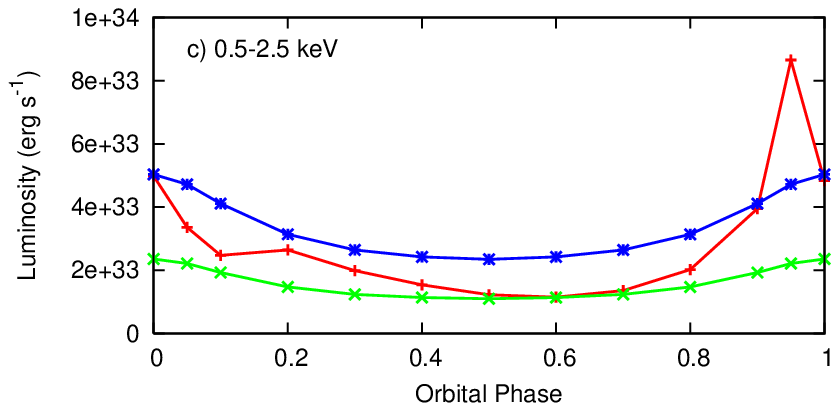,width=5.67cm}
\psfig{figure=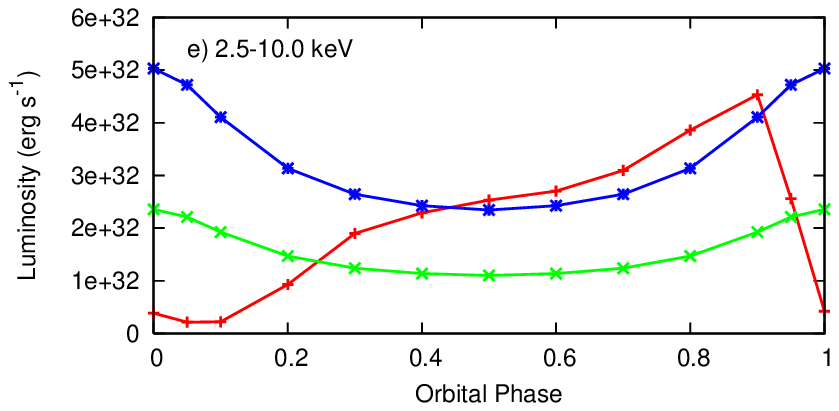,width=5.67cm}
\psfig{figure=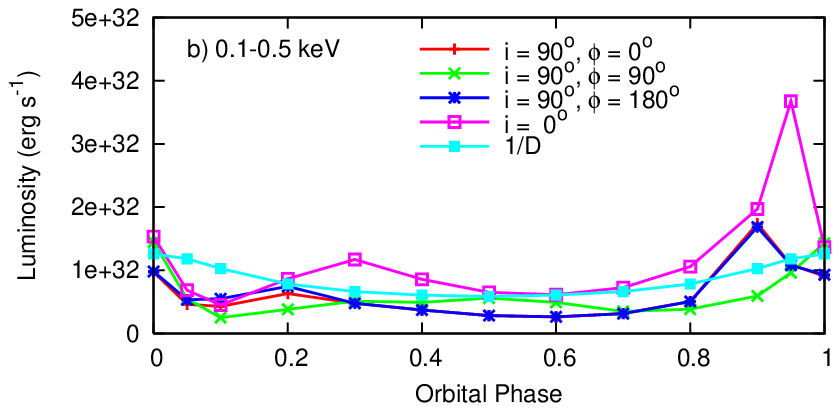,width=5.67cm}
\psfig{figure=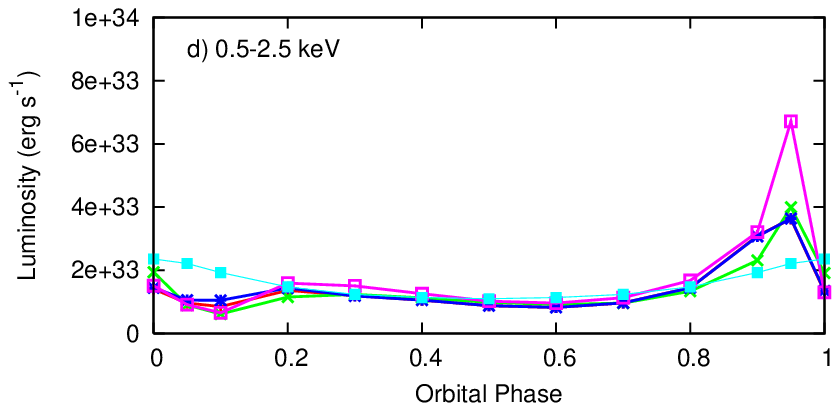,width=5.67cm}
\psfig{figure=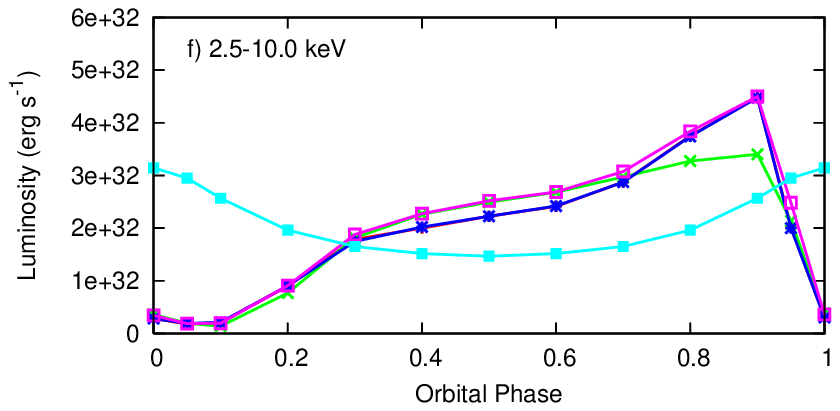,width=5.67cm}
\caption[]{X-ray lightcurves for model cwb4. The lightcurves were
calculated over the energy bands $0.1-0.5\;$keV (left), $0.5-2.5\;$keV
(middle), and $2.5-10\;$keV (right). The top panels show the
intrinsic luminosity (red curves), while the bottom panels show the attenuated
(observed) luminosity for viewing angles in the orbital plane ($i = 90^{\circ}$)
with $\phi=0^{\circ}$, $90^{\circ}$, and $180^{\circ}$ (red, green and blue
curves, respectively), and directly above/below the orbital plane
($i = 0^{\circ}$, pink curves). 
In all panels curves displaying a $1/d_{\rm sep}$ variation
(with arbitrary normalization) are over-plotted.}
\label{fig:cwb4_xray_lc}
\end{figure*}

Fig.~\ref{fig:cwb4_xray_lc} displays the lightcurves which are
obtained from model cwb4. The top panels show lightcurves of the
intrinsic emission (red curves), while the bottom panels show
attenuated lightcurves.  The intrinsic X-ray emission is no longer
constant with phase, but, at low energies, reaches a maximum at or
near periastron as the stars reach their point of closest approach
(Fig.~\ref{fig:cwb4_xray_lc}a). In contrast the $2.5-10$\,keV
intrinsic emission (Fig.~\ref{fig:cwb4_xray_lc}e) peaks at phase 0.9, 
and then undergoes a precipitous drop so that it is
actually close to its minimum value at periastron.

The phase-dependence of the intrinsic emission is complicated, and
depends on a combination of factors, including the current separation
of the stars, their separation in the recent past, and the variation
of the pre-shock wind densities and speeds and the post-shock cooling
efficiency over this interval. The intrinsic emission from systems
where the WCR is largely adiabatic should scale as $1/d_{\rm sep}$
\citep{Stevens:1992}. This relationship breaks if cooling within
the WCR becomes important \citep[cf. Fig.~9
in][]{Pittard:1997}, and/or if the pre-shock speeds of the winds
change. Both of these events occur in model cwb4, so therefore it is
not surprising that the intrinsic luminosity does not follow this scaling.
In fact the intrinsic luminosity varies by factors of approximately
20, 4 and 20 in the $0.1-0.5$, $0.5-2.5$ and $2.5-10$\,keV 
lightcurves, respectively, whereas for comparison a $1/d_{\rm sep}$
response predicts only a factor of 2 variation.
For an observer with viewing angles of $i = 90^{\circ}$ 
and $\phi=0^{\circ}$, the ISM corrected $0.5-10$\,keV luminosity 
varies between $1.22\times10^{33}\,\ergps$ at phase 0.1, and
$4.53\times10^{33}\,\ergps$ at phase 0.95, representing a 
change in $L_{\rm x}/L_{\rm bol}$ from $8.56\times10^{-7}$ to
$3.18\times10^{-6}$, respectively. For an observer located 
pole-on ($i,\phi = 0^{\circ},0^{\circ}$), $L_{\rm x}/L_{\rm bol}$
varies between $7.01\times10^{-7}$ at phase 0.1 to $5.88\times10^{-6}$
at phase 0.95. The peak values of $L_{\rm x}/L_{\rm bol}$ are
the highest obtained from any of the models in this work, and indicate not only
the strength of the colliding winds emission in this system, but also
the relative ease with which the X-ray photons escape the system at
favourable orientations because of the nature of the WCR.

\begin{figure*}
\psfig{figure=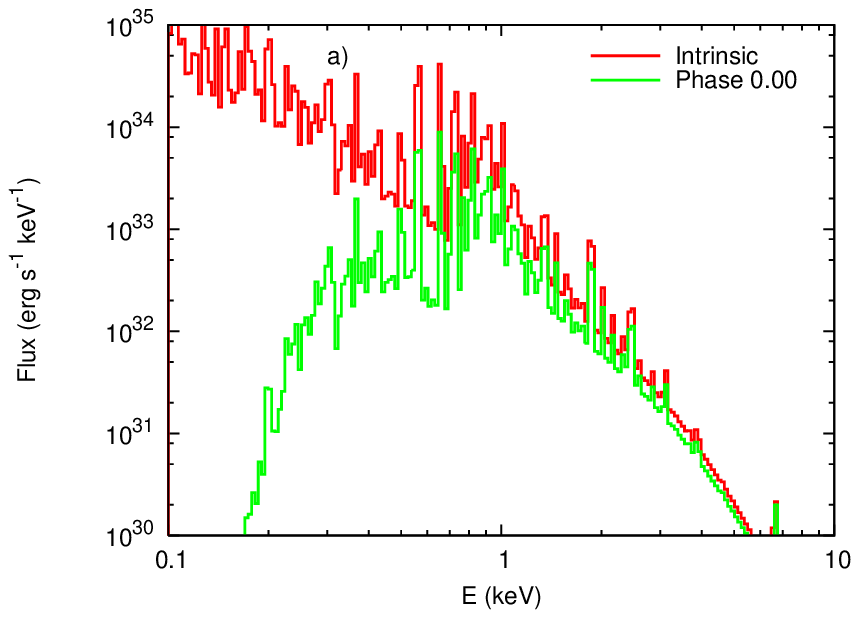,width=7.5cm}
\psfig{figure=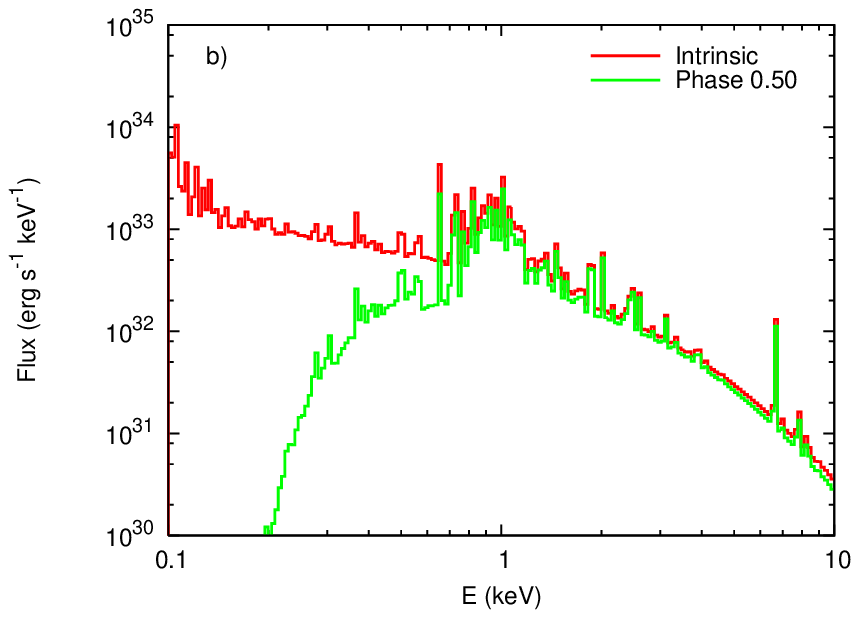,width=7.5cm}
\caption[]{Intrinsic and attenuated X-ray spectra from model cwb4 
for viewing angles $i=90^{\circ}$ and $\phi=0^{\circ}$
at a) periastron, and b) apastron.}
\label{fig:cwb4_xray_spec}
\end{figure*}

To understand the behaviour of the intrinsic lightcurves of model
cwb4, one needs to examine the intrinsic spectra, which display strong
phase-locked variation.  Fig.~\ref{fig:cwb4_xray_spec} shows that the
spectral hardness of the intrinsic X-ray emission changes by a huge
amount between periastron and apastron. At periastron the emission is
very soft, reflecting the low pre-shock wind speeds at this phase
($710 \kmps$ along the line of centres), whereas the emission is much
harder at apastron, since the increase in the stellar separation
allows the winds to accelerate to higher speeds before their collision
($1665 \kmps$ along the line of centres), and reduces the effects of
radiative inhibition (see Paper I).

\begin{figure*}
\psfig{figure=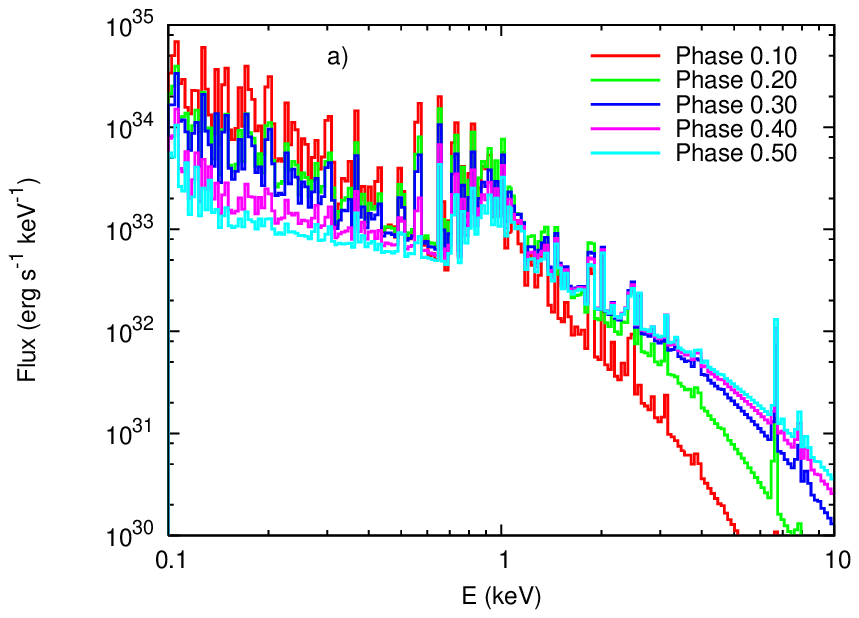,width=7.5cm}
\psfig{figure=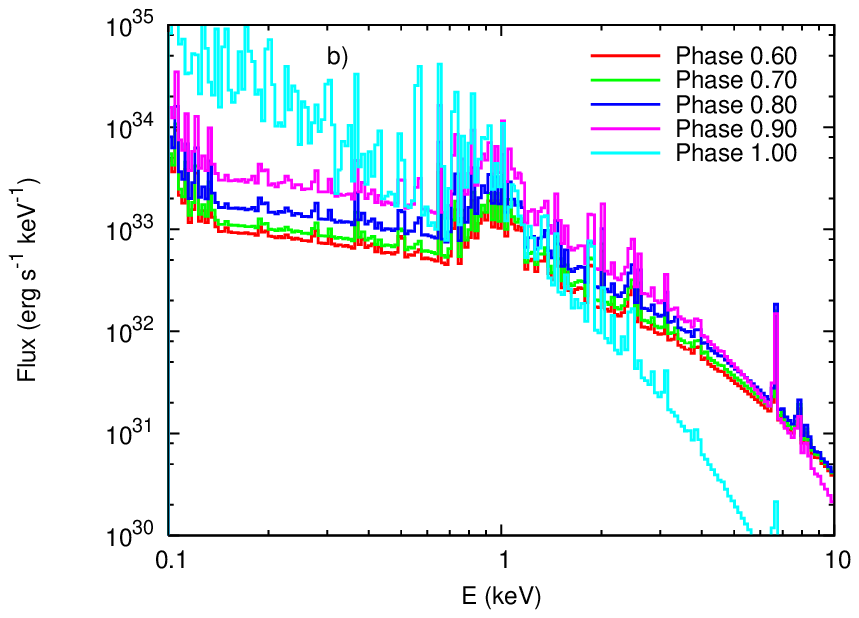,width=7.5cm}
\caption[]{Intrinsic X-ray spectra from model cwb4 as a function of 
orbital phase. a) Phase $0.1-0.5$. b) Phase $0.6-1.0$.}
\label{fig:cwb4_xray_spec1.5}
\end{figure*}

Interestingly, the intrinsic X-ray spectrum is at its hardest at phase
0.6, rather than at apastron. Thereafter, the intrinsic emission
begins to soften as the stars move closer together, the preshock wind
speeds decline, and the WCR becomes increasingly radiative (see
Fig.~\ref{fig:cwb4_xray_spec1.5}b).  The softening is initially
manifest as an increase in the soft emission, while the harder
emission (e.g. $E > 4\;$keV) remains at a relatively constant flux
until phase 0.9. Up to this point the flux at high energies
appears to be finely balanced between the intrinsic softening of the
spectrum and the increasing luminosity as the stars approach each
other.  However, this balancing act is over by phase 0.9, after which
the spectral softening rapidly accelerates. Between phase 0.9 and 1.0
(periastron) there is a precipitous collapse in the hard X-ray
emission, as the stars move deep within the acceleration zone of the
other's wind. The intrinsic emission is of comparable softness at
phase 1.0 and 1.1 (but brighter at phase 1.0 due to the reduced
stellar separation), and is markedly harder by phase 0.2 (see
Fig.~\ref{fig:cwb4_xray_spec1.5}a). These variations in the intrinsic
spectra explain the intrinsic lightcurves shown in the top panels of
Fig.~\ref{fig:cwb4_xray_lc}.

\begin{figure*}
\psfig{figure=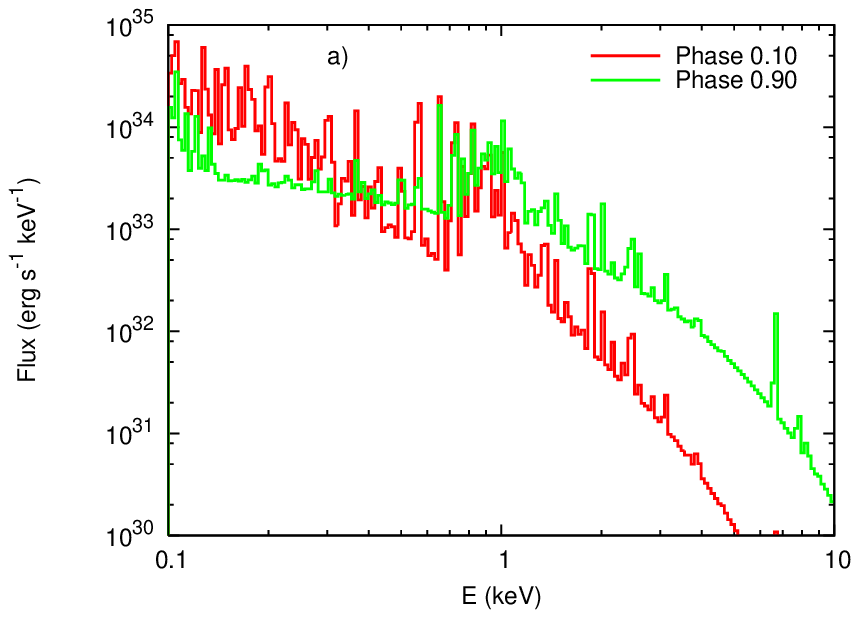,width=7.5cm}
\psfig{figure=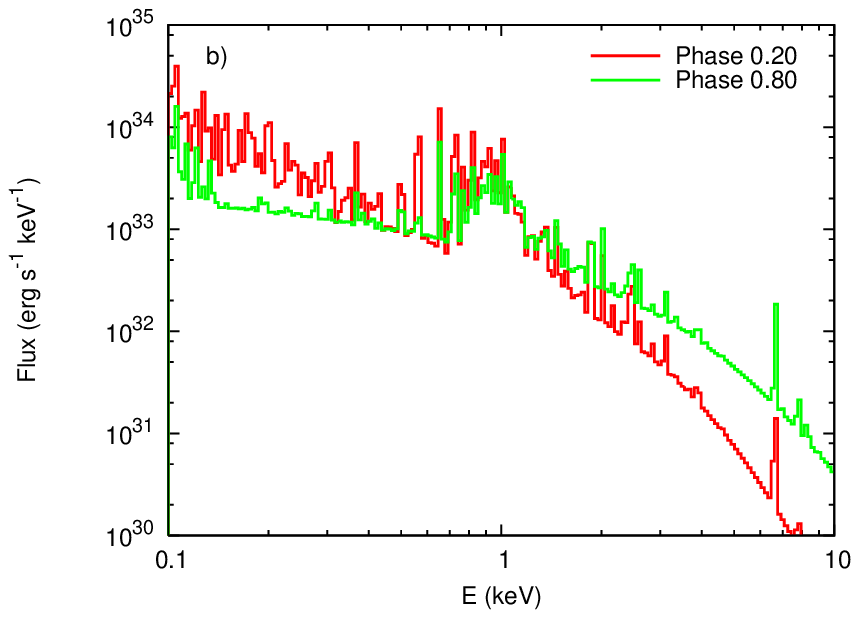,width=7.5cm}
\psfig{figure=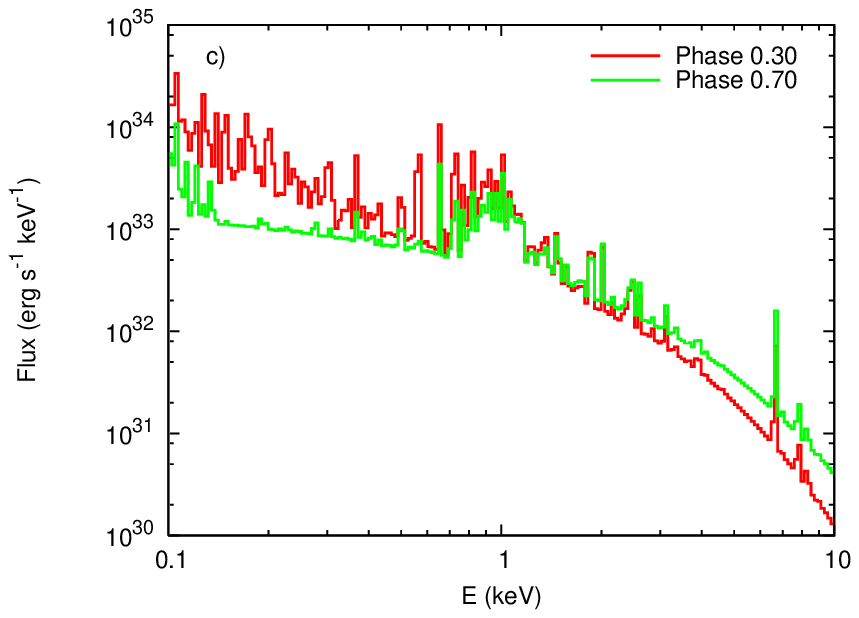,width=7.5cm}
\psfig{figure=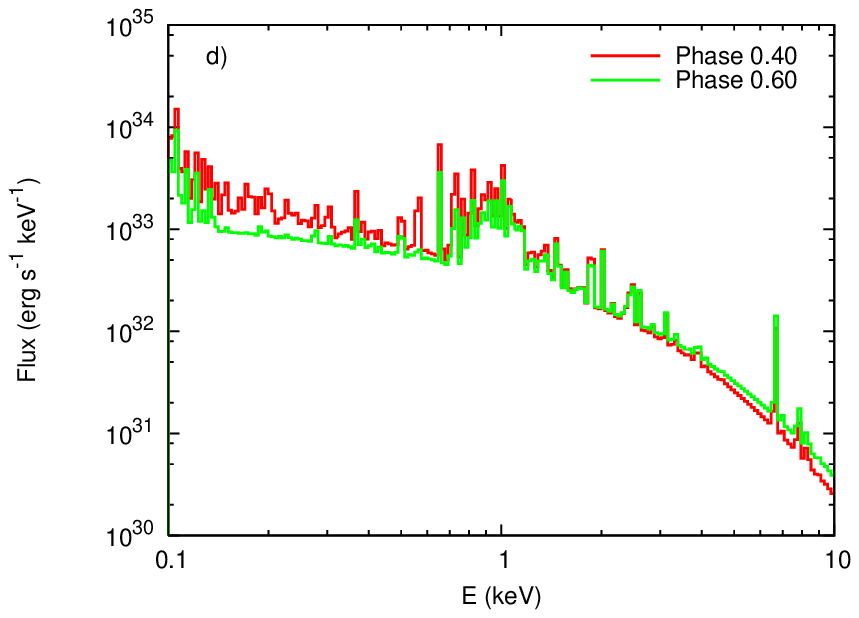,width=7.5cm}
\caption[]{Intrinsic X-ray spectra from model cwb4 as a function of 
orbital phase. In each panel the stars are at identical separations,
and either receding (phase $< 0.5$) or approaching (phase $> 0.5$). 
The hysteresis of the spectra is clearly evident.}
\label{fig:cwb4_xray_spec2}
\end{figure*}

Fig.~\ref{fig:cwb4_xray_spec2} reveals that there is a strong
hysteresis to the intrinsic emission, with large differences in the
spectrum at identical stellar separations depending on whether the
stars are approaching or receding from each other. The emission is
much harder as the stars move together compared to when they separate:
this is a natural consequence of the higher pre-shock wind speeds that
are attained prior to reductions in the stellar separation. As the
stars approach each other the conditions in the WCR reflect, to some
extent, the hot and rarefied plasma created at earlier orbital
phases. Similarly, as the stars recede the downstream conditions in
the WCR reflect still the lower preshock velocities at earlier orbital
phases, and in extremum the cold, dense plasma created during
periastron passage. The hysteresis is largest nearest periastron, when
changes in the pre-shock conditions are at their most rapid, and
smallest near apastron when the rate of change in the stellar
separation is most sedate. The observed hysteresis is also partly due
to variations in the relative wind speeds towards each star - when the
stellar separation is decreasing, the stars (and thus also their
winds) have a component of their orbital velocity directed towards each
other, which augments the wind speeds in the centre of mass frame. The
opposite effect occurs when the stars recede from each other. This
mechanism enhances the post-shock temperature in the WCR after
apastron, and reduces it at comparable orbital phases prior to
apastron. Hysteresis of the thermal radio and sub-mm emission also occurs (see
Paper~II).

\begin{figure*}
\psfig{figure=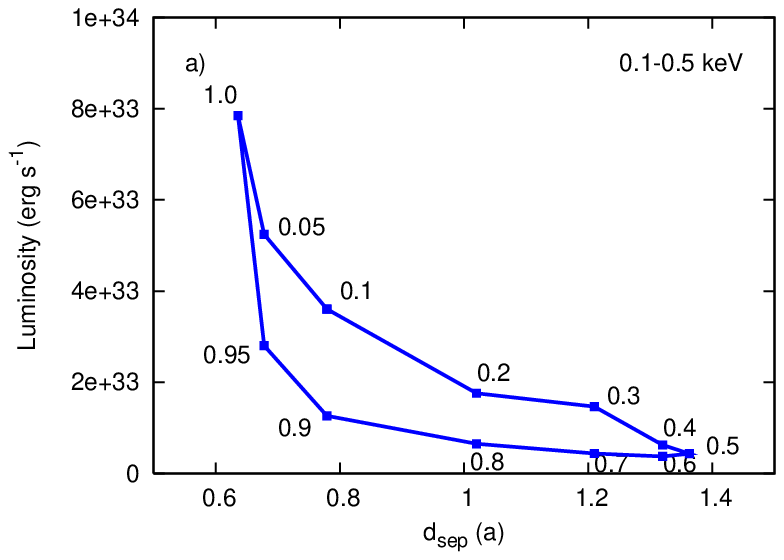,width=5.67cm}
\psfig{figure=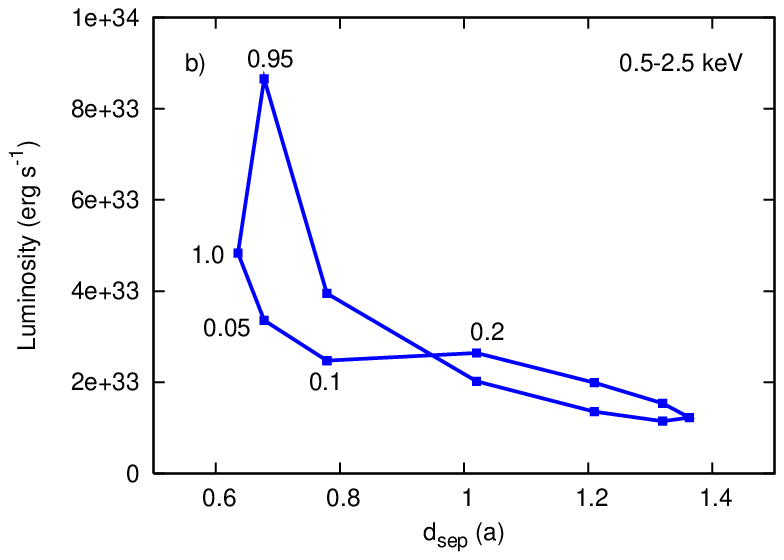,width=5.67cm}
\psfig{figure=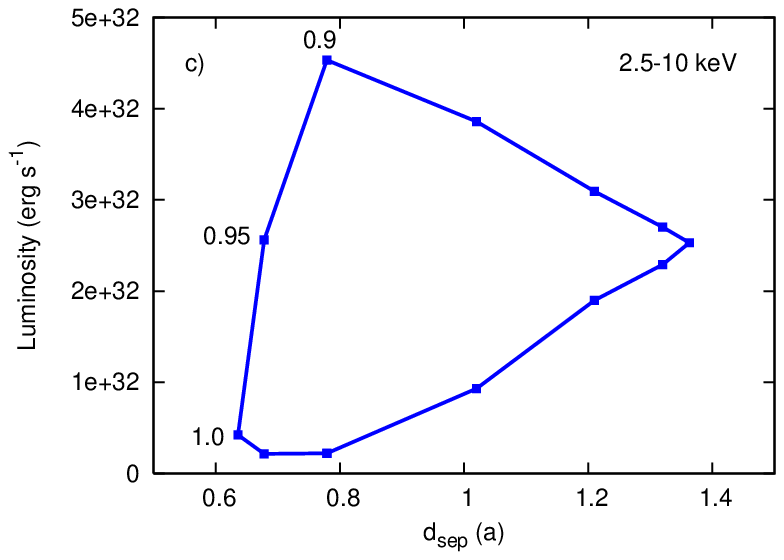,width=5.67cm}
\psfig{figure=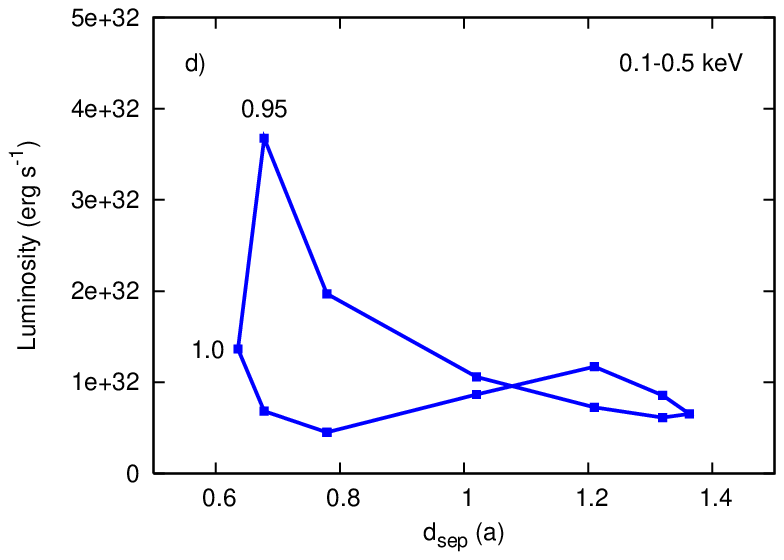,width=5.67cm}
\psfig{figure=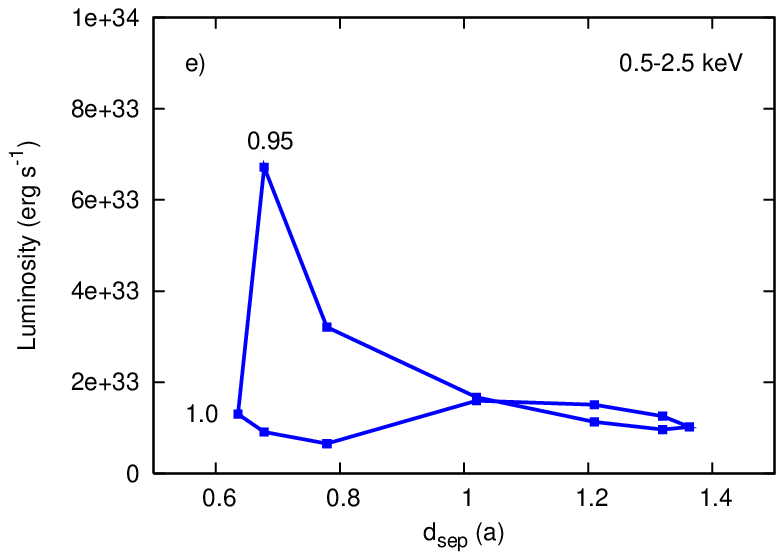,width=5.67cm}
\psfig{figure=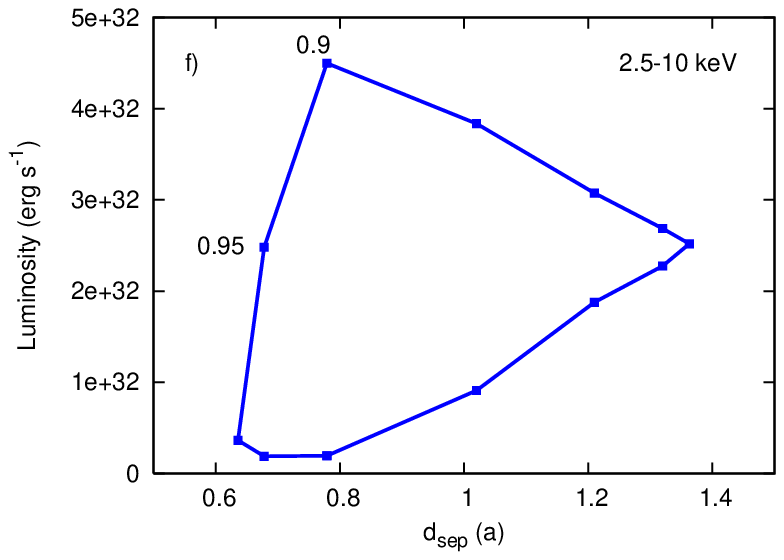,width=5.67cm}
\caption[]{Intrinsic (top row) and attenuated (bottom row) X-ray
luminosities from model cwb4 as a function of orbital separation. The
energy bands $0.1-0.5$\,keV (left), $0.5-2.5$\,keV (middle), and
$2.5-10$\,keV (right), are shown.  The attenuated lightcurves are for
an observer located directly above the orbital plane
($i=0^{\circ}$). Some orbital phases are marked on the plots.}
\label{fig:hysteresis_lc}
\end{figure*}

The hysteresis of the intrinsic emission is also clear when the
luminosities are plotted against stellar separation, as shown in
Fig.~\ref{fig:hysteresis_lc}(a)-(c). Interestingly, the intrinsic
emission in the $2.5-10$\,keV band is stronger as the stars approach
periastron, whereas the emission in the $0.1-0.5$\,keV band is
stronger as the stars recede. The hard emission requires high
temperature plasma, which the WCR is full of in the second half of the
orbit, but which is comparatively lacking in the first half of the
orbit. In contrast, emission in the soft band does not require such
high temperatures. Instead it is strongest when the postshock
densities are high, such as during the formation of cold clumps within
the WCR. Some of this emission will be from intermediate temperature
interface regions where hot plasma surrounds cooler clumps.  In
comparison the intrinsic emission in the $0.5-2.5$\,keV band displays
a transitional state: the emission is brighter in some parts of the
orbit when the stars are receding, and fainter in other parts.

Armed with an understanding of how the intrinsic X-ray emission varies
with orbital phase, we now examine the attenuated emission.  The
attenuated lightcurves shown in the bottom panels of
Fig.~\ref{fig:cwb4_xray_lc} display behaviour which depends on the
viewing angle of the observer. The lightcurves for observers at
$i=90^{\circ}$ and $\phi=0$ or $180^{\circ}$ are almost identical,
particularly in the hard $2.5-10$\,keV band. Such symmetry is
expected, given the identical stellar parameters. 
The $2.5-10$\,keV attenuated lightcurves show the closest
behaviour to their intrinsic counterpart, highlighting the ability of
hard X-rays to stream through the circumstellar environment relatively
unaffected by absorption.  For this same reason the $2.5-10$\,keV
attenuated lightcurves also show very little change with the viewing
angle of the observer, with the largest difference occuring at phases
$0.8-0.9$ when there is more attenuation for an observer at
$i=90^{\circ}$ and $\phi=90^{\circ}$ than for other orientations,
because the stars are eclipsing the apex of the WCR at this time (see
Paper~I).

The attenuated $0.5-2.5$\,keV lightcurves all peak at
phase 0.95, irrespective of the orientation of the observer, in
agreement with the timing of the peak in the intrinsic lightcurve.
However, the height and shape of the maximum in the attenuated
lightcurves is dependent on the orientation. The greatest luminosity
in the attenuated $0.5-2.5$\,keV lightcurves occurs for an observer
viewing the system face on ($i = 0^{\circ}$). In contrast, the timing
of the maximum in the attenuated $0.1-0.5$\,keV lightcurves is
strongly dependent on the viewer's orientation, ranging from phase 0.9
for observers in the orbital plane at viewing angles of
$\phi=0^{\circ}$ and $180^{\circ}$, to phase 0.0 (periastron) for a
viewing angle of $\phi=90^{\circ}$. For an observer viewing the system
face on the maximum occurs at phase 0.95 - incidentally, this is also
the highest maximum seen in the $0.1-0.5$\,keV lightcurves. These
differences in the timing of the maxima reflect the propensity for
soft X-rays to be attenuated by the circumstellar environment, and the
dependence of the strength of this attenuation on the orientation of
the observer.  The strong circumstellar absorption near periastron
arises from the enhanced wind densities around the WCR due to the
reduced stellar separation and pre-shock wind speeds.

Broad minima which are roughly centered on apastron occur in most of
the attenuated $0.1-0.5$\,keV and $0.5-2.5$\,keV lightcurves, though
there is a slight maximum at apastron in the $0.1-0.5$\,keV lightcurve for an
observer with $i=90^{\circ}$ and $\phi=90^{\circ}$ (since
lines-of-sight from the apex of the WCR initially pass through the low
opacity WCR). The minima following periastron are generally deepest at
phase 0.1.  The shape of the minimum is also generally quite smooth,
though the $0.1-0.5$\,keV lightcurves for $i=90^{\circ}$ and
$\phi=0^{\circ}$ and $180^{\circ}$ are noticeable for showing more
structure (Fig.~\ref{fig:cwb4_xray_lc}(b) shows the luminosity
levelling out between phases $0.95-1.0$, before falling more steeply
between phases $0.0-0.05$).

Panels (d)-(f) of Fig.~\ref{fig:hysteresis_lc} show the attenuated
luminosities in the three energy bands for an observer at
$i=90^{\circ}$ as a function of orbital separation. The $2.5-10$\,keV
emission is most similar to its intrinsic counterpart
(Fig.~\ref{fig:hysteresis_lc}c), again illustrating the relative ease
at which the hard X-rays travel through the circumstellar
environment. In contrast, the emission in the $0.1-0.5$\,keV band
suffers severe attenuation, and this has a large impact on the shape
of its hysteresis curve (compare Figs.~\ref{fig:hysteresis_lc}a and d).
As we have already seen, the attenuation is particularly severe after
periastron, when there is not much hot, low opacity, plasma in the
WCR. 

The eccentric orbit means that the emission from model cwb4 at various
times resembles that from models cwb1 and cwb2.  Careful examination
reveals that the periastron spectrum (Fig.~\ref{fig:cwb4_xray_spec}a)
is slightly harder than the phase 0.0 spectrum from model cwb1
(cf. Fig.~\ref{fig:cwb123_xray_spec}a), with the Fe~K emission visible in
the former plot. This reflects the fact that the downstream flow in
model cwb4 contains hotter plasma (which was shocked when the winds
previously collided at a higher speed) than in model cwb1. That they
are otherwise so similar reflects the fact that emission at the apex
of the WCR (which responds much quicker to changing pre-shock conditions
than emission from far downstream) dominates the total emission in
this model as the post-shock gas at the apex of the WCR rapidly
becomes extremely radiative.

Likewise, there is a high degree of similarity between the apastron
spectrum (Fig.~\ref{fig:cwb4_xray_spec}b) and the phase 0.5 spectrum
from model cwb2 (cf. Fig.~\ref{fig:cwb123_xray_spec}c), the former being
slightly softer.  This is again consistent with the recent history of the
WCR. Their likeness reflects the fact that in model cwb4, the rate of
change in the stellar separation is at its most sedate at apastron.
The dynamical timescale for flow out of the system is then short
compared to the timescale for significant orbital change. At phase 0.5,
$t_{\rm flow} \sim \frac{d_{\rm sep}}{v_{\infty}/2} \approx 0.5\;$d,
while the time for the orbital separation to change 10 per cent from its
apastron value is $0.19\;P_{\rm orb} = 1.2\;$d.

\begin{figure*}
\psfig{figure=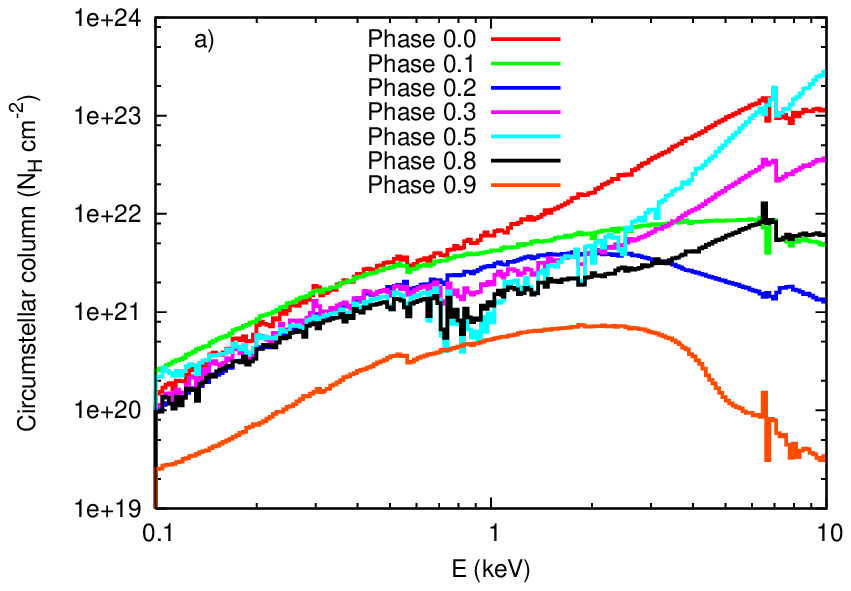,width=7.5cm}
\psfig{figure=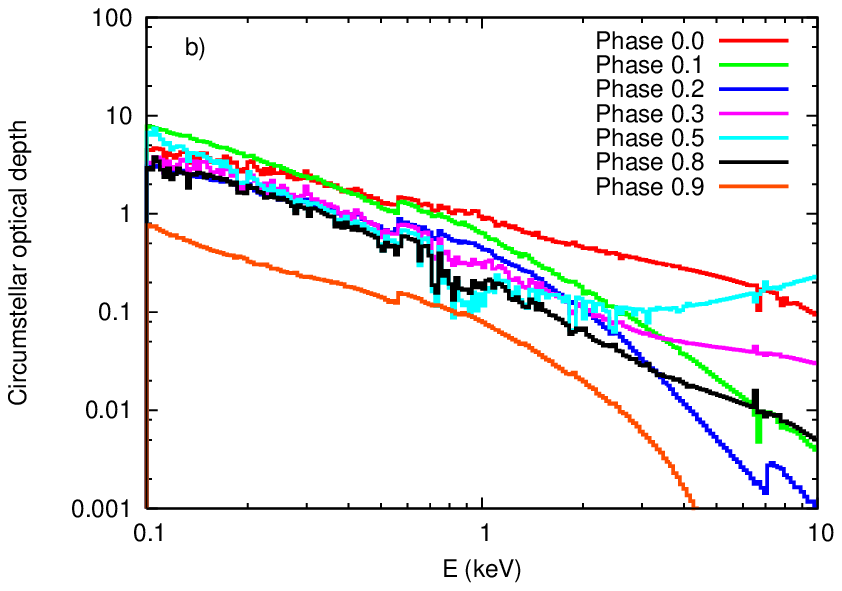,width=7.5cm}
\caption[]{a) ``Effective'' circumstellar column (a) and optical depth (b)
from model cwb4 for an observer in the
orbital plane ($i=90^{\circ}$) as a function of energy and
orbital phase. Note that occultation can severely affect
the value of the ``effective'' column. It is therefore not
always an accurate reflection of the column that the observed
X-rays pass through (see Sec.~\ref{sec:cwb1_spectra} for further details).}
\label{fig:cwb4_xray_circum}
\end{figure*}

Fig.~\ref{fig:cwb4_xray_circum} shows the effective circumstellar
column and optical depth as a function of energy and orbital phase for
an observer with $i=90^{\circ}$ and $\phi=0^{\circ}$. The columns at
phase 0.0 and 0.5 bear some similarity to those obtained from models
cwb1 and cwb2, in the same way that the attenuated spectra do. Thus,
to first order there is rough agreement between the emission and
absorption characteristics of an eccentric system at periastron and
apastron and circular systems of identical stellar separation, at 
least for the region of parameter space covered in these models.

Fig.~\ref{fig:cwb4_xray_circum}(a) also shows that there are extremely
large phase-dependent variations in the energy-dependent column. The
variation of the column to the high energy (e.g. 5\,keV) emission is
largely due to changes in the degree of occultation to this emission,
with high occultation at conjunction (phase 0.0 and 0.5), and lesser
occultation near quadrature (phase 0.14 and 0.86). There is a severe
decline in the column to the high energy emission (due largely to
changes in the degree of occultation) between periastron and phase
0.1.  At phase 0.2 the observer views the WCR apex through hot plasma
further downstream (see Fig.~10 in Paper~I), and the circumstellar
column declines at all energies. By phase 0.3 one of the stars is
already positioning itself in front of parts of the apex of the WCR,
and the column to the high energy emission increases from its value at
phase 0.2. The high energy column further increases to a maximum near
phase 0.5. The column to the high energy emission eases again by
phase 0.8, while at phase 0.9 the observer again views the WCR apex through
hot plasma further downstream, which results in the lowest effective
column and optical depth at all energies and phases.

In contrast to the 4 orders of magnitude variation in the column at
10\,keV, the column to the low energy emission is surprisingly steady
during the majority of the orbit, which reflects the large volume from
which this emission arises. However, there is again a
significant reduction in the low energy circumstellar column at phase
0.9 due to the reasons previously given.

The phase dependent variation in the optical depth shown in
Fig.~\ref{fig:cwb4_xray_circum}(b) to a large part reflects the
changes in the circumstellar columns commented on above.
The variation in the optical depth as a function of phase
spans the range $0.8-7$ at 0.1\,keV, $0.1-1$ at 1\,keV, and 
$<10^{-3} - 0.2$ at 10\,keV.

\subsubsection{Spectral fits}
\begin{figure*}
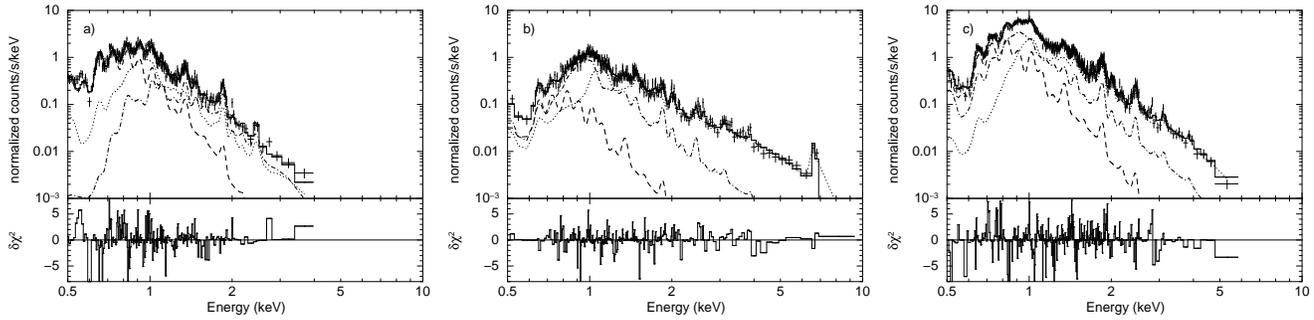

\psfig{figure=cwb4_suzaku_phase0.0_fit1_good.ps,width=4.2cm,angle=-90}
\psfig{figure=cwb4_suzaku_phase0.60_fit3_good.ps,width=4.2cm,angle=-90}
\psfig{figure=cwb4_suzaku_phase0.95_fit2_good.ps,width=4.2cm,angle=-90}
\caption[]{Three-temperature fits to ``fake'' {\em Suzaku} spectra from model cwb4
for an observer in the orbital plane viewing the system at conjunction
($\phi=0^{\circ}$). a) Phase 0.0, b) phase 0.6, c) phase 0.95. 
The dramatic variation of the spectra with orbital phase is clearly shown.
The assumed exposure time is 20\,ksec in each case.}
\label{fig:cwb4_fits}
\end{figure*}

Table~\ref{tab:fitresults_cwb4} and Fig.~\ref{fig:cwb4_fits} show the
results of spectral fits to ``fake'' {\em Suzaku} spectra with a
nominal exposure time of 20\,ksec generated from model cwb4.  The
spectra in Fig.~\ref{fig:cwb4_fits} were specifically chosen to
highlight the large spectral variations which occur over the course of
the stellar orbit. The spectrum at phase 0.0 is almost as soft as the
emission gets in this model (the spectrum at phase 0.05 is marginally
softer), reflecting the strong cooling of the plasma in the WCR at
this phase. By phase 0.2 (not shown) the spectrum is noticeably harder (and more
luminous). The spectrum at phase 0.6 is about as hard as the emission
gets, and shows a prominent Fe\,K line. At phase 0.95 the spectrum
is at its most luminous, and is again softer, reflecting the lower
pre-shock wind speeds at this phase. There are now not enough counts
at high energies to detect the Fe\,K line in the binned spectra, 
though it is of course seen in our theoretical spectra.
%(though it clearly
%still exists - see Fig.~\ref{fig:cwb4_xray_spec1.5}). 

\begin{figure*}
\psfig{figure=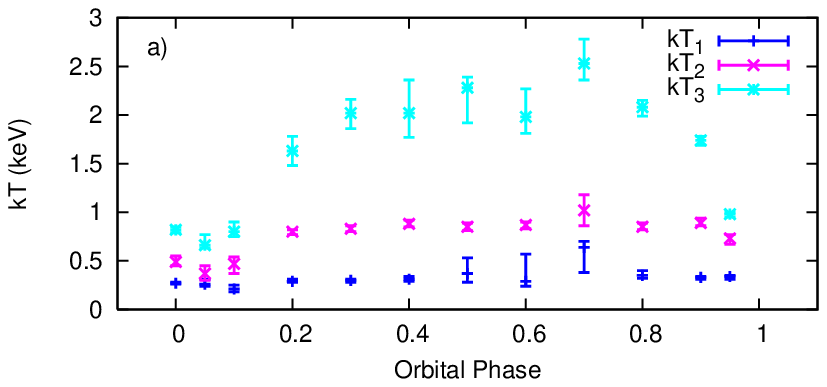,width=8.50cm}
\psfig{figure=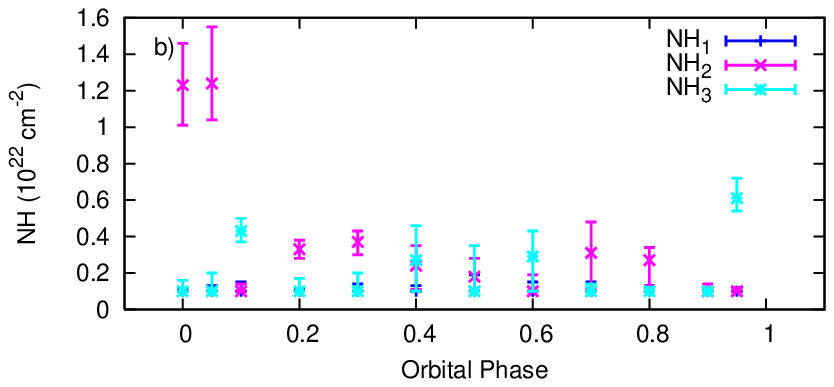,width=8.50cm}
\psfig{figure=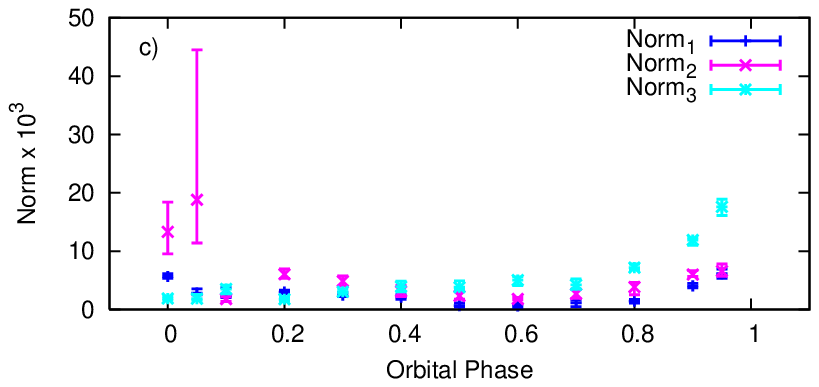,width=8.50cm}
\psfig{figure=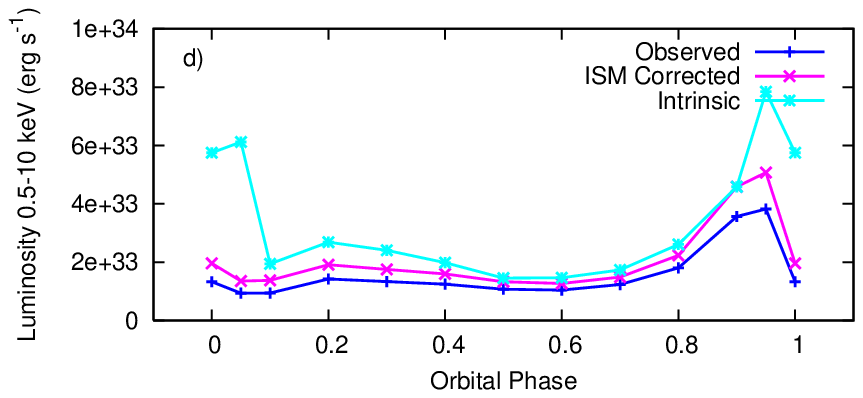,width=8.50cm}
\caption[]{Orbital phase variability of the fit parameters to the ``fake''
{\em Suzaku} spectra from model cwb4
for an observer with $i=90^{\circ}$ and $\phi=0^{\circ}$.
Panel a) shows the temperature variation, panel b) the variation in
the absorbing column (the displayed values include the assumed ISM
column of $N_{\rm H} = 10^{21}\,{\rm cm^{-2}}$), and panel c) the variation of the normalization,
of each fit component. Panel d) shows the observed, ISM
corrected and intrinsic luminosity of each component combined.}
\label{fig:fit_statistics}
\end{figure*}

The spectral variability shown in Fig.~\ref{fig:cwb4_fits} is
reflected in changes in the values of the fit parameters (see
Table~\ref{tab:fitresults_cwb4}), which are plotted in
Fig.~\ref{fig:fit_statistics}. The temperature of the hottest mekal
component shows significant variation, changing from
$0.66^{+0.11}_{-0.03}$\,keV at phase 0.05, to $2.53^{+0.25}_{-0.17}$\,keV
at phase 0.7. In addition, significant enhancements in the normalization
of the components occur as periastron is approached. Between apastron
and phase 0.95 the normalization of the hot component increases by a
factor of 4.6, far above the corresponding $1/d_{\rm sep}$ value.
Panel d) shows the combined luminosity of the three mekal components.
A comparison with Fig.~\ref{fig:cwb4_xray_lc} reveals that the fits do
a good job of recovering the phase variation of the observed and also
the intrinsic luminosity. It is clear that 
the fits infer the presence of significant circumstellar absorption 
between phases $0.95-1.05$, which is responsible for the large
difference in the intrinsic and ISM corrected luminosities.

It is also noteable that the normalization of the hot component
dominates those of the cooler components from phase 0.4 to 0.95, while
the normalization to the warm (i.e. the second) component dominates at
phase 0.0 and 0.05. There is no need for additional circumstellar
absorption to the mekal components at phase 0.9, which is consistent
with the low value of the effective circumstellar column at this phase
(see Fig.~\ref{fig:cwb4_xray_circum}a). While the fits do require
substantial additional column to the warm component at phase 0.0
($1.13^{+0.23}_{-0.21}\times10^{22}\,{\rm cm^{-2}}$), this
extra absorption is about 3 times higher than the effective
circumstellar column at 0.49\,keV, as shown in
Fig.~\ref{fig:cwb4_xray_circum}a).  It is also puzzling why the
spectral fitting did not require extra absorption (above the ISM
value) to the cold and hot mekal components at this phase, despite these
making significant contributions to the overall observed emission (see
the top left panel in Fig.~\ref{fig:cwb4_fits}).

\begin{table*}
\begin{center}
\caption[]{Spectral fitting results to simulated spectra from models
cwb1, cwb2, and cwb3. The normalization of the mekal components (norm)
is defined as $[10^{-14}/(4 \pi D^{2})] \int n_{\rm e}\,n_{\rm
H}\,dV$, where $D$ is the distance to the source (in cm), and $n_{\rm
e}$ and $n_{\rm H}$ are the electron and hydrogen number densities (in
${\rm cm^{-3}}$), respectively. The indicated range in each fit
parameter is the 90 per cent confidence interval for one interesting
parameter. The last three columns give the observed flux, the flux
corrected for ISM absorption, and the intrinsic flux, between 0.5 and
10\,keV. All fits are to ``fake'' observations of 10\,ksec duration,
except the {\em Suzaku} observations of models cwb2 and cwb3, where the
assumed exposure times are 40\,ksec and 20\,ksec, respectively.}
\label{tab:fitresults}
\begin{tabular}{cccccccccc}
\hline
\hline
Model & Phase & Model fit & $kT$ & $N_{\rm H}$ & Norm & $\chi^{2}_{\nu}$\,(d.o.f.) & Observed flux & ISM corrected flux & Intrinsic flux\\
 & & & (keV) & ($10^{22}\,{\rm cm^{-2}}$) & ($10^{-3}$) & & \multicolumn{3}{c}{(${\rm 10^{-12} erg\,cm^{-2}\,s^{-1}}$)} \\
\hline
\multicolumn{2}{l}{Chandra fits} \\
cwb1 & 0.0 & 2T & $0.28^{0.31}_{0.25}$ & $0.1^{0.18}_{0.1}$ & $2.02^{3.45}_{1.90}$ & 1.27 (93) & $2.27$ & $3.70$ & $37.0$ \\ %fit cwb1fit3
     & & & $0.65^{0.80}_{0.61}$ & $0.59^{0.67}_{0.52}$ & $4.07^{4.58}_{2.86}$ &  & $2.21$ & $2.74$& $10.4$\\
cwb1 & 0.25 & 2T & $0.25^{0.27}_{0.24}$ & $0.1^{0.13}_{0.1}$ & $4.19^{5.27}_{3.98}$ & 1.71 (110) & $4.21$ & $71.3$ & $71.3$ \\%fit cwb1fit14
     & & & $0.67^{0.70}_{0.64}$ & $0.39^{0.42}_{0.35}$ & $6.09^{6.32}_{5.62}$ & & $5.31$& $6.81$& $15.5$ \\
cwb2 & 0.0 & 3T & $0.33^{0.36}_{0.31}$ & $0.1^{0.15}_{0.1}$ & $2.69^{3.66}_{2.26}$ & 1.13 (165) & $3.51$ & $5.46$ & $5.46$ \\%fit cwb2_fit13
     & & & $1.01^{1.05}_{0.94}$ & $0.25^{0.34}_{0.1}$ & $3.45^{4.55}_{2.15}$ & & $4.01$ & $4.90$ & $6.99$ \\
     & & & $2.47^{3.25}_{2.03}$ & $0.31^{0.56}_{0.1}$ & $3.58^{4.81}_{2.55}$ & & $3.68$ & $4.00$ & $5.14$ \\ 
cwb2 & 0.25 & 3T & $0.32^{0.34}_{0.26}$ & $0.10^{0.32}_{0.1}$ & $2.18^{5.67}_{1.92}$ & 1.04 (171) & $2.75$& $4.31$ & $4.34$ \\ %fit cwb2_fit16
     & & & $0.84^{0.93}_{0.78}$ & $0.1^{0.24}_{0.1}$ & $1.63^{2.45}_{1.44}$ & & $2.86$ & $3.75$ & $3.75$\\
     & & & $2.02^{2.12}_{1.92}$ & $0.1^{0.15}_{0.1}$ & $5.55^{5.84}_{5.09}$ & & $6.65$ & $7.76$ & $7.76$\\
\multicolumn{2}{l}{Suzaku fits} \\
cwb1 & 0.0 & 3T & $0.11^{0.14}_{0.0}$ & $0.17^{0.36}_{0.1}$ & $4.21^{69.6}_{0.84}$ & 1.50 (324) & $0.30$ & $0.64$ & $1.08$ \\ %fit cwb1_suzaku_fit4
     & & & $0.31^{0.33}_{0.29}$ & $0.15^{0.23}_{0.1}$ & $2.17^{3.63}_{1.57}$ & & $2.19$ & $3.40$ & $4.24$ \\
     & & & $0.69^{0.74}_{0.66}$ & $0.57^{0.62}_{0.53}$ & $3.34^{3.65}_{2.78}$ & & $1.99$ & $2.45$ & $8.46$ \\
cwb1 & 0.25 & 3T & $0.12^{0.14}_{0.14}$ & $0.102^{0.21}_{0.1}$ & $3.17^{7.11}_{0.0}$ & 2.28 (401) & $0.58$ & $1.26$ & $1.28$ \\%fit cwb1_suzaku_fit5
     & & & $0.31^{0.31}_{0.30}$ & $0.14^{0.14}_{0.1}$ & $4.64^{5.47}_{3.80}$ & & $4.85$ & $7.57$ & $9.03$ \\ %error estimated from fit, not from error command
     & & & $0.74^{0.74}_{0.73}$ & $0.41^{0.43}_{0.40}$ & $4.60^{4.80}_{4.38}$ & & $3.91$ & $4.89$ & $11.4$ \\ 
%cwb2 & 0.0 & 2T & $0.31^{}_{}$ & $0.1^{}_{}$ & $3.11^{}_{}$ & 2.34 (298) & & & \\ %fit cwb2_suzaku_fit2
%     & & & $1.07^{}_{}$ & $0.59^{}_{}$ & $8.86^{}_{}$ & & & \\
cwb2 & 0.0 & 3T & $0.33^{0.35}_{0.31}$ & $0.1^{0.14}_{0.1}$ & $1.89^{2.43}_{1.59}$ & 1.18 (295) & $2.46$ & $3.82$ & $3.82$\\ %fit cwb2_suzaku4.ps
     & & & $0.88^{0.94}_{0.85}$ & $0.17^{0.28}_{0.1}$ & $2.23^{3.04}_{1.74}$ & & $3.26$ & $4.16$ & $4.96$ \\
     & & & $2.38^{2.55}_{2.10}$ & $0.1^{0.27}_{0.1}$ & $4.10^{4.86}_{3.69}$ & & $5.08$ & $5.84$ & $5.84$ \\
cwb2 & 0.25 & 3T & $0.31^{0.33}_{0.29}$ & $0.1^{0.13}_{0.1}$ & $2.17^{2.81}_{1.91}$ & 1.17 (319) & $2.70$ & $4.26$ & $4.26$ \\%fit cwb2_suzaku8.ps
     & & & $0.83^{0.87}_{0.79}$ & $0.13^{0.31}_{0.1}$ & $1.91^{3.15}_{1.63}$ & & $3.15$ & $4.10$ & $4.41$\\
     & & & $2.14^{2.26}_{2.02}$ & $0.1^{0.17}_{0.1}$ & $5.16^{5.56}_{4.60}$ & & $6.24$ & $7.25$ & $7.25$ \\
%cwb3 & 0.0 & 3T (10ksec) & $0.30^{0.32}_{0.28}$ & $0.1^{0.15}_{0.1}$ & $1.73^{2.53}_{1.52}$ & 1.13 (221) & & & \\ %fit cwb3_suzaku_fit2
%     & & & $0.81^{0.85}_{0.77}$ & $0.10^{0.33}_{0.1}$(NP) & $1.59^{2.74}_{1.45}$ & & & \\
%     & & & $1.68^{2.25}_{1.52}$ & $0.34^{0.50}_{0.1}$ & $2.65^{3.09}_{1.52}$ & & & \\
cwb3 & 0.0 & 3T & $0.30^{0.31}_{0.29}$ & $0.1^{0.12}_{0.1}$ & $1.82^{2.05}_{1.61}$ & 1.21 (339) & $2.20$ & $3.50$ & $3.50$ \\ %fit cwb3_suzaku_fit3
     & & & $0.84^{0.87}_{0.82}$ & $0.15^{0.24}_{0.1}$ & $1.71^{2.21}_{1.43}$ & & $2.65$ & $3.43$ & $3.93$ \\
     & & & $2.09^{2.21}_{1.69}$ & $0.1^{0.41}_{0.1}$ & $2.12^{2.86}_{1.90}$ & & $2.55$ & $2.97$ & $2.97$ \\
cwb3 & 0.25 & 3T & $0.30^{0.31}_{0.28}$ & $0.1^{0.11}_{0.1}$ & $1.69^{1.87}_{1.54}$ & 1.35 (353) & $2.02$ & $3.23$ & $3.23$ \\ %fit cwb3_suzaku_fit6
     & & & $0.85^{0.88}_{0.81}$ & $0.10^{0.18}_{0.1}$ & $1.50^{1.85}_{1.37}$ & & $2.62$ & $3.43$ & $3.23$ \\
     & & & $1.79^{1.89}_{1.71}$ & $0.1^{0.12}_{0.1}$ & $2.60^{2.77}_{2.47}$ & & $3.08$ & $3.63$ & $3.63$\\
cwb3 & 0.5 & 3T & $0.29^{0.31}_{0.28}$ & $0.1^{0.12}_{0.1}$ & $1.75^{1.97}_{1.57}$ & 1.14 (324) & $2.05$ & $3.30$ & $3.30$ \\ %fit cwb3_suzaku_fit5
     & & & $0.85^{0.88}_{0.83}$ & $0.14^{0.24}_{0.1}$ & $1.57^{2.09}_{1.32}$ & & $2.48$ & $3.21$ & $3.60$ \\
     & & & $1.66^{1.86}_{1.54}$ & $0.46^{0.58}_{0.31}$ & $2.67^{3.07}_{2.16}$ & & $2.14$ & $2.35$ & $3.84$ \\
cwb3 & 0.75 & 3T & $0.30^{0.31}_{0.29}$ & $0.1^{0.12}_{0.1}$ & $1.75^{2.01}_{1.63}$ & 0.94 (356) & $2.12$ & $3.38$ & $3.38$ \\ %fit cwb3_suzaku_fit4
     & & & $0.85^{0.88}_{0.83}$ & $0.1^{0.14}_{0.1}$ & $1.66^{1.83}_{1.55}$ & & $2.89$ & $3.78$ & $3.78$ \\
     & & & $1.99^{2.08}_{1.90}$ & $0.1^{0.12}_{0.1}$ & $2.51^{2.66}_{2.40}$ & & $3.02$ & $3.52$ & $3.52$ \\ 
\hline
\end{tabular}
\end{center}
\end{table*}

\begin{table}
\begin{center}
\caption[]{Comparison of the intrinsic $0.5-10$\,keV luminosity of the theoretical spectra with that obtained from the spectral fits.
All values are in units of $10^{33}\ergps$. The percentage ratio of the inferred to the actual intrinsic luminosity is shown in brackets for each case.}
\label{tab:lumerror}
\begin{tabular}{lll}
\hline
\hline
 & Model cwb1 & Model cwb2 \\
\hline
Intrinsic luminosity & $3.11$ & $1.81$ \\
{\em Chandra} phase 0.0 & $1.69$ (54\%) & $2.10$ (116\%) \\
{\em Chandra} phase 0.25 & $2.71$ (87\%) & $1.90$ (105\%) \\
{\em Suzaku} phase 0.0 & $1.65$ (53\%) & $1.75$ (97\%) \\
{\em Suzaku} phase 0.25 & $2.59$ (83\%) & $1.90$ (105\%) \\
\hline
\end{tabular}
\end{center}
\end{table}

\begin{table*}
\begin{center}
\caption[]{As Table~\ref{tab:fitresults} but for model cwb4. All the fits are of 3-temperature mekal models to simulated {\em Suzaku} spectra
of 20\,ksec duration, for an assumed inclination $i=90^{\circ}$.}
\label{tab:fitresults_cwb4}
\begin{tabular}{cccccccc}
\hline
\hline
Phase & $kT$ & $N_{\rm H}$ & Norm & $\chi^{2}_{\nu}$\,(d.o.f.) & Observed flux & ISM corrected flux & Intrinsic flux\\
 & (keV) & ($10^{22}\,{\rm cm^{-2}}$) & ($10^{-3}$) & & \multicolumn{3}{c}{(${\rm 10^{-12} erg\,cm^{-2}\,s^{-1}}$)} \\
\hline
0.00 & $0.27^{0.28}_{0.26}$ & $0.1^{0.11}_{0.1}$ & $5.65^{6.13}_{5.35}$ & 1.51 (253) & $6.04$ & $10.0$ & $10.0$\\ %fit cwb4_suzaku_phase0.0_fit1
     & $0.49^{0.55}_{0.45}$ & $1.23^{1.46}_{1.01}$ & $13.3^{18.4}_{9.52}$ & & $1.72$ & $1.99$ & $33.7$\\
     & $0.82^{0.85}_{0.79}$ & $0.1^{0.16}_{0.1}$ & $1.88^{2.25}_{1.62}$ & & $3.33$ & $4.38$ & $4.38$ \\
0.05 & $0.26^{0.31}_{0.24}$ & $0.1^{0.13}_{0.1}$ & $2.81^{3.53}_{2.49}$ & 1.41 (218) & $2.94$& $4.91$ & $4.91$\\ %fit cwb4_suzaku_phase0.05_fit1
     & $0.37^{0.45}_{0.30}$ & $1.24^{1.55}_{1.04}$ & $18.8^{44.5}_{11.4}$ & & $1.39$ & $1.64$ & $41.5$\\
     & $0.66^{0.77}_{0.63}$ & $0.1^{0.20}_{0.1}$ & $1.86^{2.59}_{1.59}$ & & $3.48$ & $4.76$ & $4.76$\\
0.10 & $0.21^{0.25}_{0.18}$ & $0.1^{0.15}_{0.1}$ & $2.35^{3.75}_{2.02}$ & 1.32 (209) & $1.91$ & $3.48$ & $3.48$\\ %fit cwb4_suzaku_phase0.1_fit1
     & $0.47^{0.54}_{0.37}$ & $0.1^{0.14}_{0.1}$ & $1.78^{2.17}_{1.36}$ & & $3.07$ & $4.44$ & $4.44$\\
     & $0.80^{0.90}_{0.75}$ & $0.43^{0.50}_{0.37}$ & $3.50^{4.17}_{2.68}$ & & $2.90$ & $3.57$ & $8.30$\\
0.20 & $0.29^{0.28}_{0.31}$ & $0.1^{0.11}_{0.1}$ & $2.97^{3.29}_{2.75}$ & 1.35 (290) & $3.51$ & $5.62$ & $5.62$ \\%fit cwb4_suzaku_phase0.20_fit1
     & $0.80^{0.82}_{0.77}$ & $0.33^{0.38}_{0.28}$ & $6.03^{6.92}_{5.33}$ & & $6.23$ & $7.83$ & $14.3$\\
     & $1.63^{1.78}_{1.48}$ & $0.1^{0.17}_{0.1}$ & $1.75^{2.11}_{1.41}$ & & $2.13$ & $2.53$ & $2.53$ \\
0.30 & $0.30^{0.31}_{0.28}$ & $0.1^{0.14}_{0.1}$ & $2.45^{2.91}_{2.23}$ & 1.12 (298) & $2.94$ & $4.69$& $4.69$\\%fit cwb4_suzaku_phase0.30_fit1
     & $0.83^{0.86}_{0.80}$ & $0.37^{0.43}_{0.30}$ & $4.84^{5.75}_{4.02}$ & & $4.59$& $5.68$ & $11.2$\\
     & $2.02^{2.16}_{1.86}$ & $0.1^{0.20}_{0.1}$ & $3.03^{3.49}_{2.86}$ & & $3.64$& $4.24$ & $4.24$\\ 
0.40 & $0.31^{0.34}_{0.29}$ & $0.1^{0.13}_{0.1}$ & $2.10^{2.54}_{1.75}$ & 1.05 (287) & $2.61$ & $4.12$& $4.12$\\%fit cwb4_suzaku_phase0.40_fit1.ps
     & $0.88^{0.92}_{0.85}$ & $0.24^{0.35}_{0.11}$ & $3.11^{4.12}_{2.24}$ & & $3.91$& $4.92$ & $6.96$\\
     & $2.02^{2.36}_{1.77}$ & $0.27^{0.46}_{0.1}$ & $3.92^{4.86}_{3.04}$ & & $3.86$& $4.28$ & $5.49$\\
0.50 & $0.37^{0.53}_{0.28}$ & $0.1^{0.19}_{0.1}$ & $0.65^{1.11}_{0.42}$ & 1.02 (267) & $0.94$ & $1.42$ & $1.42$\\ %fit cwb4_suzaku_phase0.50_fit1
     & $0.85^{0.89}_{0.81}$ & $0.18^{0.28}_{0.1}$ & $2.33^{3.06}_{1.75}$ & & $3.33$ & $4.27$ & $5.32$\\
     & $2.28^{2.39}_{1.92}$ & $0.1^{0.35}_{0.1}$ & $3.82^{4.90}_{3.52}$ & & $4.68$& $5.40$ & $5.40$\\
0.60 & $0.29^{0.57}_{0.24}$ & $0.1^{0.15}_{0.1}$ & $0.62^{0.93}_{0.23}$ & 1.02 (267) & $0.73$& $1.17$ & $1.17$\\ %fit cwb4_suzaku_phase0.60_fit3
     & $0.87^{0.90}_{0.83}$ & $0.1^{0.19}_{0.1}$ & $1.79^{2.04}_{1.63}$ & & $3.09$& $4.03$& $4.03$\\
     & $1.98^{2.27}_{1.81}$ & $0.29^{0.43}_{0.1}$ & $5.04^{5.59}_{4.10}$ & & $4.87$ & $5.40$ & $7.04$\\
0.70 & $0.64^{0.70}_{0.38}$ & $0.1^{0.15}_{0.1}$ & $1.17^{1.64}_{0.49}$ & 1.00 (303) & $2.19$ & $3.01$ & $3.01$\\%fit cwb4_suzaku_phase0.70_fit1
     & $1.02^{1.18}_{0.86}$ & $0.31^{0.48}_{0.11}$ & $2.68^{3.40}_{1.91}$ & & $2.77$ & $3.33$ & $5.34$\\
     & $2.53^{2.78}_{2.36}$ & $0.1^{0.14}_{0.1}$ & $4.23^{5.25}_{3.51}$ & & $5.35$& $6.12$ & $6.12$\\
0.80 & $0.35^{0.40}_{0.32}$ & $0.1^{0.13}_{0.1}$ & $1.43^{1.73}_{1.10}$ & 1.14 (378) & $1.99$& $3.04$& $3.04$\\ %fit cwb4_suzaku_phase0.80
     & $0.85^{0.89}_{0.82}$ & $0.27^{0.34}_{0.13}$ & $3.85^{4.64}_{2.53}$ & & $4.47$& $5.60$ &$8.77$\\
     & $2.08^{2.15}_{1.99}$ & $0.1^{0.12}_{0.1}$ & $7.16^{7.79}_{6.83}$ & & $8.63$& $10.0$ & $10.0$\\
0.90 & $0.33^{0.34}_{0.31}$ & $0.1^{0.11}_{0.1}$ & $4.04^{4.47}_{3.72}$ & 1.24 (490) & $5.30$ & $8.23$ & $8.23$\\ %fit cwb4_suzaku_phase0.90_fit1
     & $0.89^{0.94}_{0.86}$ & $0.1^{0.14}_{0.1}$ & $5.99^{6.66}_{5.64}$ & & $10.3$ & $13.3$ & $13.3$ \\
     & $1.74^{1.78}_{1.69}$ & $0.1^{0.11}_{0.1}$ & $11.9^{12.3}_{11.0}$ & & $14.2$ & $16.8$ & $16.8$ \\
0.95 & $0.34^{0.35}_{0.31}$ & $0.1^{0.12}_{0.1}$ & $6.14^{6.94}_{5.34}$ & 1.47 (427) & $8.24$ & $12.7$ & $12.7$ \\ %fit cwb4_suzaku_phase0.95_fit2
     & $0.73^{0.77}_{0.67}$ & $0.1^{0.12}_{0.1}$ & $6.58^{7.81}_{5.68}$ & & $12.2$ & $16.4$ & $16.4$ \\
     & $0.98^{1.00}_{0.95}$ & $0.61^{0.72}_{0.54}$ & $17.6^{18.9}_{16.1}$ & & $11.5$ & $13.3$ & $36.5$ \\
\hline
\end{tabular}
\end{center}
\end{table*}

\section{Comparison to other numerical models}
\label{sec:discuss_models}
The X-ray emission from O+O-star CWBs has been investigated using fully
hydrodynamical models \citep{Pittard:1997,Pittard:2000} and
``hydrid'' models \citep{Antokhin:2004,Parkin:2008}. The
analysis in the current work is a major improvement 
from that in \citet{Pittard:1997}, where the X-ray calculations
were based on 2D axisymmetric hydrodynamical models in which the
winds instantaneously accelerated to their terminal speeds.
As such, the plasma temperatures returned from model cwb1 are
much lower than those from model A in
\citet{Pittard:1997}. The hydrodynamical model underlying the analysis in
\citet{Pittard:2000} did consider the radiatively-driven 
acceleration of the winds, but remained 2D \citep{Pittard:1998}.
This has subsequently been shown to introduce an incorrect
phase dependence to the volume of hot gas in the WCR 
\citep{Lemaster:2007}.

In order to spatially resolve the post-shock cooling in highly
radiative WCRs, \citet{Antokhin:2004} introduced a model for the X-ray
emission in which the local post-shock cooling was decoupled from a
global solution of the ram-pressure balance. This work has the
advantage of a low computational cost, allowing the rapid exploration
of parameter space. However, its drawbacks include the neglect of
orbital motion and the use of simple ``beta'' velocity-laws to mimic
the winds' acceleration. The X-ray lightcurve of an equal winds system
(model A) shows two minima per orbit, with dual symmetry about both
conjunction and quadrature (this symmetry is broken in our models due
to the aberration of the WCR). Strong absorption occurs when the
observer's line-of-sight is down one of the arms of the WCR. This
signature is also seen from our model cwb1, though is greatly reduced
in strength, because of the downstream curvature of the WCR.

Three-dimensional colliding winds simulations of longer-period O+O
systems (with orbital periods of 1 month and 1 year), and eccentricity
of 0.3, have recently been presented by \citet{Parkin:2008}. This work
solved the equations of ram-pressure balance assuming that the winds
collide at their terminal velocities, and adopted an analytical
expression for the abberation angle, to construct a 3D surface of the
head of the WCR.  Further downstream, the gas in the WCR was assumed
to behave ballistically, which with the orbital motion of the stars
results in a downstream curvature of the WCR. Again this
approach benefits from its computational speed. The X-ray emission from this
model was then calculated by mapping onto the surface of the WCR the
emission calculated from a 2D hydrodynamical model. In all cases the
WCR was assumed to be adiabatic. Radiative transfer through the
computational volume with the inclusion of opacity then allowed the
generation of synthetic spectra and lightcurves. It was found that the
X-ray flux generally followed a $1/d_{\rm sep}$ scaling, but could
also display absorption related variations. While in principle the
inclination and orientation of the system can be constrained from
the shape of the X-ray lightcurve, in practice the lack of significant
circumstellar absorption in O+O star systems with periods of order 1
yr will make this very difficult.  The situation of course improves in
shorter period systems, and in systems with reasonably disparate wind
densities (e.g. WR+O, LBV+O, and LBV+WR systems).

\section{Comparison to observations}
\label{sec:discuss}

In this section the results from our models are compared to recent
X-ray observations of short period O+O-star systems. We focus first on
systems where the WCR is expected to be highly radiative, then on
systems where the WCR is expected to be more adiabatic, and then on
systems with eccentric orbits and unequal winds. 
We finish by discussing the observational evidence
for non-thermal X-ray emission.

\subsection{Highly radiative systems}
Model cwb1 is similar to HD\,215835 (DH\,Cep), HD\,165052, and HD\,159176.  The
X-ray emission from each of these systems is discussed below. 

\subsubsection{HD\,165052}
HD\,165052 is a double-lined binary which has recently been classified
as O6.5V + O7.5V \citep{Arias:2002}. The orbit is slightly eccentric
(e=0.09), with an inclination thought to be around $i=20^{\circ}$.
This system was first considered as a CWB by \citet{Luo:1990}.
The nature of the WCR remains unclear. An analysis of the wind ram
pressures reveals that there may be no balance at periastron, though
this might be possible at apastron. The shocked winds are likely to
both be radiative ($\chi \sim 1$).

A {\it ROSAT} lightcurve of HD\,165052 presented by
\citet{Corcoran:1996} was rephased to the correct 2.95\,d orbital
period by \citet{Arias:2002}.  The lightcurve shows two minima per
orbit, roughly centered on phases 0.0 and 0.5 (conjunction), in good
agreement with the lightcurves presented in
Fig.~\ref{fig:cwb123_xray_lc} from model cwb1. However, the amplitude
of variation of the X-ray flux is a factor of 2 or so, which is
slightly greater than the variation in the $i=30^{\circ}$ lightcurves
of model cwb1 in Fig.~\ref{fig:cwb123_xray_lc}. This difference may be
caused by the slight orbital eccentricity, and/or may reflect the
absence of a wind-wind balance or the unequal wind strengths.

Unfortunately, HD\,165052 lies just outside the field-of-view of
recent X-ray observations: a serendipitous {\em XMM-Newton}
observation of the Lagoon Nebula (M8) and the very young open cluster
NGC\,6530 \citep{Rauw:2002c} centered on the non-thermal radio
emitting O-star 9\,Sgr \citep{Rauw:2002b}, and {\em Chandra}
observations of NGC\,6530 \citep{Damiani:2004} and 9\,Sgr (PI Gagne,
not published). Dedicated observations of HD\,165052 are clearly
necessary.

\subsubsection{HD\,159176}
\label{sec:hd159176}
A single observation of the O7V+O7V system HD\,159176 was made with
{\it XMM-Newton} in March 2001 \citep{DeBecker:2004b}.  HD\,159176 has
an orbital period of 3.367\,d, and a suspected inclination angle $i
\sim 50^{\circ}$. Since the stars (and winds) are identical the winds
should collide at the system centre of mass, half-way between the
stars.  The collision should be radiative, even if the mass-loss rates
are an order of magnitude lower than the values quoted in
\citet{Pachoulakis:1996}.  The {\it XMM-Newton} observation was taken
just after quadrature, corresponding to phases $0.28-0.41$ (or,
alternatively, phase $0.78-0.91$) in our models (note that the phase
reported in De Becker et al. is $0.53-0.66$, but in their paper phase
0.0 corresponds to the maximum radial velocity of the primary
component). The observed luminosity in the $0.4-10$\,keV band is
$1.5\times10^{33}\;\ergps$.  Fig.~\ref{fig:cwb123_xray_lc} shows that
the attenuated luminosity near quadrature in model cwb1 is, in fact,
very similar to that observed from HD\,159176. This is important,
because theoretical models presented by \citet{DeBecker:2004b} which
decoupled the local postshock cooling from the global dynamics of the
WCR overpredicted the luminosity by at least a factor of 4.

A two-temperature mekal fit to the combined EPIC and RGS data yielded
$kT_{1} = 0.21\pm0.02$\,keV and $kT_{2} = 0.96\pm0.01$\,keV, with a
circumstellar column $N_{\rm H} = 0.40\pm0.01 \times 10^{22}\;{\rm
cm^{-2}}$. The latter is in good agreement with the energy dependent
column at quadrature from model cwb1 as shown in
Fig.~\ref{fig:cwb123_xray_circum}(a). We further find that for an
inclination angle $i=90^{\circ}$, at phase 0.25 the characteristic
mekal temperatures from spectral fits to model cwb1 are
$0.25^{+0.02}_{-0.01}$\,keV and $0.67\pm0.03$\,keV, and the
circumstellar column to the hot component is $0.29^{+0.03}_{-0.04}
\times 10^{22}\;{\rm cm^{-2}}$.  Therefore, the agreement is quite
good. The slightly higher temperature of the hot component in the fit
to the real data may indicate slightly higher pre-shock wind
speeds than obtained in our model.

\begin{figure}
\begin{center}
\psfig{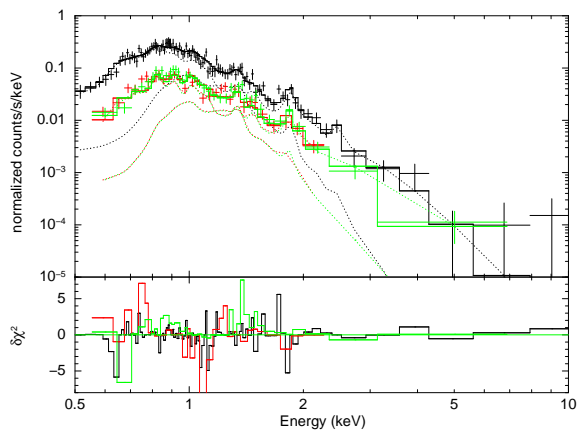}
\caption[]{Two-temperature mekal fit to combined {\em XMM-Newton} 
PN and MOS1+2 spectra of DH\,Cep. An interstellar column of
$N_{\rm H} = 0.667\times10^{22}\,{\rm cm^{-2}}$ has been assumed.}
\label{fig:dhcep_fit}
\end{center}
\end{figure}

\subsubsection{HD\,215835 (DH\,Cep)}
HD\,215835 (DH\,Cep), located in the young open cluster NGC\,7380, is
an extremely tight O6V\,+\,O7V binary with a circular orbit of just
2.1109\,d period \citep*{Penny:1997}. The stars are tidally distorted,
but at present are fully detached. The inclination remains slightly
uncertain ($35^{\circ}< i < 51^{\circ}$). The significant optical
polarization is consistent with scattering within a WCR, while its
short-term variability may be related to instabilities in the WCR
\citep{Corcoran:1991}.  An analysis of the wind ram pressures reveals
that there is unlikely to be any balance, and the primary wind should
directly impact the secondary star. However, once again a detailed
simulation of this system will be necessary to place this hypothesis
on a firmer footing.

DH\,Cep was observed by {\em XMM-Newton} on $19^{\rm th}$\,December, 2003,
for $\approx 30$\,ksec with the MOS1 and MOS2 cameras, and $\approx 28$\,ksec
with the PN camera. It was reduced using 
standard scripts and version 9.0.0 of SAS, and the spectrum was
rebinned to obtain a minimum of 25 counts per bin. Fig.~\ref{fig:dhcep_fit}
shows a two-temperature mekal fit, with an assumed ISM column of
$0.667\times10^{22}\,{\rm cm^{-2}}$, the weighted average returned
by the NH tool at the HEASARC website. The fit is good ($\chi^{2}_{\nu} = 1.18$),
and has the following parameters:
$kT_{1} = 0.24^{+0.02}_{-0.01}$\,keV, $kT_{2} = 0.76^{+0.09}_{-0.06}$\,keV,
$N_{\rm H_{1}} = 0.67^{+0.04}_{-0.00}\times10^{22}\,{\rm cm^{-2}}$,
$N_{\rm H_{2}} = 1.05^{+0.17}_{-0.14}\times10^{22}\,{\rm cm^{-2}}$,
${\rm norm_{1}} = 2.07^{+0.73}_{-0.24}\times10^{-3}$, and 
${\rm norm_{2}} = 4.29^{+0.80}_{-0.74}\times10^{-4}$.
The returned temperatures are similar to those from model cwb1. The fit
needed no extra circumstellar absorption to the soft component (as also
found for model cwb1), but favoured a circumstellar column of about 
$0.4\times10^{22}\,{\rm cm^{-2}}$ to the hard component (again in good
agreement with model cwb1). Assuming a distance of 3.73\,kpc \citep*{Massey:1995},
the ISM corrected $0.5-10$\,keV luminosity is $6.3\times10^{33}\,\ergps$.
With ${\rm log}\,L_{\rm bol} = 5.85$ \citep{Penny:1997}, 
we obtain $L_{\rm x}/L_{\rm bol} = 2.3\times10^{-6}$. The X-ray
luminosity and $L_{\rm x}/L_{\rm bol}$ values are slightly higher
than obtained from model cwb1. It is entirely plausible that
such a high X-ray luminosity could be obtained even if the primary
wind crushes that of its companion, because in such a scenario 
the secondary star would nonetheless intercept a large fraction of the
primary's wind due to its proximity.

The slightly hotter temperature of the 2nd mekal component compared to
model cwb1 may indicate that the primary wind indeed collides directly
with the surface of the companion star, since it then has more room to
accelerate before its collision, and thus the ability (ignoring
potential braking and inhibition effects - see \citet*{Gayley:1997}
and \citet*{Stevens:1994}, respectively) to reach higher pre-shock
speeds.

%intrinsic $0.5-10$\,keV luminosity is $7.4\times10^{33}\,\ergps$.
%Some X-ray obs of DH Cep are also discussed in Corcoran:1991

\subsection{Intermediate and adiabatic systems}
Systems with slightly longer orbital periods and wider stellar
separations, some of which bear similarities to model cwb2, are now
considered. Specifically, observations of HD\,93161A
and HD\,47129 (Plaskett's star) are discussed.

\subsubsection{HD\,93161A}
HD\,93161A is an O8V\,+\,O9V system with an orbital period of
8.566\,d. The mimimum masses from the orbital solution are quite
large, which suggests that the inclination of the orbit is high, most
probably $> 75^{\circ}$ \citep{Naze:2005}. The winds achieve a
ram pressure balance, and both shocked winds are largely adiabatic
(the primary's more so: we estimate that $\chi_{1} \approx 30$ and
$\chi_{2} \approx 5$).

The system was observed five times with {\it XMM-Newton} during 2000
and 2001, but no significant variations at energies above 0.5\,keV
were detected. This is nicely consistent with the theoretical
lightcurves from model cwb2 shown in Fig.~\ref{fig:cwb123_xray_lc},
which display little orbital variation in energy bands above 0.5\,keV
due to: i) the low attenuation through the circumstellar winds caused
by the relatively low mass-loss rates and the relatively wide stellar
separation, and ii) low occultation due to the large size of the WCR
relative to the stars.

The observed luminosity in the $0.4-10$\,keV band is
$1.2\times10^{32}\;\ergps$, which is an order of magnitude fainter
than model cwb2. At first glance this would appear to be a big
problem, but there are two important factors which can reconcile this
issue. Firstly, the interstellar column to this system ($N_{\rm H} =
4.5 \times 10^{21}\;{\rm cm^{-2}}$) is significantly higher than
assumed for our models. Secondly, the winds in this system are likely
to be much more feeble compared to the O6V winds assumed in our
model. Since the X-ray luminosity $L_{\rm x} \propto \Mdot^{2}$ in
adiabatic systems, a reduction of a factor of 3 or so in the mass-loss
rates would bring the observed and theoretical luminosities into
better agreement. Such a reduction is consistent with the expected
change in the mass-loss rate between an O6 and an O8/9 main-sequence
star.

\citet{Naze:2005} find that two-temperature mekal fits to the X-ray
spectra of HD\,93161A yield average values of $kT_{1} =
0.28\pm0.02$\,keV and $kT_{2} = 0.76\pm0.17$\,keV.  The derived
circumstellar column varies between $0.45\times10^{22}\;{\rm cm^{-2}}$
(at phases $0.18-0.53$) and $0.86\times10^{22}\;{\rm cm^{-2}}$ (at
phase 0.75). More recently, \citet{Antokhin:2008} performed a
2-temperature fit to the combined data, finding temperatures of
$0.27^{+0.02}_{-0.03}$\,keV and $1.16^{+0.39}_{-0.24}$\,keV.  The
hotter temperature from this fit is greater than the temperature of
the hot component in any of the 5 individual spectra analyzed by
\citet{Naze:2005}. This highlights some of the non-uniqueness issues
of spectral fits to medium-resolution spectra which is well documented
in the literature.

In comparison, two-temperature mekal fits to model cwb2 are generally
pretty poor (typically $\chi^{2}_{\nu} > 2$), with three-temperatures
needed to provide satisfactory fits.  With this in mind, typical
temperatures obtained from two-component fits are 0.25\,keV and 1.4\,keV,
with the analysis of simulated {\em Chandra} spectra returning higher
temperatures for the hotter component ($1.64$\,keV and $1.68$\,keV, 
at phase 0.0 and 0.25, respectively) than the corresponding analysis
of a ``fake'' {\em Suzaku} dataset ($1.07$\,keV and $1.35$\,keV, 
at phase 0.0 and 0.25, respectively). 
%compare Chandra cwb2_fit1 and fit15 to Suzaku cwb2_fit2 and cwb2_fit7
%(respectively kT2=1.64\pm0.06,1.68\pm0.08,1.07+/-0.02,1.35\pm0.02)
Excluding one of the {\em Suzaku} analyzes, the values of $kT_{2}$
considerably exceed those reported from HD\,93161A, even the higher
value found by \citet{Antokhin:2008}. However, this is not too
surprising given that the wind speeds of the O8V and O9V stars in
HD\,93161A are likely to be considerably lower than those assumed from
the O6V stars in model cwb2.

\subsubsection{HD\,47129 (Plaskett's star)}
X-ray observations of systems containing evolved O-stars have also
been presented in the recent literature. One notable analysis concerns
Plaskett's star (HD\,47129), an O8III/I + O7.5III system with a
circular orbit of period 14.4\,d \citep{Linder:2008}.  The stars are
very massive ($54\,\Msol$ and $56\,\Msol$ for the primary and
secondary star respectively, assuming an orbital inclination of
$71^{\circ}$). \citet{Linder:2008} note that the secondary star is
deformed due to its large rotational velocity, which results in a
non-uniform temperature distribution, and speculate that the wind is
confined near the equatorial plane. In this way the apex of the WCR
can occur closer to the primary star (even if its wind has the overall
greater mass-loss rate), as suggested from the analysis of optical
emission lines \citep{Wiggs:1992,Linder:2008}.

{\it XMM-Newton} observations of Plaskett's star were presented by
\citet{Linder:2006}. While there appears to be a minimum in the
observed count rate when the primary star is in front, the amplitude
of variability remains uncertain, as it is quite small between two
{\it XMM-Newton} observations, but larger in archival {\it ROSAT} HRI
observations. There is a pressing need for additional data to resolve
this issue, and to examine whether there is an additional minimum
during each orbit. The interstellar absorbing column is
$1.5\times10^{21}\;{\rm cm^{-2}}$ \citep{Diplas:1994}, similar to the
value assumed for our models. The observed $0.5-10$\,keV luminosity
from the {\it XMM-Newton} observations ($6.9 \times 10^{32}\;\ergps$)
is also similar to the luminosities obtained in models cwb2 and cwb3,
while a minimum in the luminosity near phase 0.0 is again consistent
with the models.

There is some evidence (e.g. a very faint Fe\,K line) that the hot
X-ray emitting plasma may not be in thermal equilibrium, or that there
is a power-law tail to the hard emission. While the latter may
indicate that non-thermal processes, such as particle acceleration,
are occuring, it could also be an artifact of the fitting method:
additional fits using a bremsstrahlung model to fit the continuum plus
individual delta functions to fit the lines did not require a
power-law component. The temperature returned for the bremsstrahlung
component is $2.22\pm0.1$\,keV. Three-component mekal fits return
temperatures of $0.31^{+0.02}_{-0.01}$\,keV,
$0.74^{+0.03}_{-0.02}$\,keV and $2.42^{+0.39}_{-0.16}$\,keV. No extra
column was needed to the softest component, but circumstellar columns of
$4.7^{+0.6}_{-0.7}\times10^{21}\;{\rm cm^{-2}}$ and
$3.0^{+1.3}_{-1.5}\times10^{21}\;{\rm cm^{-2}}$ were needed for the
intermediate and hard components, respectively.

The temperatures reported above are comparable to those obtained by
three-temperature mekal fits to the synthetic spectra generated from
models cwb2 and cwb3, and noted in Table~\ref{tab:fitresults}.  In
addition, a lower column to the harder than to the intermediate
component is consistent with the phase 0.25 results in
Fig.~\ref{fig:cwb123_xray_circum}(e).  However, a specific model of
this system is needed to determine whether all the observational data
can be accurately reproduced.

\subsection{Eccentric/unequal wind systems}
We now discuss systems with eccentric orbits and/or unequal winds,
bearing some similarities to models cwb3 and cwb4.

\subsubsection{HD\,93403}
Four {\it XMM-Newton} observations of HD\,93403, an O5.5I\,+O7V binary with
a 15.093\,d orbit of eccentricity $e=0.234$, reveal flux variations of
20 per cent amplitude in the $0.5-2.5$\,keV band, with a minimum
centered on apastron and a peak at periastron \citep{Rauw:2002}. In
the softer $0.5-1.0$\,keV band the minimum occurs at phase 0.75, when
the denser wind of the O5.5I primary is in front. Variations in the
harder $2.5-10$\,keV band are less significant. \citet{Rauw:2002}
found that HD\,93403 appears less overluminous in X-rays than
previously thought, and suggested that a significant fraction of the
total X-ray emission may arise in line-instability shocks intrinsic to
the winds.  The orbital inclination is thought to be $\sim
30^{\circ}$.

The spectrum obtained nearest to periastron is not markedly softer
than the spectra obtained at other phases.  A two-temperature mekal
fit yields temperatures of $\approx 0.23$\,keV and 1.06\,keV, with the
harder component showing more variability. The observed luminosity (assuming a
distance of 3.2\,kpc) in the $0.5-10$\,keV band is $\approx 1.6 \times
10^{33}\,{\rm erg\;s^{-1}}$.  The circumstellar column 
was found to be $\approx 8 \times 10^{20}\,{\rm cm^{-2}}$
and $\approx 4.6 \times 10^{21}\,{\rm cm^{-2}}$ to the soft and hard
mekal components, respectively.

This system combines features of both models cwb3 and cwb4 (unequal
winds and an eccentric orbit). However, unlike model cwb4, analysis of
the wind ram pressures suggests that the winds may not be able to
sustain a stable balance at periastron, with the result that the more
powerful wind from the O5.5I star overwhelms that of its companion,
and crashes directly onto its surface. Having said this, a full
examination of the potential effects of radiative inhibition and
braking is necessary to strengthen this hypothesis. Assuming that the
primary wind does impact the companion star at periastron, it will be
shocked to high temperatures, and should radiate reasonably
efficiently ($\chi \sim 1$).  A stable wind balance should be possible
at apastron, but whether the O7V star can reestablish its wind towards
its larger companion also remains to be seen.  If a wind balance
occurs, we estimate that both the shocked primary and secondary winds
should be marginally adiabatic ($\chi_{1}\approx2.5, \chi_{2} \approx
5$).

Bearing in mind that the nature of the WCR in this system could be
quite different to that in models cwb3 and cwb4, we compare in the
following the X-ray properties of this system with those from our
models. We first note that the observed luminosity is approximately a
factor of 2 greater than the luminosity from model cwb3, but is
comparable to the luminosity from model cwb4 over most of the orbit,
though the interstellar absorption column to HD\,93403 at
$3.67\times10^{21}\,{\rm cm^{-2}}$ is higher than that assumed in our
models. The circumstellar columns to the mekal components are
comparable to the energy-dependent columns shown in
Fig.~\ref{fig:cwb123_xray_circum}(e). We can also make a direct
comparison to the results of a two-temperature mekal fit to a
simulated 10\,ksec {\em Suzaku} observation of model cwb3 at
$i=90^{\circ}$ and phase 0.0. Though such a fit is poor
($\chi^{2}_{\nu} = 1.81$, with 224 d.o.f.), we find that the fit
temperatures ($kT_{1}=0.26\pm0.02$\,keV, $kT_{2} = 1.21\pm0.05$\,keV)
are nevertheless roughly comparable to the values obtained from
analysis of the observations of HD\,93403. The {\em circumstellar}
column to the cooler component is not in very good agreement ($N_{\rm
H_1}=0.23^{+0.08}_{-0.07} \times10^{22}\,{\rm cm^{-2}}$), though the
column to the hotter component is a better match ($N_{\rm
H_2}=0.31\pm0.08\times10^{22}\,{\rm cm^{-2}}$).

\subsubsection{Cyg~OB2\#8A}
Cyg~OB2\#8A is a longer period (21.9\,d) O6If + 05.5III(f) binary
with an orbital eccentricity $e=0.24$ \citep{DeBecker:2004}. It has
long been known as a very bright X-ray source. {\it XMM-Newton}
observations, together with archive {\it ROSAT} and {\it ASCA} data,
reveal that the observed emission reaches a maximum near phase 0.75, with
a minimum likely to occur shortly after periastron passage
\citep{DeBecker:2006}.  The intrinsic X-ray luminosity varies between
$1.0-1.9 \times 10^{34}\;\ergps$, giving an excess of $\sim 20$
compared to the canonical relationship for single stars\footnote{Note, however,
that this estimate may need to be substantially revised downwards if the
recent distance determination to Cyg~OB2\#5 of $925\pm25$\,pc
\citep{Linder:2009} is found to also be applicable to the other
members of the association.}. Clearly, this
is caused by the collision of two very powerful winds.  The orbital
modulation of the emission matches roughly the $1/d_{\rm sep}$
variation expected if the winds are adiabatic. 
The emission is best fit by a 3-temperature mekal model with
plasma temperatures of approximately $0.24\,$keV, $0.80\,$keV, and
$1.76\,$keV, though the hottest component shows a significant decline
in temperature (by about 15 per cent) between apastron and periastron.

While \citet{DeBecker:2006} estimated that the postshock winds would
be radiative around the entire orbit, the assumed mass-loss rates may
have been too high, and the terminal velocities too low. With
mass-loss rates and terminal velocities more typical of recent
determinations for the stellar types
\citep[e.g.][]{Repolust:2004,Martins:2005a}, we estimate that the
primary wind possibly becomes radiative at periastron though is
marginally adiabatic at apastron, while the secondary wind should
remain adiabatic throughout the orbit.

Since the nature of the WCR in Cyg~OB2\#8A is clearly different to
that in any of the models calculated in Paper~I, it is unsurprising
that the behaviour and characteristics of its X-ray emission do not
closely match those from any of our models.  Nevertheless, it is
interesting to compare this system to our model cwb4 which also
displays variations in its fit parameters (see
Table~\ref{tab:fitresults_cwb4} and Fig.~\ref{fig:fit_statistics}). In
model cwb4 the hot component from 3-temperature fits declines by 65
per cent, a much larger fall than seen from the actual observations of
Cyg~OB2\#8A. That the variability is much lower in Cyg~OB2\#8A is
likely due to smaller changes in the pre-shock wind speeds around the
orbit \citep[see Table~11 in][]{DeBecker:2006} compared to model
cwb4. In addition, the minimum temperature of the hot component in the
fits to model cwb4 actually occurs {\em after} periastron (phase
0.05), this ``time delay'' being due to the history of the plasma
temperature within the WCR.
%being 71 per cent lower than the apastron value. 
It would be interesting to obtain further X-ray observations of
Cyg~OB2\#8A, with better phase coverage, to search for a similar
effect, although the reduced eccentricity and longer orbital period
may shorten or prevent a similar time delay from being seen.  A
specific hydrodynamical model of this system, which would allow
further comparison of its emission in both the X-ray and radio domains
(Cyg~OB2\#8A is also a strongly variable non-thermal radio source),
would clearly be of interest.

\subsubsection{HD\,37043 ($\iota$\,Ori)}
$\iota$\,Ori is a highly eccentric ($e=0.764$), O9III\,+B1III binary
with a 29.134\,d orbital period \citep{Marchenko:2000}, which may have
formed via an exchange of binary components in a binary-binary
collision \citep{Bagnuolo:2001}.  It was observed twice by {\it ASCA}
during 1997, with the observations timed to coincide with periastron
and apastron \citep{Pittard:2000}. The X-ray emission is bright
because of its relative proximity ($D \sim 450$\,pc).  The observed
$0.5-10$\,keV luminosities are $\approx 1.0 \times 10^{32}\,\ergps$ in
both pointings. Since the wind attenuation is negligible, and the
interstellar column is low ($N_{\rm H} = 2\times 10^{20}\,{\rm
cm^{-2}}$), the intrinsic X-ray luminosity of this system is also
relatively low, consistent with the relatively feeble winds in this
system.  Surprisingly, there is no significant variation in the
luminosity, spectral shape, or absorption between the periastron and
apastron observations. Possible explanations for this lack of
variability are that the WCR stays pinned to the surface of the
secondary star throughout the entire orbit, or that intrinsic shocks
within the winds generated by the radiatively driven line deshadowing
instability dominate the emission
\citep{Pittard:1998}. 
%\citet{Lemaster:2007} has shown that
%hydrodynamic models with no perpendicular motion to the line of
%centres produce less X-ray variability and phase-offsets.

\subsection{Non-thermal X-ray emission}
Since accelerated particles exist in many CWBs, the possibility of
inverse Compton X-ray emission has previously been raised by a number
of authors \citep[e.g.][]{Pollock:1987}. Others have argued that
non-thermal X-rays can be produced in the intrinsic wind shocks formed
through the line-deshadowing instability \citep{Chen:1991}. A recent
review of non-thermal emission processes in massive binaries can be
found in \citet{DeBecker:2007}. However non-thermal X-rays might be
produced, their detection requires that their flux become significant
relative to the thermal X-ray flux. Unfortunately, a campaign focused
on O-stars with known non-thermal radio emission failed to
unambiguously detect any non-thermal X-ray's at energies below 10\,keV
\citep[][]{Rauw:2002b,Rauw:2005,DeBecker:2004c,DeBecker:2005,DeBecker:2006}.
Only from HD\,159176 was possible evidence of a non-thermal component
found \citep[]{DeBecker:2004b}. As noted in Sec.~\ref{sec:hd159176},
this system has a very short period. The stellar winds likely collide
at relatively low speeds, which limits the hardness of the thermal
X-ray emission, allowing easier detection of a possible non-thermal
component.

The chance of unambiguously detecting non-thermal X-ray's increases
with energy, as the emission from the thermal component falls off. It
is partly for this reason that non-thermal X-ray emission has been
reported from a {\em Suzaku} observation of $\eta$~Car
\citep{Sekiguchi:2009}.  However, it appears that, at least in the
soft ($E \ltsimm 10\,$keV) domain, non-thermal X-rays are at best a
relatively insignificant component\footnote{A summary of detections
and upper limits of CWBs at MeV, GeV and TeV energies can be found in
\citet{Pittard:2009c}.}

\section{Summary and conclusions}
\label{sec:summary}
This work investigates the X-ray emission from short-period O+O-star
binaries where orbital effects and the acceleration of the winds are
important. The structure and dynamics of the stellar winds and the
wind-wind collision region for four distinct systems were previously
calculated using a 3D hydrodynamical code (Paper~I). We explore the
emission arising from these models, under conditions where the WCR is
radiative (model cwb1) or adiabatic (model cwb2), where the winds have
unequal strengths (model cwb3), and where the orbit is eccentric
(model cwb4).

We find that model cwb1 shows the greatest X-ray variability of the
three models with circular orbits (models cwb1, cwb2 and cwb3), due to
the closer stellar separation and the higher densities surrounding the
WCR. The lightcurves from each of these models are asymmetrical
because of orbit induced aberration and curvature of the WCR. The
asymmetry is greatest for model cwb1, which has the highest ratio of
orbital to wind speeds, $v_{\rm orb}/v_{\rm w}$, of the 3
simulations. The lightcurve asymmetry is always greatest for an
observer located in the orbital plane ($i=90^{\circ}$), and generally
decreases with decreasing $i$ (as does the level of variability).
Model cwb4 shows that the X-ray emission from systems with eccentric
orbits can be spectacularly variable, and can display a strong
hysteresis around the orbit. In general, the emission is softer while
the stars are separating after periastron passage, and harder when the
stars are approaching. The degree of hysteresis, and its duration,
likely depends on the amplitude of changes to the ratio of the orbital
period to the flow time of the shocked plasma out of the system.  As
expected, the increase in the luminosity as periastron is approached
does not follow the $1/d_{\rm sep}$ scaling of an adiabatic, terminal
wind speeds, WCR. The observed change also depends on the energy band and
the observer's orientation. The ratio of the ISM corrected
$0.5-10$\,keV X-ray luminosity to the system bolometric luminosity,
$L_{\rm x}/L_{\rm bol}$, reaches values as high as $5.9\times10^{-6}$
under favourable viewing angles, consistent with a very strong
colliding winds signature. All-in-all, the diversity and richness of
the simulated X-ray lightcurves matches well their observational
counterparts.

We use the difference between the intrinsic and attenuated spectra
from the models to calculate ``effective'' circumstellar absorbing
columns which we demonstrate can be extremely energy
dependent. Therefore, simple spectral fits which assume energy
independent columns are over-simplified in the same way that modelling
the spectrum with only a handful of different plasma temperatures is,
and risk compromising the subsequent interpretation. We also fold our
theoretical spectra through the response files of the {\em Chandra}
and {\em Suzaku} observatories, add poisson noise, and then analyze
using the same method used for the analysis of real data.  In many
cases the fits return circumstellar columns in good agreement with the
energy-dependent ``effective'' columns calculated directly from the
theoretical spectra. However, there are also many instances where the
returned columns have values which are not so close, and/or do not
properly capture the trend of the effective circumstellar column with
energy. We also find that despite our theoretical spectra being
calculated with solar abundances, exclusively sub-solar values
are returned by the spectral fitting package to our simulated spectra
when the global abundance is allowed to fit freely. The global
abundances become closer to the true (solar) value as the complexity
of the spectral model increases. These issues highlight some of the
problems associated with fitting overly simplified models to spectra,
including non-uniqueness of the returned values, and of course
are of wider significance than just CWBs.  Moreover, the intrinsic
luminosities inferred from very short period systems are often
underestimated (by up to a factor of two), since the spectral models
as used are unable to account for occultation losses. Despite these
failings, in many cases the resulting fit parameters are, at least to
first order, similar to those obtained from the analysis of real
observations. More stringent comparisons between theory and
observations will require the modelling of specific systems, such as
work currently being conducted on $\eta$\,Car.

In this paper we have examined the emission from main-sequence
O+O-star systems, in which circumstellar absorption is relatively
minor. Circumstellar absorption becomes more important as the stellar
mass-loss rates increase, so that CWBs which contain a Wolf-Rayet (WR)
star should display more significant orbital modulation of their
lightcurves and spectra than seen in this work.  For this reason,
future work will explore the dynamics and emission from WR+O systems,
where dynamical effects like radiative braking \citep{Gayley:1997}
will also be important. It will also be interesting to explore other
parts of the CWB parameter space. This could include: i) systems where
a weaker wind is completely overwhelmed by a stronger wind (relevant
to HD\,93205, amongst others), ii) systems with an eccentric orbit
where all (or one side) of the shocked plasma in the WCR is radiative
throughout the entire orbit (relevant to systems like HD\,155248, an
O7.5(f)III\,+\,O7(f)III binary with a 5.816\,d orbit of eccentricity
0.127 \citep{Mayer:2001}, which shows strong phase-locked X-ray
emission with an asymmetric modulation \citep{Sana:2004}), iii)
systems with more than two stars (relevant to HD\,167971 and 
QZ\,Car, amongst others).

This work has also been limited to simulating medium resolution X-ray
spectra.  Future work will examine X-ray line profiles, which can
directly probe the dynamics deep in the WCR
\citep*{Henley:2003,Henley:2005,Henley:2008}, and interesting physics
such as non-equilibrium ionization \citep[e.g.][]{Pollock:2005} and
non-equilibrium electron and ion temperatures \citep{Zhekov:2000}.
Further papers will examine the high energy non-thermal emission at
X-ray and $\gamma$-ray energies up to the TeV range.

Despite this long to-do list, the present work has provided further
essential groundwork, highlighting the factors which affect various
aspects of the X-ray emission from short period O+O star binaries.
It lays the foundations for a better understanding of some of the
key physics of such systems (e.g. radiatively driven winds, high
Mach number shocks), which will be studied with future models of
particular systems.

\section*{acknowledgements}
We would like to thank Dave Strickland for providing a code to
generate ``fake'' spectra, and the referee for a constructive and
timely report.  JMP is grateful to the Royal Society for a University
Research Fellowship, while ERP thanks the University of Leeds for
funding.

\label{lastpage}

\end{document}